\documentclass[11pt,a4paper]{article}

\usepackage{setspace}

\usepackage[left=2cm,right=2cm,top=2.5cm,bottom=3.5cm]{geometry}

\usepackage{xcolor}
\usepackage{colortbl}
\definecolor{dark-red}{rgb}{0.80,0.12,0.12} 
\definecolor{dark-green}{rgb}{0,0.60,0.30} 
\definecolor{dark-blue}{rgb}{0,0.1,0.7} 

\usepackage{hyperref}
\hypersetup{
    colorlinks=true,
    linkcolor=dark-red,
    filecolor=blue,      
    urlcolor=cyan,
    citecolor=dark-green
  }

\usepackage{comment}
\usepackage{textcomp}
\usepackage{amssymb,amsmath,amsfonts,mathrsfs,braket}
\usepackage[utf8]{inputenc}
\usepackage{graphicx,longtable,afterpage}
\usepackage{cleveref}
\usepackage{genyoungtabtikz}

\usepackage{tikz}
\usepackage[numbers,sort&compress]{natbib}

\usepackage{floatrow}
\newfloatcommand{capbtabbox}{table}[][\FBwidth]

\def\one{{\,\hbox{1\kern-.8mm l}}}

\newcommand{\SU}{\mathrm {SU}}
\newcommand{\OSp}{\mathrm {OSp}}

\newcommand{\ip}{\mathrm {ip}}

\newcommand{\CH}{\mathcal{H}}
\newcommand{\CC}{\mathcal{C}}

\newcommand{\cO}{\mathcal{O}}

\newcommand{\cN}{\mathcal{N}}

\newcommand{\ula}{\underline{\lambda}}
\newcommand{\umu}{\underline{\mu}}
\newcommand{\unu}{\underline{\nu}}

\allowdisplaybreaks

\def\makeatletter{\catcode`\@=11}
\makeatletter
\def\mathbox#1{\hbox{$\m@th#1$}}%
\def\math@ccstyles#1#2#3#4#5#6#7{{\leavevmode
      \setbox0\mathbox{#6#7}%
      \setbox2\mathbox{#4#5}%
      \dimen@ #3%
      \baselineskip\z@\lineskiplimit#1\lineskip\z@
      \vbox{\ialign{##\crcr
             \hfil \kern #2\box2 \hfil\crcr
             \noalign{\kern\dimen@}%
             \hfil\box0\hfil\crcr}}}}
\def\mathaccstyles{\math@ccstyles\maxdimen}
\def\maththroughstyles{\math@ccstyles{-\maxdimen}}
\def\unity%
 {\maththroughstyles{.45\ht0}\z@\displaystyle {\mathchar"006C}\displaystyle 1}

\def\AdS{\text{AdS}}

\def\CC{{\cal C}}
\def\DD{{\cal D}}

\def\FF{{\cal F}}
 
\def\HH{{\cal H}}

\def\MM{{\cal M}}
\def\NN{{\cal N}}
\def\OO{{\cal O}}
\def\PP{{\cal P}}

\def\WW{{\cal W}}

\def\R{{\RR}}

\def\beq{\begin{equation}}
\def\eeq{\end{equation}}
\newcommand{\bea}{\begin{eqnarray}}
\newcommand{\eea}{\end{eqnarray}}
\def\bal{\begin{align}}
\def\eal{\end{align}}

\numberwithin{equation}{section}

\begin{document}

\thispagestyle{empty}

~\\[-2.25cm]

\begin{center}
\hfill{QMUL-PH-26-41}
\vskip 0.5truecm 
{\LARGE\bf
    {Exploring M-theory on AdS$_7\times$S$^4$: \\[.2cm] Mellin amplitudes with composite operators }}
\vskip 1.25truecm
	{\bf Francesco Aprile$^{1}$, Hector Puerta-Ramisa$^{2}$, Mitchell Woolley$^{3,4}$ \\
	}
		\vskip 0.4truecm
	{\it
		$^{{1})}$ Departamento de F\'{i}sica Te\'{o}rica \& IPARCOS, Facultad de Ciencias F\'{i}sicas,\\
		Universidad Complutense, 28040 Madrid, Spain\\\vskip .2truecm
		$^{{2})}$ Department of Mathematical Sciences, \\ Durham University, DH1 3LE Durham, UK \\\vskip .2truecm
        $^{{3})}$ 
        Centre for Theoretical Physics, Department of Physics and Astronomy, \\
        Queen Mary University of London, E1 4NS London, UK \\\vskip .2truecm
		$^{{4})}$ Deutsches Elektronen-Synchrotron DESY, \\
        Notkestr. 85, 22607 Hamburg, Germany  ~ \\\vskip .2truecm	
		\vskip .2truecm   }
	\vskip .5truecm

\textit{E-mail:} ~\href{mailto:faprile@ucm.es}{{\tt faprile@ucm.es}},~\href{mailto:hector.puerta-ramisa@durham.ac.uk}{{\tt hector.puerta-ramisa@durham.ac.uk}},~\href{mailto:woolleymitchell@gmail.com}{{\tt woolleymitchell@gmail.com}}
\end{center}

\vskip 1.25truecm

\centerline{\bf Abstract}
{\baselineskip=15pt
\vskip .4truecm
\noindent

We bootstrap mixed correlators in M-theory on $\mathrm{AdS}_7\times \mathrm{S}^4$, focusing on tree-level correlation functions involving one double-particle operator and three single-particle operators. Combining inputs from AdS/CFT, the chiral algebra, 
and the OPE  
at large central charge, we uniquely determine the Mellin amplitude for
$
\langle [{\cal O}_{2}{\cal O}_{q}]{\cal O}_{2}{\cal O}_{p}{\cal O}_{p+q-2}\rangle
$
for generic $p$ and $q$. While our main focus is the $\mathcal N=(2,0)$  SCFT in six dimensions, we also derive the analogous result in $\mathrm{AdS}_5\times \mathrm{S}^5$, applying the same bootstrap approach to $\mathcal N=4$ SYM. Despite their different dynamics, the resulting Mellin amplitudes exhibit a strikingly similar structure, with non-trivial OPE predictions for low-twist protected operators providing a key input.
Our work is motivated by recent developments in the study of Kaluza–Klein correlators in bubbling geometries, where four-point functions involving single- and multi-particle operators are computed as response functions to smooth deformations of the vacuum geometry. While well tested in $\mathrm{AdS}_5\times \mathrm{S}^5$, we aim to extend this framework to M-theory.

}

{		
\hypersetup{linkcolor=black}
\newpage
\setcounter{page}{1}\setcounter{footnote}{0}
\setcounter{tocdepth}{2}
\tableofcontents
}

		\baselineskip=17pt
		\parskip=5pt

\newpage

\setcounter{tocdepth}{2}

\section{Introduction} \label{intro}

Analytic and numerical bootstrap techniques have significantly advanced our 
understanding of interacting conformal field theories, particularly in the strongly 
coupled regime. In this regard, the conformal bootstrap offers a complementary 
perspective on the microscopic foundations of the AdS/CFT correspondence, 
providing a (non-Lagrangian) framework for investigating how quantum gravity emerges 
from strongly coupled CFTs. In this context, theories  with maximal supersymmetry 
play a particularly important role, offering concrete realisations of this duality.

In recent years, the main focus has been on the study of four-point functions of half-BPS single-particle operators,
\begin{equation}\label{introintro4pt}
\langle {\cal O}_{p_1} {\cal O}_{p_2} {\cal O}_{p_3} {\cal O}_{p_4}\rangle.
\end{equation} 
The ${\cal O}_{p}$ are superconformal primary operators dual to the Kaluza-Klein 
fields of the gravity theory and provide the natural starting point.  
The four-point functions above are then the simplest correlation functions 
that are not protected by supersymmetry and therefore receive quantum corrections.
The tree-level correlators \eqref{introintro4pt} are known 
for all maximally supersymmetric theories \cite{Alday:2020dtb}.
Although these results were not obtained through explicit bulk computations, 
at least for some theories, it is understood how the corresponding bulk calculations can be 
organised diagrammatically \cite{Arutyunov:2018tvn}. Moreover, thanks to the flat-space 
limit established in \cite{Penedones:2010ue}, these bulk computations admit a natural 
interpretation in the language of scattering amplitudes.

However, the spectrum of the theory is much richer than that of single-particle operators 
alone. Owing to the special properties of half-BPS operators, analytic functions of the 
single-particle operators define states in the Hilbert space of the theory and correspond 
to valid half-BPS operators. The gravitational counterpart of this statement is provided 
by the beautiful classification of Lin, Lunin and Maldacena (LLM) in \cite{Lin:2004nb}, which showed that the 
full spectrum of half-BPS states admits a holographic description in terms of smooth bubbling geometries.

A particularly interesting class of geometries consists of deformations of 
the vacuum AdS$\times$S solution, represented by a circular droplet.  In the CFT, single- 
and multi-particle operators provide the building blocks for these shape deformations. 
Prototypical examples are superpositions such as 
\begin{equation}
\begin{tikzpicture}
  \colorlet{ink}{blue!70!cyan}
  \def\R{2}   
  \def\A{0.45} 
  \def\N{3}    
  \filldraw[ink, thick, fill=ink!25,scale=.16 ]
    plot[domain=0:360, samples=400, smooth, variable=\t]
      (\t:{\R+\A*(1-cos(\N*\t))/2});
   \draw (2.4,0) node {$\Big\rangle \simeq\, \alpha\, {\cal O}_{q} + \alpha^2\, {\cal O}_q^2 +\ldots$};
   \draw (-.5,0) node {$\Big|$};
\end{tikzpicture} 
\end{equation}
Examples include ripples \cite{Grant:2005qc,Turton:2025svk} and 
exponential/coherent operators \cite{Aprile:2025hlt,Anempodistov:2025maj}. 
From this perspective, the four-point correlators in \eqref{introintro4pt} 
can be understood as heavy-heavy-light-light correlators \cite{Abajian:2023jye,Aprile:2024lwy}, in which two of the 
single-particle operators serve as building blocks of the heavy operators, 
in a regime of small parameters,  while the remaining two single-particle operators act as light probes. 
The next-to-leading-order response to a bubble deformation is captured by the new class of correlators: 
\begin{equation}\label{class_with_oned}
\langle [{\cal O}_{r_1}{\cal O}_{r_2}] {\cal O}_{p_2} {\cal O}_{p_3} {\cal O}_{p_4}\rangle\,
\end{equation}
involving the insertion of a double-particle operator $[{\cal O}_{r_1}{\cal O}_{r_2}]$.

On the CFT side, the correlators \eqref{class_with_oned} are also particularly 
interesting for two reasons. First, they can be viewed as five-point functions 
of single-particle correlators in disguise, obtained by taking the coincident-point 
limit where ${\cal O}_{r_1}$ approaches ${\cal O}_{r_2}$. Second, it was pointed out 
in \cite{Aprile:2026uxe} that, at tree-level, these correlators satisfy OPE constraints 
similar to those obeyed by four-point functions of single-particle operators,
albeit with crucial differences. We will elaborate on these OPE constraints 
and the associated bootstrap approach in the following.

To summarise our introduction so far:
the study of the correlators $\langle [{\cal O}_{r_1}{\cal O}_{r_2}] {\cal O}_{p_2} {\cal O}_{p_3} {\cal O}_{p_4}\rangle$ 
is motivated by several complementary approaches. On the one hand, there is the 
geometric picture provided by LLM, while on the other, there is the CFT perspective. 
Having access to both is particularly valuable when one of the corresponding 
solution techniques is unavailable.

We study both type IIB string theory and M-theory through their respective 
dualities with four-dimensional ${\cal N}=4$ SYM and the non-Lagrangian six-dimensional $\cN=(2,0)$ SCFT. 
This parallels the discussion of LLM, where both holographic settings are considered. 
While ${\cal N}=4$ SYM provides a well-established framework, for which many results are 
already known, our main focus will be on M-theory. To put this motivation into context, 
let us first review the current state of the art in type IIB supergravity before 
comparing it with the situation in M-theory.

In the four-dimensional ${\cal N}=4$ SYM theory 
and its type IIB supergravity dual, the works of \cite{Aprile:2024lwy,Aprile:2025hlt,Aprile:2026uxe} 
showed that both the supergravity and bootstrap approaches are viable and yield consistent results. 
To illustrate both, these works focused primarily on the simplest LLM geometry, an ellipse
with major and minor axes given by \cite{Aprile:2026urq}
\begin{equation}
L_{+}^2= \frac{1+\sqrt{2}\alpha}{1-\sqrt{2}\alpha},\qquad\qquad L_{-}^2= \frac{1-\sqrt{2}\alpha}{1+\sqrt{2}\alpha}\,.
\end{equation}
This background can be described by a consistent truncation ansatz \cite{Cvetic:2000nc, Liu:2007xj,Giusto:2024trt} and corresponds 
in the CFT to the state 
$\exp(\frac{\alpha}{\sqrt{2}}\,\! {\cal O}_2)$. 
As discussed above, the small-$\alpha$ expansion of light-light two-point functions 
in this background generates four-point correlators 
with insertions of the multi-particle operators $({\cal O}_2)^n$.\footnote{The gravity 
approach \cite{Aprile:2025hlt}  was also applied to solutions in AdS$_3\times$S$^3$ holography:
KST \cite{Kanitscheider:2007wq} and the Special Locus 
\cite{Ganchev:2021pgs}. The study of multi-particle correlators in AdS$_3\times$S$^3$ 
was initiated in \cite{Ceplak:2021wzz}.} With one $({\cal O}_2)^2$ insertion, the supergravity
and the bootstrap approaches were shown to be consistent with each other, and with the corresponding tree-level 
five-point functions of single-particle operators, computed in \cite{Fernandes:2025eqe}, 
upon taking a coincident limit.

In M-theory, by contrast, no corresponding computations have been carried out to date.
Even in the simplest case, based on the seven-dimensional consistent truncation 
of \cite{Chong:2004ce}, the situation is different: one simplification in the 
supergravity approach of \cite{Aprile:2024lwy,Aprile:2025hlt} was the use of 
a Ward identity connecting the single-particle 
operators to the Kaluza-Klein modes of the dilaton $\Phi$, whose bulk e.o.m.~is $\Box\Phi=0$.
There is no analogue of this in the six-dimensional $(2,0)$ theory. Consequently, 
the bulk computation must be approached differently from the outset. Although 
this obstruction is technical in nature,\footnote{It could be resolved 
with additional work, largely mechanical.}  establishing a set of simple 
benchmark results at this stage would 
be conceptually valuable and could guide future investigations of precision holography in M-theory.

Fortunately, the aforementioned bootstrap approach provides a complementary 
avenue: with one double-particle insertion $[{\cal O}_{r_1}{\cal O}_{r_2}]$, it 
applies equally well to both the four-dimensional ${\cal N}=4$ SYM theory and 
the six-dimensional $(2,0)$ theory. The underlying CFT logic is simple.
At large central charge, the leading order OPE of single-particle operators closes on the 
double-particle sector. This is a key feature of the AdS/CFT 
correspondence and allows the leading-order three-point couplings and anomalous 
dimensions to be extracted from four-point functions of single-particle operators. 
Consequently, whenever a double-particle operator is exchanged in another correlator, 
its contribution can be predicted. 
We refer to this methodology as \emph{the double-particle bootstrap}.

{\bf In this paper}, we establish our benchmark for future studies of bubbling geometries 
in M-theory by focusing on the next-to-next-to-extremal (N$^2$E) correlators
\begin{equation}\label{we_boot_these}
\langle [{\cal O}_{2} {\cal O}_{q}] {\cal O}_{p+q-2}  {\cal O}_{2}  {\cal O}_{p}  \rangle\,,\qquad\qquad q\ge 2\,.
\end{equation}
We shall see that the tree-level correlators, while retaining the simplicity of N$^2$E four-point kinematics, 
exhibit a richer dynamical structure than their single-particle counterparts.  

Although these correlators with one double-particle insertion 
receive contributions from the exchange of triple-particle operators, the double-particle 
bootstrap uniquely determines the OPE within a twist ``window'', that extends 
from the unitarity bound to the onset of triple-particle exchanges. At the unitarity bound, the 
relevant operators that are key to our approach are the protected semi-short double-particle operators. In the long sector, 
they are the generalization of those studied in \cite{Aprile:2018efk}. 
Remarkably, we shall see that the window data for the correlators in \eqref{we_boot_these} are sufficient to uniquely bootstrap 
the tree-level contribution.

{\bf Outline.}
In section \ref{sec_4pt_generalities} we set our conventions and describe general 
features of correlation functions  of half-BPS operators in the four-dimensional
$\NN=4$ theory and the six-dimensional $\cN=(2,0)$ theory. We review the consequences 
of the superconformal Ward identities and describe the kinematics of the reduced 
correlator formalism. This is done in a unified manner for both theories by introducing the parameter $\theta=\frac{d-2}{2}$.\footnote{Our discussion also applies to the three-dimensional 
theory although we leave it for future work.} 

In section~\ref{sec:doublebootstrap}, we describe the double-particle bootstrap and the 
predictions for our correlators. 
Unlike the case 
with four single-particle external operators, originally studied in \cite{Dolan:2006ec}, semi-short operators 
at the unitarity bound contribute non-trivially and they are key to our bootstrap approach. 
This feature was first identified in \cite{Aprile:2026uxe,Aprile:2025nta} in the context of the four-dimensional 
$\NN=4$ SYM theory. 
However, the six-dimensional $\cN=(2,0)$ theory exhibits dynamical features with no 
analogue in $\NN=4$ SYM, making it particularly satisfying to see that the double-particle 
bootstrap nevertheless applies in a similar way. 

In section~\ref{formulae_freet}, we develop the first part of our bootstrap programme. 
Our main interest is in the reduced correlator, which encodes the dynamical information. 
As an intermediate step though, we compute various free-like contributions 
at strong coupling. We do so using the half-BPS OPE, and in particular we present a 
non-trivial application of the double-particle limit in the six-dimensional $\cN=(2,0)$ theory.
As an independent check, we derive the same result using the chiral algebra projection of \cite{Beem:2014kka}.
In section~\ref{sec:cauchy}, we explain how to decompose these free-like contributions into 
superconformal blocks in order to assemble the OPE predictions. We further show that the traditional brute-force 
approach is \emph{unnecessary}: the mathematical structure underlying superconformal 
blocks allows us to determine the relevant block coefficients using 
the simpler and more efficient combinatorics of a Cauchy-like identity.

Finally, in section~\ref{sec_results} we put together the OPE predictions and we 
bootstrap the Mellin amplitudes
corresponding to the reduced correlators in \eqref{we_boot_these}, namely,
\begin{equation}
{\cal H}_{[2q],2,p+q-2,p},\qquad\qquad q\ge 2\,.
\end{equation}
Despite the dynamical differences between four-dimensional ${\cal N}=4$ SYM 
and the six-dimensional $(2,0)$ theory, we have been able to present the results 
in a unified form across dimensions by introducing the parameter $\theta$ in 
the Mellin amplitude. Remarkably, the final Mellin amplitudes for $\theta=1,2$ 
exhibit striking similarities. While the flat-space limit of Penedones \cite{Penedones:2010ue} 
provides some intuition for a similar result involving single-particle operators \cite{Alday:2020dtb}, 
it is not obvious that the same reasoning extends to correlators involving a 
double-particle operator. We conclude in section \ref{outlook} with a discussion 
of the main features of our findings and an outlook on future directions.

\section{Four-point correlators of BPS operators}\label{sec_4pt_generalities}

We will be studying four-point functions of half-BPS superconformal primary operators 
in the maximally supersymmetric theories in four and six dimensions: ${\cal N}=4$ 
SYM theory in $d=4$ and the ${\cal N}=(2,0)$ theory in $d=6$. 
The half-BPS operators we are interested in are scalars under the Lorentz group and 
transform in the rank-$p$ symmetric traceless representation
of the  $R$-symmetry group. Their dimension is protected by supersymmetry and is given by
$\Delta= \theta p$, where we introduce the parameter
\begin{equation}
    \theta=\frac{d-2}{2},
\end{equation}
which will be ubiquitous throughout the paper. 

To deal with the traceless symmetric $R$-symmetry tensors, we introduce an auxiliary null complex vector 
$Y^I$ and define the index-free scalar operator
\begin{align}\label{eq:singleparticle}
    O_p(X,Y)\equiv O_{I_1\dots I_p}(X)\;Y^{I_1}\cdots Y^{I_p}.
\end{align}
In the free theory, the half-BPS superconformal primary operators $O_{I_1\dots I_p}$ are described by
gauge invariant operators built out of the scalar fields (bottom components) of the corresponding supersymmetric multiplets. 
In $d=4$ we have the six scalars of the ${\cal N}=4$  
multiplet while in $d=6$ we have the five scalars of  ${\cal N}=(2,0)$ tensor multiplet.
The whole multiplet construction can be carried out in superspace 
by identifying a unique fundamental analytic superfield $W$. 
Half-BPS superconformal multiplets 
are then constructed from polynomials in $W$, e.g.~$W^p$  
\cite{Howe:2000nq,Heslop:2001zm,Heslop:2004du,Doobary:2015gia,Heslop:2022xgp}, and more general multiplets are given by applying differential operators to $W$ \cite{Doobary:2015gia,Hansen:2025lig}.

In the regime where the conformal field theories are holographic, half-BPS operators describe both \emph{single-} and \emph{multi-particle} states in the bulk.
The former, denoted hereafter by $\cO_p$, are dual to the Kaluza-Klein modes of the relevant $\mathrm{S}^5$ or $\mathrm{S}^4$ compactifications of string or M-theory.
The latter are defined from the Operator Product Expansion (OPE) of two 
single-particle operators $\cO_p\times \cO_q$, appropriately projected onto the half-BPS sector. 
The space of all half-BPS operators can be described by considering 
arbitrary analytic functions on the basis of single and multi-particle operators. 
A particularly nice geometrical realization of the space of all half-BPS operators in the holographic description is given by the LLM construction \cite{Lin:2004nb}.

For a given $R$-symmetry representation, the half-BPS operators that belong to that representation are degenerate. 
The only exceptions are the stress tensor multiplet, corresponding to ${\cal O}_{2}$, and the operator ${\cal O}_3$.\footnote{Also $\cO_1$ though we will consider the strongly-coupled regime in which it decouples.}
From $p\ge 4$, the space of half-BPS operators contains the single-particle operators ${\cal O}_p$ 
and the multi-particle operators labelled by a partition of $p$.  

The discussions of this section are kinematical and thus apply to both single- and multi-particle operators.
We will use $O_p$ to label a generic half-BPS operator as defined in \eqref{eq:singleparticle}.

\subsection{Structure of the correlators}

Our notation for the two-point functions is 
\begin{align}\label{two_point_conventions}
    \langle O_{p}(X_1,Y_1)O_{q}(X_2,Y_2)\rangle= \delta_{p,q}\left(g_{12}\right)^pf_{p}\left(c\right) ,
    \hspace{1.5cm}g_{ij}=\frac{Y_{ij}}{(X^2_{ij})^\theta},
\end{align}
where $X_{ij}=X_i-X_j$, $Y_{ij}=Y_i\cdot Y_j$ and
 we defined the two-point function normalization as follows,
\begin{equation}
\left|O_{p}\right|^2\equiv f_{p}\left(c\right).
\end{equation}

Our notation for the four-point functions is
\begin{align}
    \langle O_{p_1}(X_1,Y_1)O_{p_2}(X_2,Y_2)O_{p_3}(X_3,Y_3)O_{p_4}(X_4,Y_4) \rangle =&\;
    \mathcal{P}_{\vec{p}}\left[g_{ij}\right]\;\mathcal{C}_{\vec{p}}\left(U,V;\hat{\sigma},\hat{\tau}\right),
\label{eq:full4pt}
\end{align}
where $\vec{p}=(p_1,p_2,p_3,p_4)$, $\mathcal{P}_{\vec{p}}\left[g_{ij}\right]$ is a given prefactor of two-point function normalisations $|O_{p_i}|$ and propagators $g_{ij}$ as follows\footnote{We assumed $p_{12},p_{43}\ge 0$ without loss of generality}
\begin{equation}
    \PP_{\vec{p}}= |O_{p_1}| |O_{p_2}||O_{p_3}||O_{p_4}|\times g_{12}^{\frac12 (p_1+p_2)} g_{34}^{\frac12 (p_3+p_4)} \left(\frac{g_{14}}{g_{24}}\right)^{\frac12 p_{12}}\left(\frac{g_{14}}{g_{13}}\right)^{\frac12 p_{43}},
\end{equation}
and
 $\mathcal{C}_{\vec{p}}\left(U,V;\hat{\sigma},\hat{\tau}\right)$ is the \textit{normalised} four-point correlator. 
 By conformal and $R$-symmetry invariance, the latter is only a function of cross-ratios 
\begin{align}\label{eq:crossratios}
    U=\frac{X_{12}^2X_{34}^2}{X_{13}^2X_{24}^2},\qquad V=\frac{X_{14}^2X_{23}^2}{X_{13}^2X_{24}^2},\qquad \hat\sigma=\frac{Y_{13}Y_{24}}{Y_{12}Y_{34}},\qquad \hat\tau=\frac{Y_{14}Y_{23}}{Y_{12}Y_{34}}.
\end{align}
Note also that, by construction, $\mathcal{C}_{\vec{p}}\left(U,V;\hat{\sigma},\hat{\tau}\right)$ is a polynomial in $\hat\sigma, \hat\tau$ of degree 
 \begin{align}\label{eq:kappa}
     \kappa=\sum_{ij=\{12,13,14\}}\min\left[\frac{p_i+p_j}{2},\Sigma-\frac{p_i+p_j}{2}\right]-\Sigma,
\end{align}
with $\Sigma\equiv\frac{p_1+p_2+p_3+p_4}{2}$. We will refer to $\kappa$ as the degree of extremality of the correlator.

It will be useful to introduce symmetric variables $x_i$, $y_i$ to describe the conformal and $R$-symmetry cross-ratios:
\begin{align}
    U=x_1x_2,\hspace{1cm} V=(1-x_1)(1-x_2), \hspace{1cm}\hat\sigma=\frac{1}{y_1y_2},\hspace{1cm}\hat\tau=\frac{(1-y_1)(1-y_2)}{y_1y_2}.
\end{align}
The symmetric variables $x_i$, $y_i$ have a natural interpretation in analytic superspace as eigenvalues of a cross-ratio (super)matrix, see e.g.~\cite{Doobary:2015gia,Aprile:2021pwd,Hansen:2025lig}. 

It is well known that superconformal invariance puts stronger constraints on the four-point functions. 
In particular, it was shown in \cite{Eden:2000bk,Heslop:2003xu,Dolan:2004mu,Heslop:2004du} 
that any four-point function of half-BPS operators must satisfy, non-perturbatively, the Ward identities
\begin{align}\label{Ward_Identity_Constraints}
    \left(x_i\frac{\partial}{\partial x_i}  + \theta y_j\frac{\partial}{\partial y_j}\right)\CC_{\vec{p}}\bigg|_{x_i=y_j}=0\,.
\end{align}
A useful way to solve \eqref{Ward_Identity_Constraints} is to split the correlator into two contributions:
\begin{align}
    \mathcal{C}_{\vec{p}}\left(U,V;\hat{\sigma},\hat{\tau}\right)=\FF_{\vec{p}}\left(U,V;\hat{\sigma},\hat{\tau}\right)+\Upsilon_{\theta}\circ\HH_{\vec{p}}\left(U,V;\hat{\sigma},\hat{\tau}\right),
\end{align}
where the first contribution, $\FF_{\vec{p}}$, contains the ``free-like" monomials in the cross-ratios, while the second contribution is seeded by the so-called reduced correlator, 
 $\HH_{\vec{p}}$. The latter can be written as a polynomial in $\hat{\sigma},\hat{\tau}$ 
of degree $\kappa-2$, i.e.~two units lower than that of the full correlator. The coefficient functions of each monomial $\hat\sigma^a\hat\tau^b$ contributing to $\HH_{\vec{p}}$ are unconstrained functions of $U$ and $V$, and
the operator $\Upsilon^{(\theta)}$ is 
constructed precisely so as to solve the kinematics of the Ward-identity constraints
 \eqref{Ward_Identity_Constraints}, see e.g.~\cite{Nirschl:2004pa,Beem:2015aoa,Woolley:2026cii}.

In this paper we focus on correlators satisfying $\kappa=2$ \eqref{eq:kappa}, known as next-to-next-to-extremal (N$^2$E) correlators.
In this case, the kinematics are well understood: 
the reduced correlator does not depend on the $R$-symmetry, that is,
\begin{equation}
   \left. {\cal H}_{\vec{p}}\left(U,V;\hat{\sigma},\hat{\tau}\right)\right|_{\kappa=2}={\cal H}_{\vec{p}}\left(U,V\right),
\end{equation}
and as a result the operator $\Upsilon_{\theta}$ takes the following explicit form,
\begin{align}\label{Upsilon}
    \left.\Upsilon_{\theta}\circ \HH_{\vec p} \right|_{\kappa=2}=\frac{(x_1 x_2)^{2\theta}}{(y_1y_2)^2}
    D_\theta \Bigg[ 
    \left(x_1-y_1\right)\left(x_1-y_2\right)\left(x_2-y_1\right)\left(x_2-y_2\right)\HH_{\vec p}\Bigg],
\end{align}
where 
\begin{equation}\label{eq:D2}
    D_1 =1
    \qquad\qquad
    D_2 = \partial_{x_1} \partial_{x_2} - \frac{2}{x_1-x_2} \left(\partial_{x_1} - \partial_{x_2}\right).
\end{equation}
We further note the relation
\begin{equation}\label{intriligator}
\frac{ \prod_{i\neq j}(x_i-y_j) }{ (y_1 y_2)^2} = V+V(V-U-1)\hat\sigma +(1-U-V)\hat\tau+ U V \hat\sigma^2 +U(U-V-1)\hat\sigma\hat\tau+U\hat\tau^2.
\end{equation}

Permuting the operators $O_{p_i}(x_i,y_i)$ in the correlator \eqref{eq:full4pt} 
imposes non-trivial crossing relations: 
\begin{equation}
    \begin{aligned}
\mathcal{C}_{p_1,p_2,p_3,p_4}(U,V;\hat{\sigma},\hat\tau)=\;\mathcal{C}_{p_2,p_1,p_3,p_4}\left(\frac{U}{V},\frac{1}{V};\hat\tau,\hat\sigma\right)
&=\;\left(\frac{U^{\theta}}{V^\theta}\hat\tau\right)^{\!\!\kappa}\mathcal{C}_{p_1,p_4,p_3,p_2}\left(V,U;\frac{\hat\sigma}{\hat\tau},\frac{1}{\hat\tau}\right) \\
&=\;\left(U^\theta\hat\sigma\right)^{\!\!\kappa}\mathcal{C}_{p_1,p_3,p_2,p_4}\left(\frac{1}{U},\frac{V}{U};\frac{1}{\hat\sigma},\frac{\hat\tau}{\hat\sigma}\right).
\label{eq:fullcross}
\end{aligned} 
\end{equation}
The above relations can be lifted to relations for the reduced correlator. 
In our case, $\kappa=2$,  
we can use the invariance of $D_\theta$ under $2\leftrightarrow4$ crossing, as well as 
\begin{align}
    \left(D_\theta\right)\Big|_{1\leftrightarrow2}=V^{2\theta}\left(D_\theta\right)V^{-2},\hspace{1.5cm}\left(D_\theta\right)\Big|_{2\leftrightarrow3}=U^{2\theta}\left(D_\theta\right)U^{-2},
\end{align}
to arrive at 
\begin{align}
    \HH_{p_1,p_2,p_3,p_4}\left(U,V\right)=\frac{1}{V^{4}}\HH_{p_2,p_1,p_3,p_4}\left(\frac{U}{V},\frac{1}{V}\right)=\HH_{p_1,p_4,p_3,p_2}\left(V,U\right)=\frac{1}{U^{4}}\HH_{p_1,p_3,p_2,p_4}\left(\frac{1}{U},\frac{V}{U}\right).
\end{align}

Before moving on, let us mention two important facts. 

First, 
solving \eqref{Ward_Identity_Constraints} for correlators with higher degree 
of extremality, $\kappa>2$, is straightforward in four dimensions ($\theta=1$). 
The solution has the same structure as in \eqref{Upsilon} owing to the fact that $D_1=1$ always. Thus, the polynomial \eqref{intriligator} 
provides the main kinematical block for all $\kappa\geq2$. 
In six dimensions ($\theta=2$),  
the solution is more involved due to the non-trivial form of $D_2$. A form of the solution can be found in \cite{Dolan:2004mu,Rastelli:2017ymc}.  

Second, the Ward identities \eqref{Ward_Identity_Constraints} play a distinctive role 
in the study of superconformal blocks. In particular, in the polynomial literature they are known as 
quasi-invariance conditions and they are used to define the so-called symmetric 
super-Jack polynomials \cite{SergeevVeselov2005GeneralisedDiscriminants}. 
Writing superconformal blocks in terms of these polynomials \cite{Aprile:2021pwd} plays a crucial role in the CFT analysis performed in this paper.
We include the explicit form of the relevant superconformal blocks in appendix \ref{app:blocks}.
The super-Jack expansion will be key in section \ref{sec:cauchy}.

\subsection{Half-BPS OPE}\label{subsec:OPE}

In this section, we tabulate the operators exchanged in the OPE of two half-BPS operators. Our approach is based on the analytic superspace techniques developed in \cite{Heslop:2001zm,Heslop:2004du,Doobary:2015gia,Aprile:2021pwd,Hansen:2025lig}, which allow us to describe all maximally supersymmetric theories in a unified framework (including the theory in three dimensions although we will not discuss it in this paper). We will see that this provides a key input for the bootstrap calculations in later sections. Here, we focus exclusively on the representation content; the space of exchanged operators that will be relevant to our bootstrap computations is described in section \ref{sec:doublebootstrap}.

All superconformal multiplets in the OPE of two half-BPS operators are labelled by a charge $\gamma$ and a Young diagram $\ula$:
\begin{equation}
    O^\gamma_{\ula}\, .
\end{equation}
In maximally supersymmetric theories, the Young diagram can have up to two rows and up to two columns of arbitrary length, but the length of the third row cannot exceed 2, that is, $\ula$ is of the following form,
\begin{equation}\label{typicaldiagram}
   \ula \ =\  \begin{tikzpicture}[scale=0.65, x=1.3cm, y=1.3cm, baseline=(current bounding box.center)]
    \draw [thick] (-1,-1) rectangle (0,-2);
    \draw [thick] (-1,-2) rectangle (-0.5,-3);
        \draw [thick] (-1,0) rectangle (3,-0.5);
        \draw [thick] (-1,-0.5) rectangle (2,-1);
        \draw [<->,thick] (-1,0.2)--(3,0.2) node[midway,above] {$\lambda_1$};
        \draw [<->,thick] (-1.2,0)--(-1.2,-3) node[midway,left] {$\mu_1$};
        \draw [<->,thick] (3.2,0)--(3.2,-2) node[midway,left] {$\mu_2$};
        \draw [<->,thick] (-1,-3.2)--(2,-3.2) node[midway,above] {$\lambda_2$};
        \draw [dashed] (2,-1)--(2,-3.2);
        \draw [dashed] (0,-2)--(3.2,-2);
    \end{tikzpicture} \ = \ [\lambda_1,\lambda_2, 2^{\mu_2-2},1^{\mu_1-\mu_2}] .
\end{equation}
Once $\gamma$ and $\ula$ are given, the dimension $\Delta$ and twist $\tau$ of the superprimary ${\cal O}_{\ula}^{\gamma}$ are 
\begin{equation}\label{eq:dimensionandtwist}
    \Delta= \theta \gamma+ \lambda_{1}+\lambda_2-4, \qquad \tau=\theta \gamma +2\lambda_2-4,
\end{equation}
the spin is 
\begin{equation}
    l=\lambda_1 -\lambda_2,
\end{equation}
and $R$-symmetry Dynkin labels are
\begin{equation}
    \theta=1: \quad [\mu_1-\mu_2,\gamma-2\mu_1,\mu_1-\mu_2]_{\mathfrak{su}(4)}, \qquad \theta=2: \quad [\gamma-2\mu_1,2(\mu_1-\mu_2)]_{\mathfrak{usp}(4)},
\end{equation}
where we use the Dynkin labels convention of \cite{Aprile:2021pwd}.
Since both $R$-symmetry representations ultimately depend on the same two quantities, $a=\mu_1-\mu_2$ and $b=\gamma-2\mu_1$, we introduce the following unified notation,
\begin{equation}\label{eq:Rsym}
    [a,b]_\theta = \left\{\begin{aligned}
        &\ [a,b,a]_{\mathfrak{su}(4)}, & \qquad  \theta=1, \\
        &\ [b,2a]_{\mathfrak{usp}(4)}, & \theta=2. \\
    \end{aligned} \right.
\end{equation}

If $\ula$ is a subdiagram of the single row and column hook ,
\begin{equation}
   [\lambda,1^\mu]= \begin{tikzpicture}[scale=0.65, x=1.3cm, y=1.3cm, baseline=(current bounding box.center)]
    \draw [thick] (-0.5,-0.5) rectangle (0,-2);
    \node at (-0.25,-1.75) {$\mu$};
        \draw [thick] (-0.5,0) rectangle (3,-0.5);
        \node at (2.75,-0.25) {$\lambda$};
    \end{tikzpicture} ,
\end{equation}
then the superconformal multiplet satisfies $Q$-shortening conditions, classified by the table below.
\begin{equation}\label{tableQN1}
\begin{array}{|c|c|c|c|c|}
\hline
\textrm{Multiplet} & \ula & \tau &  l & R_\theta \\[.3ex]
\hline
\hline
\frac{1}{2}\textrm{-BPS}  & [\varnothing] &\rule{0pt}{.4cm}\theta\gamma & 0 & [0,\gamma]_{\theta} \\[1ex]
\frac{1}{4}\textrm{-BPS}  &  [1^{\mu}] &\theta\gamma & 0 & [\mu,\gamma-2\mu]_{\theta}\\[1ex]
\textrm{semi-short}
&  [\lambda,1^\mu] &\theta \gamma & \lambda-2& [\mu,\gamma-2\mu-2]_{\theta}\\[1ex]
\hline
\end{array}
\end{equation}

Long superconformal multiplets are instead obtained from Young diagrams $\ula$ containing the window diagram $\Yvcentermath1 \Yboxdim{6pt} \ula\supseteq\yng(2,2)\,$, i.e.~the Young diagram has the form \eqref{typicaldiagram}, known as typical.
For such multiplets, the labels $\Delta,l,R$ are invariant under shift 
\begin{equation}\label{shiftsymmetry}
    \lambda_i\to\lambda_i-\theta s, \qquad \mu_i\to \mu_i+s , \qquad \gamma \to \gamma+2s. 
\end{equation}
with $s\in\mathbb{Z}$.
For $s<0$, this shift symmetry allows us to pick a representative Young diagram with the smallest possible values of $\mu_i$ and $\gamma$, subject to $\mu_i\ge 2$. Conversely, for $s>0$ we can always choose a representative Young diagram with the smallest possible value of $\lambda_i$, subject to $\lambda_i\ge 2$, and largest value of $\gamma$. 
Furthermore, long representations in general have non-integer dimension. This is obtained in this framework by analytically continuing $\delta$ in the following Young diagram,
\begin{equation}
    [\lambda_1+\delta,\lambda_2+\delta,2^{\mu_2-2},1^{\mu_1-\mu_2}]\ = \ \begin{tikzpicture}[scale=0.65, x=1.3cm, y=1.3cm, baseline=(current bounding box.center)]
    \draw [thick] (-1,-1) rectangle (0,-2);
    \draw [thick] (-1,-2) rectangle (-0.5,-3);
    \draw [thick] (-1,0) rectangle (0,-0.5);
        \draw [thick] (-1,-0.5) rectangle (0,-1);
        \draw [thick] (0,0) rectangle (1.5,-1);
        \draw [thick] (1.5,0) rectangle (4,-0.5);
        \draw [thick] (1.5,-0.5) rectangle (3,-1);
        \draw [<->,thick] (0,-1.2)--(1.5,-1.2) node[midway,below] {$\delta$};
    \end{tikzpicture},
\end{equation}
which only contributes to the dimension $\Delta$, see \eqref{eq:dimensionandtwist}.
We refer to \cite{Doobary:2015gia,Aprile:2021pwd,Hansen:2025lig} for further details including explicit applications.

In the above notation, the half-BPS OPE admits a very compact expression for all $\theta$: 
\begin{equation}\label{eq:scalarope}
   O_{p_1} \times O_{p_2}=  \sum_{i=0}^{\min(p_1,p_2)} \sum_{h(\ula)\leq i} O^{\gamma=|p_1-p_2|+2i}_{\ula}
\end{equation}
where
$h(\ula)$ denotes the height of $\ula$, that is, 
the number of rows.
It is important to note that the above unified description 
is a simplification due to the symmetries of the half-BPS OPE. In more general 
OPEs, the relevant four-dimensional representations are labelled by two Young 
diagrams $\ula_L,\ula_R$ \cite{Hansen:2025lig,Bissi:2026awg}, and in six dimensions 
the representations are labelled by a larger Young diagram with up to four long rows \cite{Heslop:2004du}.

In this paper, we will be considering N$^2$E half-BPS correlators ($\kappa=2$), 
in which case the relevant exchanged operators are obtained by intersecting 
two of the above OPEs subject to the N$^2$E constraint on the charges $p_i$. 
The resulting intersection is simply
\begin{equation}\label{eq:N2EOPE}
    (O_{p_1} \times O_{p_2} ) \cap (O_{p_3} \times O_{p_4}) = O^{\gamma_m} + \sum_{\lambda=0}^\infty O^{\gamma_m+2}_{[\lambda]} + \sum_{\lambda_1\geq \lambda_2}^\infty O^{\gamma_m+4}_{[\lambda_1,\lambda_2]},
\end{equation}
where 
\begin{equation}\label{gammam}
\gamma_m=\max(|p_1-p_2|,|p_4-p_3|).
\end{equation}
We observe that in \eqref{eq:N2EOPE} there are three $\gamma$-sectors.
The first two only contain short and semi-short superconformal multiplets, as can be seen from \eqref{tableQN1}. 
Long multiplets only appear in the third sector. 

In general, for extremality $\kappa$ there are $\kappa+1$ $\gamma$-sectors, with the following values,
\begin{equation}
   \gamma= \gamma_m, \gamma_m+2, \gamma_m+4, \dots,  \gamma_M-2, \gamma_M.
\end{equation}
where the largest value of $\gamma$ is given by
\begin{equation}\label{gammaM}
    \gamma_{M}=\min(p_1+p_2,p_3+p_4).
\end{equation}
Long multiplets always appear starting from the third sector when counting 
from lowest value of $\gamma$.\footnote{This feature is more general and also 
holds for correlators involving quarter-BPS operators, see \cite{Bissi:2026awg}.} 
For N$^2$E correlators $\gamma_m+4=\min(p_1+p_2,p_3+p_4)$ and we recover the result in \eqref{eq:N2EOPE}.

\section{Double-particle bootstrap}\label{sec:doublebootstrap}

When the OPE of two single-particle operators $\cO_p\times \cO_q$ is considered, the operators exchanged at leading order 
in the large-$c$ expansion are double-particle operators. As shown in \cite{Aprile:2018efk}, 
the leading order CFT data of these double-particle operators is directly related to 
the block coefficients extracted from the disconnected and the  tree-level connected contributions 
to four-point correlators of  single-particle operators \cite{Aprile:2017xsp}.  
This CFT data can then be used to constrain correlators in a variety of 
settings within the large-$c$ expansion. We refer to this methodology as the double-particle bootstrap.

More recently, \cite{Aprile:2025nta,Aprile:2026uxe} have  shown that the double-particle bootstrap 
can be used to compute a new class of four-point correlators. These are the tree-level connected 
correlators involving one double-particle insertion, such as
$\langle [{\cal O}_{r_1} {\cal O}_{r_2}] {\cal O}_{p_2}  {\cal O}_{p_3}  {\cal O}_{p_4}  \rangle$. 
In this section we explain what predictions the double-particle bootstrap makes on the family of correlators
\begin{equation}\label{theones_wecompute}
\langle [{\cal O}_{2} {\cal O}_{q}] {\cal O}_{p+q-2}  {\cal O}_{2}  {\cal O}_{p}  \rangle\,.
\end{equation}
These correlators are N$^2$E extremal, and therefore maintain the particularly simple kinematics 
that we explained in equations \eqref{Ward_Identity_Constraints}-\eqref{Upsilon}.
Their tree-level dynamics, however, are considerably richer compared to the single-particle correlators obtained 
by replacing $ [{\cal O}_{2} {\cal O}_{q}]\rightarrow {\cal O}_{2+q}$.  In particular, while the tree-level  
correlators $\langle {\cal O}_{2+q} {\cal O}_{p+q-2}  {\cal O}_{2}  {\cal O}_{p}  \rangle$ are seeded by a single 
$D$-function \cite{Uruchurtu:2011wh,Zhou:2017zaw}, we do not expect a similar result here. 
Indeed, this intuition is due to the following two observations. First, 
by applying the half-BPS double-particle limit \cite{Aprile:2026uxe} 
on the ${\cal O}_{2} {\cal O}_{q}$ pair of the five-point correlator 
$\langle {\cal O}_{2} {\cal O}_{q} {\cal O}_{p+q-2}  {\cal O}_{2}  {\cal O}_{p}  \rangle$, we must recover the 
four-point correlators with $[{\cal O}_2{\cal O}_q]$, as in \eqref{theones_wecompute}. 
Second, although we do not compute the five-point correlators explicitly, the latter are described by a 
basis of superconformal invariants. This is quite different from the four-point correlators, which only admit 
a unique superconformal invariant. Together, these facts suggest that the double-particle limit retains several 
dynamical contributions, originating from distinct  five-point propagator structures. As a result, we shall see that, 
despite the simplicity of the $\kappa=2$ kinematics, the tree-level dynamics of our correlators is quite rich.

In the large-$c$ expansion, our correlators behave as follows 
\begin{equation}\label{eq:largec}
    \begin{aligned}
        \langle[\OO_2 \OO_q] \OO_{p+q-2} \OO_2 \OO_p \rangle &= 
        {\cal P}_{[2q],p+q-2,2,p}
        \Bigg[ 
        \frac{1}{c^{\frac12}}  {\cal C}^{(\frac12)}_{[2q](p+q-2)(2)(p)} + \frac{1}{c^{\frac32}} {\cal C}^{(\frac32)}_{[2q](p+q-2)(2)(p)}+ O\left(c^{-\frac52}\right)\Bigg],
    \end{aligned}
\end{equation}
where
\begin{itemize}
    \item the disconnected correlator is ${\cal C}^{(\frac12)}$,
    \item the tree-level connected correlator that we aim to bootstrap is ${\cal C}^{(\frac32)}$. 
\end{itemize} 
Our approach in the four-dimensional ${\cal N}=4$ SYM theory parallels the 
one already given in \cite{Aprile:2026uxe}, 
whereas the method for the six-dimensional $\cN=(2,0)$ theory is new.
We will show that the two cases can be unified by introducing the parameter 
$\theta=\frac{d-2}{2}$, after taking into account a few important 
aspects of the OPE predictions which we spell out below.

\subsection{Double-particle operators}\label{sec:doubledef}

To describe and count double-particle operators with given $R$-symmetry quantum numbers, 
it is convenient to work in the free-theory basis, given by

\begin{equation}\label{eq:doubleparticle}
    {\cal D}_{rs;\tau,l}= \cO_{r} \partial^l  \Box^{\frac12(\tau-\theta r-\theta s)} \cO_{s}. 
\end{equation}
with $\tau\ge \theta(r+s)$. The $R$-symmetry representation $[a,b]_{\theta}$ (see~\eqref{eq:Rsym} for conventions) 
constrains the allowed values of $r,s\geq2$. We classify this space in both the protected and unprotected sectors below.

In the protected sector, the twist must be at unitarity bound
\begin{equation}
    \tau =\theta(r+s)= \theta(2a+b+2).
\end{equation}
Therefore, we have to find $r,s\geq2$ such that $r+s=2a+b+2$, that is, $(r,s)$ is a partition of $2a+b+2$.
Assuming $r\leq s$ we thus have
\begin{equation}
    r\leq \left\lfloor\frac{2a+b+2}{2}\right\rfloor, \qquad s=2a+b+2-r.
\end{equation}

In the long sector, the lightest twist is $\tau=\theta(2a +b+4)$. 
The higher-twist operators are then obtained by increasing the twist in steps of $2\theta$, so that
\begin{equation}
    r+s=2a+b+2+2i, \qquad i\geq1.
\end{equation}
A parametrisation of $r,s$ satisfying this is
\begin{equation}\label{rectangle1}
\begin{array}{ll} r=i+a+1+j,\\ i=1,\ldots,(t-1),\end{array}\qquad\qquad \begin{array}{l} s=i+a+1+b-j,\\ j=0,\ldots,(\mu-1),\end{array}
\end{equation}
where the maximum values of $i$ and $j$ are fixed by $\tau,a,b$, through the formulae 
\begin{equation}\label{rectangle2}
t\equiv \left\lfloor \frac{\frac{1}{\theta}\tau-b-2a}{2}\right\rfloor,\qquad\qquad \mu\equiv \left\{ \begin{array}{lc} \lfloor \frac{b+2}{2}\rfloor & a+l\ \textrm{even} \\[.2cm] \lfloor \frac{b+1}{2}\rfloor & a+l\ \textrm{odd} \end{array}\right.\,.\\[.2cm]
\end{equation}
We will refer to this solution as the rectangle $R_{\tau,a,b}$.
Note that each choice of $\tau$ and $[a,b]_{\theta}$ leads to a 
rectangle $R_{\tau,a,b}$ with  $\mu(t-1)$ pairs $(r,s)$.

The main difference between the four-dimensional ${\cal N}=4$ SYM theory 
and the six-dimensional $\cN=(2,0)$ theory is a certain `doubling': this 
is implemented by the floor function in the definition of $t$  \eqref{rectangle2}. 
The value of the degeneracy $\mu$ does not change across dimensions. 

Let us illustrate the doubling with an example. Consider the case 
\begin{equation}
    [a=0,b=p-2]_\theta,
\end{equation}
which will 
appear in the bootstrap of
$\langle [{\cal O}_2{\cal O}_q]{\cal O}_{p+q-2}{\cal O}_2{\cal O}_p\rangle$. 
While for $p=2$ and $p=3$ there is no degeneracy, for $p>3$ the degeneracy grows with $p$, so we focus on this regime. 
In $\cN=4$ SYM, we have that $\theta=1$ and thus by \eqref{rectangle1} 
the operators contributing to each representation are as follows,
\begin{equation}\label{eq:4ddoubleexample}
    \begin{array}{llllllll}
        &\tau = (p+2)      &\qquad {\color{blue}\cO_2 \cO_p},\, &{\color{blue}\cO_{3}\cO_{p-1}},\,&  \dots &  & \\
        &\tau = (p+2)+2  &\qquad \cO_2 \Box\cO_p,\, & \cO_{3} \Box \cO_{p-1},\, &  \dots &{\color{blue}\cO_{3} \cO_{p+1}}, &{\color{blue}\cO_{4} \cO_{p}}, & \ldots
    \end{array}
\end{equation}
In the first row, we highlight in blue the first two degenerate operators. 
The ellipsis completes the sequence ${\cal O}_r{\cal O}_s$ with $r+s=\tau=p+2$. 
In the second row, the operators associated with the same pairs $(r,s)$ as in the first line 
acquire one power of $\Box$ to match the twist $\tau=p+4$. 
The blue terms denote the first new pairs $(r,s)$ appearing in 
the rectangle $R_{\tau=p+4,0,p-2}$, and the ellipsis indicates the continuation of the sequence. 
In the 6d $\cN=(2,0)$ theory, we instead have $\theta=2$, and thus the operators contributing 
to each representation are as follows,
\begin{equation}
    \begin{array}{llllllll}
        &\tau = 2(p+2)     &\qquad {\color{blue} \cO_2 \cO_p},\, &{\color{blue} \cO_{3}\cO_{p-1}},\,&  \dots &  \\
        &\tau = 2(p+2)+2 &\qquad \cO_2 \Box\cO_p,\, & \cO_{3} \Box \cO_{p-1},\, & \dots & \\
        &\tau = 2(p+2)+4 &\qquad \cO_2 \Box^2\cO_p,\, &\cO_{3} \Box^2 \cO_{p-1},\, & \ldots &{\color{blue}\cO_{3} \cO_{p+1}}, &{\color{blue}\cO_{4} \cO_{p}},& \  \dots \\
        &\tau = 2(p+2)+6 &\qquad \cO_2 \Box^3\cO_p,\, &\cO_{3} \Box^3 \cO_{p-1},\, & \ldots &\cO_{3} \Box \cO_{p+1}, &\cO_{4}\Box  \cO_{p},& \  \dots
    \end{array}
\end{equation}
This doubling with respect to the $\theta=1$ case \eqref{eq:4ddoubleexample} is due to the fact that
a new pair $(r,s)$ enters $R_{\tau,0,p-2}$ every $\Delta\tau=2\theta$, i.e.~$\Delta \tau=4$ in 6 dimensions versus $\Delta \tau=2$ in 4 dimensions.

\subsection{Harnessing the predictions} 
\label{subsec:predictions}

It is convenient to introduce a notation for the large-$c$ expansion of 
the three-point couplings which we will use to derive the double-particle predictions.
We do so in the more general context of the four-point function
\begin{equation}\label{eq:generic4pt}
\langle [{\cal O}_{r_1}{\cal O}_{r_2}] {\cal O}_{p_2} {\cal O}_{p_3} {\cal O}_{p_4}\rangle.
\end{equation}
For an exchanged double-particle $\DD_{pq;\tau,l}$ \eqref{eq:doubleparticle} we shall write the two relevant OPE coefficients as 
\begin{align}
  {\cal C}_{ {\cal O}_{p_3} {\cal O}_{p_4} {\cal D}_{pq;\tau l} } & =  {\cal C}^{(0)}_{p_3,p_4,[pq]_{\tau,l}} + \frac{1}{c}  {\cal C}^{(1)}_{ p_3,p_4,[pq]_{\tau,l}} + \ldots  \\[.2cm]
   {\cal C}_{ [{\cal O}_{r_1}{\cal O}_{r_2}] {\cal O}_{p_2} {\cal D}_{pq;\tau,l} } & =  \frac{1}{\sqrt{c}}{\cal C}^{(\frac{1}{2})}_{[r_1r_2],p_2,[pq]_{\tau,l} } + \frac{1}{\sqrt{c}^3}  {\cal C}^{(\frac{3}{2})}_{[r_1r_2],p_2,[pq]_{\tau,l} }  +\ldots
\end{align}
All three-point couplings are taken to be normalised with respect to the corresponding two-point function normalisations, see e.g.,~\eqref{two_point_conventions}.

We wish to understand the values of the twist $\tau$ admitting double-particle exchanges in the four-point function \eqref{eq:generic4pt}.
To this end, consider the diagram below,
\begin{equation}\label{3ptssd}
\begin{array}{c}
\begin{tikzpicture}
\def\xinterm{2.5};
\def\ystop{4.5};
\def\posC{2}

\draw (\posC+.05,+.5) -- (\posC+.05,\ystop) -- (\posC-.05,\ystop) --(\posC-.05,+.5);
\draw (-\posC+.05,+.5) -- (-\posC+.05,\ystop) -- (-\posC-.05,\ystop) --(-\posC-.05,+.5);

\foreach \x in {0,0.5,1,1.5,2,2.5,3,3.5}
\draw (-\posC+.15+\x,\ystop-1)--(-\posC+\x+.35,\ystop);
\draw (-\posC+1.6,\ystop-.8) node[scale=.7,fill=white] {${\color{red}\tau=\theta(r_1+r_2+p_2)}$};
\draw[ thick] (-\posC-.3,\ystop-1)--(+\posC+.3,\ystop-1);

\draw (-\posC+2,+.3) node[scale=.7] {${\color{red}\textrm{unitarity bound}}$};
\draw[dashed, red] (-\posC-.05,+.5)--(+\posC+.05,+.5);

\draw (+\posC+1.4,+1.5) node[scale=.7] {$\tau=\theta(p_3+p_4)$};
\draw[black, thick] (+\posC-.3,+1.5)--(+\posC+.3,+1.5);

\draw (+\posC+1.75,+.75) node[scale=.7] {$\tau_{\textrm{long}}\equiv \theta(4+2a+b)$};
\draw[black, thick] (+\posC-.3,+.75)--(+\posC+.3,+.75);

\foreach \y in {0,0.25,0.5,0.75,1,1.25,1.5,1.75,2,2.25,2.5}
\draw[fill=gray,draw=blue]  (\posC,+.75+\y) circle(1pt);

\draw[gray, line width=2pt] (\posC+.05,\ystop) -- (\posC,\ystop+.5) --(\posC-.05,\ystop);
\draw[gray, line width=2pt] (-\posC+.05,\ystop) -- (-\posC,\ystop+.5) --(-\posC-.05,\ystop);

\draw (\posC+.8,0) node {${\cal C}_{ p_3,p_4 ,[pq]_{\tau,l}}$};
\draw (-\posC-.8,0) node {${\cal C}_{ [r_1r_2],p_2,[pq]_{\tau,l} }$};

\end{tikzpicture}
\end{array}
\end{equation} 
where the vertical arrows capture the allowed values of the twist.
Double-particle predictions are available only within a finite window.  The upper bound of this window, $\tau\leq \theta(r_1+r_2+p_2)$, 
follows from the fact that $\tau=\theta(r_1+r_2+p_2)$ is the threshold twist at which triple-particle operators can appear 
in the OPE of $[{\cal O}_{r_1}{\cal O}_{r_2}]$ with  ${\cal O}_{p_2}$. The lower bound is set by the unitarity bound for 
the $R$-symmetry representation that is being considered. When  $\theta(r_1+r_2+p_2)=\tau_{\textrm{long}}$, the window 
consists only of the unitarity bound.

Let us begin from studying the three-point coupling ${\cal C}_{p_3,p_4 ,[pq]_{\tau,l}}$ 
corresponding to a double-particle operator ${\cal D}_{pq;\tau,l}$ of twist $\tau$ and spin $l$. 
In a given $R$-symmetry representation $[a,b]_\theta$, see \eqref{eq:Rsym} for conventions, 
we know that the order of the three-point coupling depends on $\tau$ as follows,
\begin{equation}
   {\cal C}_{p_3,p_4, [pq]_{\tau,l}} =  \left\{\begin{aligned}
        &O(1) &  \qquad \theta(p_3+p_4) &\le \tau , \\[.2cm]
        & O(\tfrac{1}{c}) & \theta(2a+b+2)&\leq \tau< \theta(p_3+p_4).
    \end{aligned}\right.
\end{equation}
In order to 
understand how the operator ${\cal D}_{pq;\tau,l}$ contributes to the full correlator we need information about the three-point 
coupling ${\cal C}_{ [r_1r_2],p_2,[pq]_{\tau,l} }$. A way to obtain such information is to consider the large-$c$ expansion of the five-point 
correlator where we replace ${\cal D}_{pq;\tau,l}\mapsto {\cal O}_p{\cal O}_q$. The idea is that 
the block expansion of such correlator carries information about our three-point coupling. Then, at leading order we expect a non 
vanishing result as long as 
\begin{equation}
{\cal C}^{(\frac{1}{2})}_{ [r_1r_2],p_2,[pq]_{\tau,l} }\neq 0\qquad  
\longleftrightarrow\qquad \langle {\cal O}_{r_i} {\cal O}_p\rangle 
\langle {\cal O}_{r_j} {\cal O}_{p_2} {\cal O}_{q}\rangle
+\langle {\cal O}_{r_i} {\cal O}_{q}\rangle \langle {\cal O}_{r_j} {\cal O}_{p_2} {\cal O}_p\rangle \neq 0.
\end{equation}
Assuming this holds, we distinguish the following two cases:
\begin{itemize}
\item When $\theta(p_3+p_4)\le \tau$, we can write the contributions 
\begin{equation}\label{firstC12}
\frac{1}{\sqrt{c}}{\cal C}^{(\frac{1}{2})}_{ [r_1r_2],p_2,[pq]_{\tau,l} } {\cal C}^{(0)}_{p_3,p_4, [pq]_{\tau,l}}\qquad;\qquad  
\frac{1}{\sqrt{c}^{1+2k} } {\cal C}^{(\frac{1}{2})}_{ [r_1r_2],p_2,[pq]_{\tau,l} }( \eta_{[pq]_{\tau,l}} )^{k} {\cal C}^{(0)}_{p_3,p_4, [pq]_{\tau,l}}.
\end{equation}
\item When $\theta(2a+b+2)\leq\tau< \theta(p_3+p_4)$, 
we can write the contribution 
\begin{equation}\label{secondC12}
\frac{1}{\sqrt{c}^3}{\cal C}^{(\frac{1}{2})}_{ [r_1r_2],p_2,[pq]_{\tau,l} } {\cal C}^{(1)}_{p_3,p_4, [pq]_{\tau,l}}.
\end{equation}
\end{itemize}
In \eqref{firstC12} we distinguished the case $k=0$ to emphasize 
that we will have to first extract ${\cal C}^{(\frac{1}{2})}_{ [r_1r_2],p_2,[pq]_{\tau,l}}$ 
from the disconnected contribution in order 
to predict the contributions for $k\ge 1$ using the known CFT data of the double-particle operators, 
i.e.~$\eta_{[pq]_{\tau,l}}$, ${\cal C}^{(0)}_{p_3p_4 [pq]_{\tau,l}}$ and ${\cal C}^{(1)}_{p_3p_4 [pq]_{\tau,l}}$.

\subsubsection{Triple-particle operators do not contribute}

As already shown in \eqref{3ptssd}, the leading order operators exchanged for $\tau\geq \theta(r_1+r_2+p_2)$ are triple-particle operators. 
However, there are also triple-particle contributions below this bound.
Therefore, before moving on, we explain why these operators do not contribute to \eqref{firstC12} and \eqref{secondC12}, 
as this is a crucial condition for applying the double-particle bootstrap.

Let ${\cal T}_{kmn;\tau,l}$ be a triple-particle operator of twist $\tau$, spin $l$ and $R$-symmetry representation $[a,b]_{\theta}$, such that $\tau$ belongs to the window $\theta(2a+b+2) \leq \tau\leq \theta(r_1 +r_2+p_2)$. 
In the long sector, when $\theta(p_3+p_4)\leq\tau$, 
the following triple-particle exchange could potentially contribute to our prediction in \eqref{firstC12} via
\begin{equation}\label{contr_triple_potential1}
{\cal C}_{[r_1r_2],p_2, [kmn]_{\tau,l}}  \left(\frac{\eta_{[kmn]_{\tau,l}}}{c}\right)^k\, {\cal C}_{p_3,p_4, [kmn]_{\tau,l}} .
\end{equation}
To determine whether this contribution is allowed, we need to understand the 
large-$c$ scaling of the two three-point couplings, and compare them to that of \eqref{firstC12}.
To this end, let us show that, 
\begin{equation}\label{discussing_contr_triple_potential1}
{\cal C}_{[r_1r_2],p_2, [kmn]_{\tau,l}}=  O\left(\frac{1}{{c}}\right),\qquad 
\qquad {\cal C}_{p_3,p_4, [kmn]_{\tau,l}} = O\left(\frac{1}{\sqrt{c}} \right) .
\end{equation}
This result is already enough to rule out mixing between triple-particle \eqref{contr_triple_potential1} and double-particle \eqref{firstC12} contributions since the former is sub-leading for all $k$.

We show that ${\cal C}_{[r_1r_2],p_2 ,[kmn]{\tau,l}}$ cannot be of order $O(1)$ by considering the six-point function
with 
external operators $\langle{\cal O}_{r_1}{\cal O}_{r_2}{\cal O}_{p_2}{\cal O}_k{\cal O}_m{\cal O}_n\rangle$. 
The idea is that an $O(1)$ contribution to this correlator 
would provide information about the three-point coupling under 
consideration through its block expansion. 
However, an $O(1)$ contribution corresponds to three
disconnected two-point functions. In the present twist window such a contribution 
is forbidden because $k+m+n<r_1+r_2+p_2$. 
Applying the same reasoning to ${\cal C}_{p_3p_4 [kmn]{\tau,l}}$, we consider the
five-point function $\langle {\cal O}_{p_3}{\cal O}_{p_4}
{\cal O}_k{\cal O}_m{\cal O}_n\rangle$. 
In the large-c expansion, five-point functions are already at most $O(1/\sqrt{c})$ via their disconnected part, thus completing the proof of \eqref{discussing_contr_triple_potential1}.

We can similarly rule out the  
triple-particle exchange that could contribute to our prediction in \eqref{secondC12} via
\begin{equation}\label{contr_triple_potential2}
{\cal C}_{[r_1r_2]p_2 [kmn]_{\tau,l}}  {\cal C}_{p_3p_4 [kmn]_{\tau,l}} 
\end{equation}
with $\theta(2a+b+2)\leq \tau< \theta(p_3+p_4)$. 
Repeating the arguments that led to \eqref{discussing_contr_triple_potential1}, we can prove that
\begin{equation}\label{eq:triple2}
{\cal C}_{[r_1r_2],p_2, [kmn]_{\tau,l}}= O\left(\frac{1}{{c}}\right),\qquad 
\qquad {\cal C}_{p_3,p_4, [kmn]_{\tau,l}}= O\left(\frac{1}{\sqrt{c}^3} \right) .
\end{equation}
In particular, 
we already know that ${\cal C}_{[r_1r_2]p_2 [kmn]{\tau,l}}$ cannot be of order $O(1)$. 
Now, considering
$\langle {\cal O}_{p_3}{\cal O}_{p_4} {\cal O}_{k}{\cal O}_{m}{\cal O}_n\rangle$, since $k+m+n<p_3+p_4$, this correlator cannot factorise into a product of a two- and a three-point function, and thus its disconnected $1/\sqrt{c}$ part is not present. Therefore, ${\cal C}_{p_3p_4 [kmn]_{\tau,l}}$ cannot be of leading order. This completes the proof of \eqref{eq:triple2}, so we conclude that the triple-particle contributions \eqref{contr_triple_potential2} are sub-leading and do not mix with the double-particle ones  \eqref{secondC12}.

\subsubsection{Predictions as block coefficients}

An important feature of the double-particle bootstrap is that 
the predictions in \eqref{firstC12}  and \eqref{secondC12} can be recast in terms of block coefficients extracted from four-point correlators. 
Let us describe briefly how this works. Consider the block expansion of a four-point correlator,
\begin{equation}\label{nota_blockexpa}
{\cal C}_{\vec{p}}(x_1,x_2,y_1,y_2) =\sum_{\gamma,\ula} A_{\gamma,\ula,\, \vec{p}} B_{\gamma,\ula}(x_1,x_2,y_1,y_2;\theta;p_{12},p_{43}),
\end{equation}
where $\gamma,\ula$ label the exchanged operator $O^\gamma_{\ula}$ as described in section \ref{subsec:OPE}, $B_{\gamma,\ula}$ are the relevant superblocks and $A_{\gamma,\ula,\vec{p}}$ are the corresponding block coefficients. The underlying CFT structure implies that these coefficients can be expressed as sums of products of three-point couplings,
\begin{equation}\label{blockcoeff_A}
A_{\gamma,\ula,\vec{p}}= \sum_{O} {\cal C}_{p_1,p_2, [O_{\ula}^{\gamma}]_i}\ \mathbb{G}^{ij}_{\gamma,\ula}\ {\cal C}_{p_3,p_4, [O_{\ula}^{\gamma}]_j},
\end{equation}
where $\mathbb{G}^{ij}_{\gamma,\ula}$ is the inverse two-point function metric on a basis of exchanged operators $\big\{[O_{\ula}^\gamma]_{i}\big\}$ in representation $\gamma$, $\ula$. To simplify notation we shall 
refer to the block coefficients of the four-point correlator \eqref{nota_blockexpa} by 
\begin{equation}
{\cal C}_{\vec{p}}\big|_{\gamma, \ula}\equiv A_{\gamma,\ula,\vec{p}}.
\end{equation}

Now, by \eqref{blockcoeff_A} when a single operator is exchanged in a four-point correlator, the product of the corresponding three-point couplings can be read off directly from the block coefficient of that same correlator. This is no longer possible when several exchanged operators are degenerate. The case of double-particle operators can nevertheless be treated systematically. The key observation, discussed in \cite{Aprile:2017xsp,Aprile:2018efk,Alday:2019nin} and in particular \cite{Aprile:2019rep}, is that we can solve the mixing by assembling a matrix of block coefficients built from four-point correlators with single-particle external operators only.

To extract the leading and subleading CFT data of the double-particle operators of given twist $\tau$ and $R$-symmetry representation $[a,b]_\theta$, as classified in section \ref{sec:doubledef}, we have to assemble and analyze the matrices
\begin{equation}\label{thematrix_4ptsingle}
{\cal C}_{p,q,r,s}\big|_{\gamma, \ula}\,,\qquad (pq),(rs)\in\, R_{\tau,a,b}\,,
\end{equation}
where $R_{\tau,a,b}$ is defined in \eqref{rectangle1}.
In the four-dimensional ${\cal N}=4$ theory, this analysis was carried out extensively in \cite{Aprile:2018efk,Aprile:2019rep}. In the six-dimensional $(2,0)$ theory the only unmixing computation was done by \cite{Alday:2020tgi} which focused on long operators in the singlet, i.e.~$R_{\tau,0,0}$ (see \eqref{rectangle1}). Our subsequent computations will require us to go beyond the singlet sector.

In the next sections we describe the double-particle operators and the structure 
of the predictions in the three independent orientations, namely
\begin{equation}\label{summary_channels_corr}
\langle [{\cal O}_{2} {\cal O}_{q}] {\cal O}_{p+q-2}  {\cal O}_{2}  {\cal O}_{p}  \rangle\,,\qquad\qquad
\langle [{\cal O}_{2} {\cal O}_{q}] {\cal O}_{p}  {\cal O}_{2}  {\cal O}_{p+q-2}  \rangle\,,\qquad\qquad
\langle [{\cal O}_{2} {\cal O}_{q}]  {\cal O}_{2}  {\cal O}_{p}  {\cal O}_{p+q-2}  \rangle\,,
\end{equation}
in the $(12)(34)$ OPE channels.
For the case of $\langle [{\cal O}_{2} {\cal O}_{q}] {\cal O}_{p+q-2}  {\cal O}_{2}  {\cal O}_{p}  \rangle$, 
which is the most technical, we present the methodology underlying the OPE predictions 
in a pedagogical manner. We then streamline the discussion of the other two channels, 
which are simpler.  The last channel,  $\langle [{\cal O}_{2} {\cal O}_{q}]  
{\cal O}_{2}  {\cal O}_{p}  {\cal O}_{p+q-2}  \rangle$,  
is particularly interesting, as it exhibits a novel feature compared with 
the tree-level double-particle bootstrap for correlators with single-particle 
external operators. From left to right in \eqref{summary_channels_corr},  
we will refer to each channel by ${\cal O}_{2} \times {\cal O}_{p}$, ${\cal O}_{2} \times {\cal O}_{p+q-2}$ and 
$[{\cal O}_{2} {\cal O}_{q}]  \times {\cal O}_{2}$, respectively. 
We choose these OPEs to denote each channel because they fix the values of $\gamma_m$ \eqref{gammam} and $\gamma_M$ \eqref{gammaM}.

\subsection{The $\cO_2\times \cO_{p}$ channel}\label{2_p_sect}

This channel corresponds to the intersection of OPEs 
\begin{equation}
\big([\cO_2\cO_q]\times \cO_{p+q-2}\big)\quad \cap \quad\big(\cO_2\times \cO_{p}\big).
\end{equation} 
The relevant $R$-symmetry representation $[a,b]_{\theta}$ (see \eqref{eq:Rsym} for convention) has labels
\begin{equation}\label{repchannel1}
a=0,\qquad b=p-2.
\end{equation}
Returning to the diagram in \eqref{3ptssd} we obtain the twist window 
\begin{equation}\label{repchannel1_again}
\tau=\theta p,{ \theta(p+2) ,\theta(p+2)+2,\ldots,\theta(p+2q)-2}_{},
\end{equation}
where we recognise the values 
\begin{equation}
        \tau_{\rm unitary}=\theta p,
\end{equation}
and 
\begin{equation}\label{long_gammma_min2p}
    \tau_{\rm long}=\theta(p+2)+2i,\qquad i=0,\ldots,\theta(q-1)-1.
\end{equation}
In what follows, we organise the predictions according to whether the exchanged operators are long or protected. 
Each case has several features worth highlighting, and the reasoning underlying the predictions differs slightly 
between the two. We therefore take this opportunity to explain the details.

\subsubsection{Long operators}\label{subsec:long_s}

Following our discussion around \eqref{firstC12}, these operators contribute to the $\log(U)$ 
sector of the tree-level connected correlators with one power of their anomalous dimension.
Translating the values of $a,b$ \eqref{repchannel1} and $\tau_{\rm long}$ \eqref{long_gammma_min2p} to Young diagram notation, $\ula$, 
we have 
\begin{equation}\label{gammaulaschannel1}
    \gamma=p+2, \qquad  
    \ula = [2+l+i,2+i], \qquad i=0,\dots,\theta (q-1)-1\,.
\end{equation}
As explained in \eqref{shiftsymmetry}, for long representations, both $\gamma$ and $\ula$ can 
be shifted without changing the representation. In our present case, the superconformal representation \eqref{gammaulaschannel1} is equivalent to 
\begin{equation}\label{gaugedYD}
    \gamma=p+2+2j, \qquad  \ula=[2+l+i-\theta j,\, 2+i-\theta j,\, 2^{j}]
\end{equation}
for all $j.$
We will shortly use this freedom.

For each $\tau_{\text{long}}$ \eqref{long_gammma_min2p}, the prediction takes the following form
\begin{equation}\label{whatpredsection3}
\sum_{(rs)\in R_{\tau,0,p-2}} {\cal C}^{(\frac{1}{2})}_{ [2q](p+q-2)[rs]_{\tau,l} }( \eta_{[rs]_{\tau,l}} )^{} {\cal C}^{(0)}_{2p [rs]_{\tau,l}}.
\end{equation}
In this channel there are $\theta(q-1)$ such predictions.
From the argument in \eqref{thematrix_4ptsingle} and the work of \cite{Aprile:2018efk,Aprile:2019rep}
, we know that 
the information about the leading order three-point couplings $\CC^{(0)}_{2,p,[rs]}$ and the anomalous 
dimensions of the exchanged operators $\eta_{[rs]}$ is contained in the matrices 
\begin{equation}
\left(\CC^{(0)}_{r,s,r,s}\big|_{\gamma,\ula }\right),\qquad\qquad \left( \CC^{(1)}_{r',s',r,s} \big|_{\log(U), \gamma, \ula} \right)\qquad\qquad (rs),(r's')\in R_{\tau,0,p-2},
\end{equation}
where $R_{\tau,a,b}$ is defined in \eqref{rectangle1}.
Note that $\CC^{(0)}$ is diagonal whereas $\CC^{(1)}$ is full, i.e.~it has no vanishing entry. 
The sum in \eqref{whatpredsection3} can be rewritten in the following form
\begin{equation}\label{eq:predictionlog}
    \CC^{(\frac32)}_{[2q],p+q-2,2,p}\big|_{\log(U),\gamma, \ula} = \sum_{(rs)\in R_{\tau,0,p-2}}\left( \CC^{(\frac12)}_{[2q],p+q-2,r,s} \big|_{\gamma,\ula}\right)\frac{1}{\left(\CC^{(0)}_{r,s,r,s}\big|_{\gamma,\ula }\right)}
   \left( \CC^{(1)}_{r,s,2,p} \big|_{\log(U), \gamma, \ula} \right)
\end{equation}
using only block coefficients of four-point correlators. 
Some important observations are in order:
\begin{itemize}
\item 
The prediction \eqref{eq:predictionlog} is for the full correlator, however,
\begin{equation}
    \CC^{(\frac32)}_{[2q],p+q-2,2,p}\big|_{\log(U),\gamma, \ula}=
    {\cal H}^{(\frac32)}_{[2q],p+q-2,2,p}\big|_{\log(U),\gamma, \ula}
\end{equation}
since by definition, only ${\cal H}$ has logs. 
This is precisely the function that we aim to bootstrap.
\item 
The correlators ${\cal C}^{(1)}_{r,s,2,p}$ have $\kappa=2$, always, and what actually enters the 
prediction is, 
\begin{equation}\label{intermedita_pred_1channel}
\CC^{(1)}_{r,s,2,p}\big|_{\log(U),\gamma, \ula}= {\cal H}^{(1)}_{r,s,2,p}\big|_{\log(U),\gamma, \ula}.
\end{equation}
We compute the block coefficients by performing an explicit decomposition implementing the superconformal blocks of \cite{Aprile:2021pwd} in \texttt{Mathematica}, using \eqref{gaugedYD} as input.

\item
The correlators ${\cal C}^{(0)}_{r,s,r,s}$ and ${\cal C}^{(\frac{1}{2})}_{[2q],p+q-2,r,s}$ are determined by free 
monomials, therefore
\begin{equation}\label{block2pchannel_free}
{\cal C}^{(0)}_{r,s,r,s}\big|_{\gamma,\ula }={\cal F}^{(0)}_{r,s,r,s}\big|_{\gamma,\ula },\qquad\qquad 
\CC^{(\frac12)}_{[2q],p+q-2,r,s} \big|_{\gamma,\ula}= {\cal F}^{(\frac{1}{2})}_{[2q],p+q-2,r,s}\big|_{\gamma,\ula }.
\end{equation}
Differently from \eqref{intermedita_pred_1channel},
these correlators may have higher extremality depending on the pair $(r,s)$, 
and have a value of $\gamma_M=r+s\ge p+2$, as defined in \eqref{gammaM}.
It is then convenient to gauge the Young diagram representatives, using \eqref{gaugedYD}, as follows
\begin{equation}\label{gaugedforfree}
\gamma_M=r+s\qquad\qquad 
\ula=\Big[2+l+ \tfrac{\tau-\gamma_M\theta}{2},\,2+\tfrac{\tau-\gamma_M\theta}{2},2^{\frac{\gamma_M-(p+2)}{2}}\Big].
\end{equation}
Note also that $\gamma_M<2q+p$, i.e.~it remains below the threshold of three-particle exchanges \eqref{3ptssd}.
\end{itemize}
A reason why \eqref{gaugedforfree} is useful is that we shall compute 
the block coefficients \eqref{block2pchannel_free} very efficiently 
by using the so-called superconformal Cauchy identity \cite{Aprile:2021pwd}, 
which we review in section \ref{sec:cauchy}. 
This is an evaluation formula adapted to free-theory monomials and whose input are the labels $\gamma$ 
and $\ula$ (as well as the external charges and $\theta$).

There is a key simplification at this point: $\CC^{(\frac12)}$ is a disconnected contribution 
that factorises into a product of a two-point function, either $\langle {\cal O}_2{\cal O}_2\rangle$ or 
$\langle {\cal O}_{q}{\cal O}_{q}\rangle$, and a leading order three-point function of the form 
$\langle {\cal O}_X {\cal O}_{p+q-2}{\cal O}_Y\rangle$, therefore, if the external charges 
do not allow such a factorisation, the contribution vanishes. Therefore,
\begin{equation}
\CC^{(\frac12)}_{[2q],p+q-2,r,s}\neq 0 \qquad \Longleftrightarrow \qquad 
\left\{\begin{array}{c}
{\cal C}^{(\frac{1}{2})}_{2,p+q-2,s}\neq 0
\qquad (qs)\in R_{\tau,0,p-2}\\[.2cm]
{\cal C}^{(\frac{1}{2})}_{q,p+q-2,s}\neq 0
\qquad (2s)\in R_{\tau,0,p-2}
\end{array}\right. .
\end{equation}
The number of non-vanishing three-point couplings depends on the pairs in 
$R_{\tau,0,p-2}$ and can be determined computationally. As a result, 
the sum \eqref{eq:predictionlog} is sparse. 
We further note that the three-point 
couplings above admit well-known analytic expressions given by 
\cite{Lee:1998bxa,Corrado:1999pi,Bastianelli:1999en}, 
\begin{equation}\label{eq0:3pt4d}
    \CC^{(\frac12)}_{p_1,p_2,p_3} \bigg|_{\theta=1}=  \sqrt {p_1 p_2 p_3},
\end{equation}
and
\begin{equation}\label{eq0:3pt6d}
    \CC^{(\frac12)}_{p_1,p_2,p_3} \bigg|_{\theta=2}=  
    \frac12 \Gamma\left[\frac{p_1+p_2+p_3}{2}\right] \prod_{i=1}^3 \frac{2^{p_i} \Gamma\left[\frac{p_1+p_2+p_3+1}{2} - p_i\right]}{\sqrt{\pi \Gamma[2p_i-1]}}.
\end{equation}

The arguments discussed here apply identically to the $\log(U)$ predictions of the 
${\cal O}_2\times {\cal O}_{p+q-2}$ channel, which we will discuss in
section \ref{2_pqm2_sect}.

\subsubsection{Protected semi-short operators}\label{subsec:protected_s}

The relevant exchanged operators are the double-particle operators in the same $R$-symmetry representation as in the long sector $[a,b]_{\theta}=[0,p-2]_{\theta}=\eqref{repchannel1}$. 
They are of  the form ${\cal O}_r\partial^l{\cal O}_s$ with $2a+b+2=r+s$, thus
\begin{equation}
    r+s=p,\qquad r,s \ge2.
\end{equation} 
Their Young diagram notation \eqref{tableQN1} reads
\begin{equation}\label{gammaulaschannel1prot}
    \gamma=p, \qquad\qquad  
    \ula = [2+l]. 
\end{equation}
Here, since the representation is at the unitarity bound, the above is the unique choice of $\gamma,\ula$.
We recall that by equation \eqref{secondC12}, these operators contribute with subleading three-point couplings, 
and in principle, the prediction involves a sum over degenerate operators,
\begin{equation}
\sum_{r+s=p} {\cal C}^{(\frac{1}{2})}_{ [2q],(p+q-2),[rs]_{\tau,l} } {\cal C}^{(1)}_{2,p, [rs]_{\tau,l}}\qquad;\qquad \tau_{\rm unitarity}=\theta p\,.
\end{equation}
As in the previous case, we want to derive this prediction using four-point correlators only.  We follow the 
approach described in \cite{Aprile:2019rep,Aprile:2025nta}, in which the starting point is to construct the Gram matrix 
\begin{equation}\label{Gram_matrix}
\left(
\begin{array}{c|c}
{\cal C}_{r',s',r,s}^{(0)}\big|_{\gamma,\ula} & {\cal C}_{r',s',2,p}^{(1)}\big|_{\gamma,\ula} \rule{.2cm}{0pt} \\[.2cm]
\hline
\rule{0pt}{.7cm}\rule{.8cm}{0pt}{\cal C}_{[2q],p+q-2,r,s}^{(\frac{1}{2})}\big|_{\gamma,\ula} & \rule{.8cm}{0pt} {\cal C}^{(\frac{3}{2})}_{[2q],p+q-2,2,p}\big|_{\gamma,\ula} \rule{.2cm}{0pt}
    \end{array}\right),
\end{equation}
where 
\begin{itemize}
\item the upper-left entry is the leading order contribution of the matrix introduced in \eqref{thematrix_4ptsingle},
\item the lower-right entry is the correlator we aim to bootstrap,
\item the off-diagonal entries form two vectors indexed by the pairs $(2p)$ and $[2q](p+q-2)$.
\end{itemize}
By substituting in the expression for the block coefficient as a sum over three-point couplings \eqref{blockcoeff_A}, one can verify that rows and columns of \eqref{Gram_matrix} are 
linearly dependent. Again this is obvious if there is only 
one exchanged operator, and crucially, the statement holds true in the case of degenerate 
double-particles operators. Consequently, the determinant of \eqref{Gram_matrix} vanishes. 

The vanishing determinant of the Gram matrix yields 
an equation for the lower-right entry, i.e.~the correlator that we aim to bootstrap. 
Using the fact that ${\cal C}^{(0)}$ is diagonal, we can write the prediction in the following form,
\begin{equation}
   \CC^{(\frac32)}_{[2q],p+q-2,2,p}\big|_{\gamma, \ula} = \sum_{\substack{r,s\geq2\\
   r+s=p}}\left( \CC^{(\frac12)}_{[2q],p+q-2,r,s} \big|_{\gamma,\ula}\right)\frac{1}{\left(\CC^{(0)}_{r,s,r,s}\big|_{\gamma,\ula }\right)}
   \left( \CC^{(1)}_{r,s,2,p} \big|_{\gamma, \ula} \right).
\end{equation}
This equation has similar structure as the one we gave in the previous section, see \eqref{eq:predictionlog},
however there are some key facts and differences that are important to highlight. 

The first fact is that the single-particle correlators enter the prediction as
\begin{equation}\label{novelty6dunitarybound}
{\cal C}^{(1)}_{r,s,2,p}\Big|_{\gamma,\ula}= {\cal F}^{(1)}_{r,s,2,p}\Big|_{\gamma,\ula}+ \Upsilon_\theta \circ {\cal H}^{(1)}_{r,s,2,p}\Big|_{\gamma,\ula},
\end{equation}
thus, in principle, both tree-level contributions ${\cal F}$ and ${\cal H}$ are needed in order to compute the prediction.
In the four-dimensional ${\cal N}=4$ theory, the contribution 
from ${\cal H}$ vanishes. However this is not the case the six-dimensional theory. 

The second fact concerns ${\cal C}^{(\frac{1}{2})}_{[2q],p+q-2,r,s}$. 
We already pointed out that a disconnected contribution is entirely determined by the product 
of two and three-point functions. However, 
since $r+s=p$, the only way to disconnect the four-point contribution is through 
three-point functions, but the latter actually vanish, i.e.,
\begin{equation}
 \CC^{(\frac12)}_{q,p+q-2,p-2}=0,\qquad\qquad \CC^{(\frac12)}_{2,p+q-2,p-q}=0.
\end{equation}

In fact, $\CC^{(\frac32)}_{[2q],p+q-2,2,p}$ can also be shown to vanish 
from a closer inspection of the four-point 
functions  entering \eqref{Gram_matrix}. In particular, this is done by considering the multi-point orthogonality 
theorem for single-particle correlators proven  in the ${\cal N}=4$ SYM 
theory in \cite{Aprile:2020uxk}. Since the proof only uses the abstract definition 
of single-particle operators via their two-point function orthogonality with 
multi-particle operators, it applies also to the $\AdS_7\times \mathrm{S}^4$ Kaluza-Klein spectrum 
and its dual six-dimensional theory. We can then introduce
the concept of \emph{$n$-point near-extremal correlators} in both theories, defined as, 
\begin{equation}\label{eq:nearextremal}
\tfrac{1}{2}(-P_j+\sum_{i\neq j} p_i)\leq n-3, \qquad\qquad P_j=\max\{p_1,\ldots,p_n\}.
\end{equation}
The statement of the theorem is that a single-particle correlator $\langle \cO_{p_1} \cO_{p_2} \dots \cO_{p_n}\rangle$ satisfying \eqref{eq:nearextremal} vanishes. 
Applying this to the correlators in \eqref{Gram_matrix} yields ${\cal C}_{r',s',2,p}=0$,
and by point-splitting the double-particle operator, we also find that ${\cal C}^{}_{[2q],p+q-2,r,s}=0$. 

As a result, in this channel, 
\begin{equation}\label{prediction_protected_semishort_2p}
{\cal C}^{(\frac{3}{2})}_{[2q],p+q-2,2,p}\Big|_{\gamma=p,\ula=[2+l]}=0
\end{equation}
In other words, in this OPE channel there are no semi-short operators exchanged at this order. 

The prediction \eqref{prediction_protected_semishort_2p} is a constraint for the full correlator, which therefore 
implies that the combination of ${\cal F}$ and ${\cal H}$ gives a vanishing result, 
\begin{equation}
{\cal F}^{(\frac{3}{2})}_{[2q],p+q-2,2,p}\Big|_{\gamma=p,\ula=[2+l]} + \Upsilon_\theta \circ {\cal H}^{(\frac{3}{2})}_{[2q],p+q-2,2,p}\Big|_{\gamma=p,\ula=[2+l]}=0
\end{equation}
We shall exploit this equation as a constraint on ${\cal H}$ given the knowledge of ${\cal F}$, namely
\begin{equation}
\Upsilon_\theta \circ{\cal H}^{(\frac{3}{2})}_{[2q],p+q-2,2,p}\Big|_{\gamma=p,\ula=[2+l]}=-{\cal F}^{(\frac{3}{2})}_{[2q],p+q-2,2,p}\Big|_{\gamma=p,\ula=[2+l]} 
\end{equation}
Computing the ${\cal F}^{(\frac{3}{2})}_{[2q],p+q-2,2,p}$ is therefore an essential step in deriving the semi-short predictions.
We will explain in section \ref{formulae_freet},  and more specifically in section \ref{formulae_freet_db}, how the 
latter is computed. In ${\cal N}=4$ SYM, thanks to the partial non-renormalization theorem, the computation 
can be done entirely in the free theory. But the situation in the six-dimensional $\cN=(2,0)$ theory is different.

\subsection{The $\cO_2\times \cO_{p+q-2}$ channel}\label{2_pqm2_sect}

This channel corresponds to the intersection of OPEs 
\begin{equation}
   \left( [\cO_2\cO_q]\times \cO_p\right) \quad \cap \quad \left(\cO_2\times \cO_{p+q-2}\right).
\end{equation}
The relevant $R$-symmetry representation $[a,b]_{\theta}$ (see \eqref{eq:Rsym} for conventions) has labels
\begin{equation}\label{}
a=0,\qquad\qquad b=p+q-4.
\end{equation}
Returning to the diagram in \eqref{3ptssd} we obtain the window 
\begin{equation}\label{repchannel2}
\tau=\theta (p+q-2),{ \theta(p+q) ,\theta(p+q+2)-2}_{}
\end{equation}
where we recognise the values 
\begin{equation}
\tau_{\rm unitary}= \theta (p+q-2)\qquad\qquad \tau_{\rm long}=\theta(p+q)+2i,\qquad i=0,\theta-1\,.
\end{equation}
Note that $i=0$ for the ${\cal N}=4$ SYM theory, $\theta=1$. Instead, $i=0,1$ in the six-dimensional $(2,0)$ theory, $\theta=2$.
In this case the predictions do not grow with the external charges. 

As in the previous channel, we discuss long and semi-short sectors separately. The methods and results are nearly identical across both channels, so we keep the discussion brief.

\subsubsection{Long operators}

The exchanged double-particle operators are indexed by pairs $(rs)$ 
in $R_{\tau,0,p+q-4}$ \eqref{rectangle1}. In Young diagram notation, a representative of the equivalence class is given by
\begin{equation}
    \gamma=p+q, \qquad \ula=[2+l+i,2+i], \qquad i=0, \theta-1.
\end{equation}
Applying the same logic as in section \ref{subsec:long_s}, the predictions in this sector are given by
\begin{equation}\label{eq:tpredictionlog}
      \CC^{(\frac32)}_{[2q],p,2,p+q-2}\big|_{\log(U),\gamma, \ula} = \sum_{(rs)\in R_{\tau,0,p+q-4}}\left( \CC^{(\frac12)}_{[2q],p,r,s} \big|_{\gamma,\ula}\right)\frac{1}{\left(\CC^{(0)}_{r,s,r,s}\big|_{\gamma,\ula }\right)}
   \left( \CC^{(1)}_{r,s,2,p+q-2} \big|_{\log(U),\gamma, \ula} \right).
\end{equation}
The sum is sparse: it is only non-vanishing for the pairs $(rs)$ for which the disconnected contribution to ${\cal C}^{(\frac{1}{2})}_{[2q],p,r,s}$ exists.
These pairs can be determined computationally, however it is simple to see that there are only two such terms, given by
\begin{equation}
    \begin{aligned}
        (2,p+q-2)&\in R_{\tau,0,p+q-4} &\qquad \quad {\cal C}^{(\frac{1}{2})}_{q,p,p+q-2}\neq 0 ,\\
        (q,p)&\in R_{\tau,0,p+q-4} & {\cal C}^{(\frac{1}{2})}_{2,p,p}\neq 0.
    \end{aligned}
\end{equation}
To see why recall that we are considering
the lowest $\tau_{\rm long}$ in the representation \eqref{repchannel2}, 
i.e.~$r+s=p+q$ with $r,s\ge 2$. In fact, $R_{\tau,0,p+q-4}=\{ (2,p+q-2),(3,p+q-1),\ldots ,(q,p),\ldots \}$ 
until the degeneracy is saturated. The degeneracy can be 
written as $\lfloor \frac{p-q-2}{2}+q\rfloor$ \eqref{rectangle2}, thus both $(rs)=(2,p+q-2)$ 
and $(rs)=(q,p)$ belong to the set.  

Summarising, the final formula for the predictions in this sector is
\begin{align}\label{summary_long_2pqm2}
     \CC^{(\frac32)}_{[2q],p,2,p+q-2}\big|_{\log(U),p+q, [2+l,2]} = \sum_{(rs)=(2,p+q-2),(q,p)}\!\!\!\!\!\!\!\frac{
     \left( \CC^{(\frac12)}_{[2q],p,r,s} \big|_{p+q,[2+l,2]}\right)
     \left( \CC^{(1)}_{r,s,2,p+q-2} \big|_{\log(U),p+q, [2+l,2]} \right)
     }{\left(\CC^{(0)}_{r,s,r,s}\big|_{p+q,[2+l,2] }\right)}.
\end{align}
Here we do not have to pick different Young diagram representatives because 
all correlators involved have the same $\gamma_M=q+p$.

\subsubsection{Protected semi-short operators}
These transform with twist $\theta(p+q-2)$ and $R$-symmetry $[0,p+q-4]_\theta$, which translate to the following analytic superspace labels,
\begin{equation}
    \gamma=p+q-2, \qquad \ula=[2+l].
\end{equation}
We index these double-particle exchanges by pairs $(rs)$ satisfying $r+s=p+q-2$, $r,s\geq2$.

The discussion here is identical to that of section \ref{subsec:protected_s}.
In particular, the Gram matrix argument \eqref{Gram_matrix} leads to
\begin{equation}\label{prediction_protected_semishort_2pq2}
   \CC^{(\frac32)}_{[2q],p,2,p+q-2}\big|_{\gamma, \ula} = \sum_{\substack{r,s\geq 2\\
   r+s=p+q-2}}\left( \CC^{(\frac12)}_{[2q],p,r,s} \big|_{\gamma,\ula}\right)\frac{1}{\left(\CC^{(0)}_{r,s,r,s}\big|_{\gamma,\ula }\right)}
   \left( \CC^{(1)}_{r,s,2,p+q-2} \big|_{\gamma, \ula} \right),
\end{equation}
similarly to \eqref{prediction_protected_semishort_2p}.
Note the single-particle correlator $\CC_{r,s,2,p+q-2}$ is again in a vanishing near-extremal configuration \eqref{eq:nearextremal}, from which we conclude,
\begin{equation}
     \CC^{(\frac32)}_{[2q],p,2,p+q-2}\big|_{\gamma, \ula} =0,
\end{equation}
leading to
\begin{equation}
\Upsilon_\theta \circ{\cal H}^{(\frac{3}{2})}_{[2q],p,2,p+q-2}\Big|_{\gamma=p+q-2,\ula=[2+l]}=-{\cal F}^{(\frac{3}{2})}_{[2q],p,2,p+q-2}\Big|_{\gamma=p+q-2,\ula=[2+l]} .
\end{equation}

\subsection{The $[\cO_2 \cO_q]\times \cO_2 $ channel}\label{2q_2_sect}

This channel corresponds to the intersection of OPEs 
\begin{equation}
   \left([\cO_2 \cO_q] \times \cO_2 \right)\quad  \cap \quad \left(\cO_{p+q-2}  \times \cO_p\right).
\end{equation}
The relevant $R$-symmetry representation $[a,b]_{\theta}$ has labels
\begin{equation}
a=0,\qquad\qquad b=q.
\end{equation}
The threshold for triple-particle exchanges is $\tau=\theta(q+4)$ and therefore this channel 
only has a single prediction in the non-$\log(U)$ sector at the unitarity bound,
\begin{equation}
\tau_{\rm unitary}= \theta(q+2)\,.
\end{equation}
Interestingly, we shall now see that the prediction is non-trivial, 
differently from the previous two cases. 
Furthermore, the predictions in this channel will be the most important for the main results of this paper.

The initial discussion is identical to that of the semi-short sectors of the other two channels.
The exchanged operators in this sector are semi-shorts operators 
${\cal O}_r\partial^l{\cal O}_s$ with $r+s=q+2$. In Young diagram notation, 
\begin{equation}
    \gamma= q+2, \qquad \ula=[2+l].
\end{equation}
and the Gram matrix argument leads to
\begin{equation}
   \CC^{(\frac32)}_{[2q],2,p,p+q-2}\big|_{\gamma, \ula} 
   = \sum_{r+s=q+2}\left( \CC^{(\frac12)}_{[2q],2,r,s} \big|_{\gamma,\ula}\right)\frac{1}{\left(\CC^{(0)}_{r,s,r,s}\big|_{\gamma,\ula }\right)}
   \left( \CC^{(1)}_{r,s,p,p+q-2} \big|_{\gamma, \ula} \right).
\end{equation}
Now, the sum localises on $(r,s)=(2,q)$ since this is the only possible 
pair that allows $\CC^{(\frac12)}_{[2q],2,r,s}$ to disconnect. 
Thus,
\begin{equation}\label{eq:prednonlog}
   \CC^{(\frac32)}_{[2q],2,p,p+q-2}\big|_{\gamma, \ula} 
   = \left( {\cal F}^{(\frac12)}_{[2q],2,2,q} \big|_{\gamma,\ula}\right)\frac{1}{\left(\CC^{(0)}_{2,q,2,q}\big|_{\gamma,\ula }\right)}
   \left( \CC^{(1)}_{2,q,p,p+q-2} \big|_{\gamma, \ula} \right)
\end{equation}
where we replaced $\CC^{(\frac12)}_{[2q],2,r,s}$ with the free theory ${\cal F}^{(\frac12)}_{[2q],2,2,q}$.

The crucial difference with the other two channels is that the four-point correlator $\CC^{(1)}_{2,q,p,p+q-2}$ is now above near-extremality 
and thus non-vanishing. Moreover,  
as pointed  out already around \eqref{novelty6dunitarybound}, the coefficient 
$\CC^{(1)}_{2,q,p,p+q-2} \big|_{\gamma, \ula}$ in the six-dimensional theory 
receives a contribution from both ${\cal F}$ and ${\cal H}$. 
Note also that 
\begin{equation}\label{corr_2q22q}
\CC^{(\frac12)}_{[2q],2,2,q} =  
\CC ^{(\frac12)}_{q,2,q} {U^\theta\sigma}+
\CC ^{(\frac12)}_{2,2,2} \frac{U^\theta\hat{\tau}}{V^\theta}\,,
\end{equation}  
where now the three-point couplings involved are non-vanishing.

The prediction \eqref{eq:prednonlog} is a constraint for the full correlator, 
and, as done before, we rewrite it as a constraint on ${\cal H}$, with the difference that we have to include $\FF$:
\begin{equation}\label{eq_prednonlog2}
\Upsilon_{\theta}\circ{\cal H}^{(\frac{3}{2})}_{[2q],2,p,p+q-2}\Big|_{\gamma=q+2,\ula=[2+l]}=-{\cal F}^{(\frac32)}_{[2q],2,p,p+q-2}\Big|_{\gamma=q+2,\ula=[2+l]} +
\CC^{(\frac32)}_{[2q],2,p,p+q-2}\Big|_{\gamma=q+2,\ula=[2+l]}.
\end{equation}
This is an explicit prediction once both ${\cal F}$ and \eqref{eq:prednonlog} are known. 

A novel feature of our tree-level bootstrap algorithm for correlators with one 
double-particle and three single-particle external operators is the need for non-trivial predictions for 
double-particle operators at the unitarity bound. This distinguishes our approach from 
the CFT bootstrap of correlators with four single-particle external operators, see e.g.~\cite{Dolan:2006ec}.

\section{Formulae for the free-like theory}\label{formulae_freet}
We determine the free-like contributions,
\begin{equation}\label{whatwewantcalF}
{\cal F}_{q,p+q-2,2,p}^{\left(1\right)}\qquad\text{and}\qquad
{\cal F}_{[2q],p+q-2,2,p}^{\left(\frac{3}{2}\right)},
\end{equation} 
required by our predictions from section \ref{sec:doublebootstrap}. 
In the four-dimensional $\cN=4$ SYM the partial non-renormalisation theorem 
\cite{Eden:2000bk} allows us to compute both correlators directly at zero coupling. 
In the six-dimensional $\cN=(2,0)$ theory, however, the splitting between ${\cal F}$ 
and ${\cal H}$ is different. 

We therefore present a unified method for bootstrapping 
these ``free-like" functions in both theories by imposing consistency with the half-BPS OPE.
The final result will need the input from the double-particle limit, which we discuss in section \ref{sec:doublelimit}. In section \ref{chiralalgebra}, we reproduce the six-dimensional result independently from the chiral algebra.

\subsection{Bootstrapping $\FF$ from the half-BPS OPE}\label{sec:F}

\subsubsection{Single-particle $\FF$}
We start with the single-particle correlator given by
\begin{equation}\label{eq:Fsingle}
{\cal F}_{q,p+q-2,2,p}^{\left(1\right)}=a U^{\theta}\hat{\sigma} + b \frac{U^{\theta}\hat\tau}{V^\theta} + c \frac{U^{2\theta} \hat{\sigma}\hat\tau}{V^{\theta}}  ,
\end{equation}
with $a,b,c$ to be determined. 
Recall from section \eqref{eq:N2EOPE}, that, as a N$^2$E correlator, the above receives contributions from three different $\gamma$ sectors, given by 
\begin{equation}
   \gamma= \gamma_m, \gamma_m+2, \gamma_m+4, \qquad \gamma_m=p-2.
\end{equation}
The sector $\gamma_{m}=p-2$  in \eqref{eq:Fsingle} has no contribution. 
For the other two sectors, $\gamma_m+2=p$ and $\gamma_m+4=p+2$, we have
\begin{equation}\label{eq:Fsingle_again}
{\cal F}_{q,p+q-2,2,p}^{\left(1\right)}\Bigg|_{\gamma=p}= U^{\theta}\hat{\sigma}\Bigg[ a + b \frac{\tau}{\hat{\sigma}}\frac{1}{V^\theta}\Bigg]\qquad\qquad
{\cal F}_{q,p+q-2,2,p}^{\left(1\right)}\Bigg|_{\gamma=p+2}=U^{2\theta}\hat{\sigma}^2\Bigg[ c \frac{\hat\tau}{\hat{\sigma}}\frac{1}{V^\theta} \Bigg].
\end{equation}
To see this we recall that $\gamma$ counts the number of propagators from points 1, 2 to points 3, 4.

Note that in \eqref{eq:Fsingle}, the sector $U^{2\theta}$ does not included the terms $U^{2\theta}\sigma^2$ and 
$U^{2\theta}\tau^2/V^{2\theta}$. The reason is that these terms are absent as a consequence of the 
definition of the single-particle operators. The argument is the same as the one given in \cite{Aprile:2020uxk},
which only uses orthogonality of the single-particle operators with respect to half BPS multi-particle operators.

The strategy that we shall adopt to solve for $a,b,c$ is to consider the 
half-BPS OPE in the $\gamma=\gamma_m+2$ sector, and repeat our analysis in each of 
the three OPE channels. 
In \eqref{eq:Fsingle} the $\gamma_m+2$ sector  corresponds to the
coefficients of the $U^\theta$ term. There observation is that there is one crossing orbit that exchanges \emph{only} 
$a \leftrightarrow b\leftrightarrow c$ so after crossing we obtain a closed linear system of equations.
This strategy is particularly advantageous in the 6d theory because the $U^{2\theta}$ term in ${\cal F}$ (from $\gamma=\gamma_m+4$) contributes 
to the half-BPS OPE together with the reduced correlator $\HH$. 
In $\cN=4$ SYM, this method is not needed since $a,b,c$ are fixed by free-theory, though we will see it remains similarly efficient.

The relevant half-BPS block is  $B_{\gamma=p,\ula=[\varnothing]}(x_i,y_i)$ and 
it is leading in the limits $x_i\rightarrow 0$, and $y_i\rightarrow 0$. More precisely,  
\begin{equation}
\lim_{x_i\rightarrow 0}\lim_{y_i\rightarrow 0}\ \frac{ 1}{ U^\theta\hat{\sigma}} B_{\gamma=p,\ula=[\varnothing]}\left(x_i,y_i;\,\theta, \,^{p_{12}=2-p}_{p_{43}=p-2}\right)=1\times \delta_{\ula,[\varnothing]}.
\end{equation}
We refer to appendix \ref{app:blocks} for explicit formulae for the superconformal blocks from which this follows.

The limits above correspond to the intersection of OPEs ${\cal O}_q\times{\cal O}_{p+q-2}$ 
and ${\cal O}_2\times{\cal O}_p$. Applying this to our correlator we find
\begin{equation}
\lim_{x_i\rightarrow 0}\lim_{y_i\rightarrow 0}
  {\cal F}_{q,p+q-2,2,p}^{\left(1\right)}\Bigg|_{U^\theta\hat{\sigma}}
   =  a   + b
\end{equation}
The key observation is that there is a single exchange with charge $\gamma=p$. 
We can therefore directly identify $a+b$ with the block coefficient corresponding 
to the exchanged single-particle half-BPS operators $\cO_p$, namely
\begin{equation}\label{eq:ab}
    a+b= \CC^{(\frac12)}_{q,p+q-2,p} \CC^{(\frac12)}_{p,2,p}.
\end{equation}

We now repeat the argument in the other two orientations using crossing symmetry. 
We first exchange points $1\leftrightarrow 4$. The correlator under study is 
${\cal F}_{p,p+q-2,2,q}$. Compared to \eqref{eq:Fsingle} we find that $b\leftrightarrow c$. 
The relevant exchange is the half-BPS single-particle with $\gamma=\gamma_m+2=q$, giving
\begin{equation}\label{eq:ac}
    a+c= \CC^{(\frac12)}_{p,p+q-2,q} \CC^{(\frac12)}_{q,2,q}. 
\end{equation}
Then, we exchange points $1\leftrightarrow 3$. The correlator under study 
is ${\cal F}_{2,p+q-2,q,p}$. Compared to \eqref{eq:Fsingle} we find that 
$a\leftrightarrow c$. The relevant exchange is the half-BPS single-particle 
with $\gamma=\gamma_m=p+q-2$, giving
\begin{equation}\label{eq:bc}
    b+c= \CC^{(\frac12)}_{p,q,p+q-2} \CC^{(\frac12)}_{p+q-2,2,p+q-2}.
\end{equation}
Solving \eqref{eq:ab}, \eqref{eq:ac}, \eqref{eq:bc} simultaneously, using that 
$\CC^{(\frac12)}_{p_1,p_2,p_3}$ is invariant under cyclic permutations, 
and that $\CC^{(\frac12)}_{p,p,2}=\sqrt 2 p$ in both theories, we obtain
\begin{equation}\label{result_forFsingleparticle}
    a= \sqrt2 \CC^{(\frac12)}_{p,q,p+q-2} , \qquad b= \sqrt2 (p-1) \CC^{(\frac12)}_{p,q,p+q-2}, \qquad c=\sqrt2  (q-1)\CC^{(\frac12)}_{p,q,p+q-2}. 
\end{equation}
These formulae are valid in both $\cN=4$ SYM and the 6d $\cN=(2,0)$ theory. 
The only difference is the value of the three-point 
coefficient, which we repeat here for reader's convenience: 
\begin{equation}\label{eq:3pt4d}
    \CC^{(\frac12)}_{p_1,p_2,p_3} \bigg|_{d=4}=  \sqrt {p_1 p_2 p_3},
\end{equation}
and
\begin{equation}\label{eq:3pt6d}
    \CC^{(\frac12)}_{p_1,p_2,p_3} \bigg|_{d=6}=  \frac12 \Gamma\left[\frac{p_1+p_2+p_3}{2}\right] \prod_{i=1}^3 \frac{2^{p_i} \Gamma\left[\frac{p_1+p_2+p_3+1}{2} - p_i\right]}{\sqrt{\pi \Gamma[2p_i-1]}}.
\end{equation}

\subsubsection{Double-particle $\FF$}\label{formulae_freet_db}

We now turn to the double-particle correlator.  For our purpose it is enough to consider 
\begin{equation}
{\cal F}_{[2q],p+q-2,2,p}^{\left(\frac{3}{2}\right)} =a U^\theta \hat{\sigma} + b \frac{U^\theta\hat\tau}{V^\theta} + c \frac{U^{2\theta} \hat{\sigma} \hat\tau }{V^\theta}.
\end{equation}
Unlike the case of single-particle operators, this tree-level correlator might have 
contributions from $U^{2\theta}\sigma^2$ or $U^{2\theta}\tau^2/V^{2\theta}$, 
however, together with unity, these three terms form a closed orbit under crossing that does not mix with $a$, $b$ or $c$. Thus
our strategy to find $a,b,c$ is identical to the one we used in the previous section. Moreover, we observe that 
${\cal F}_{[2q],p+q-2,2,p}$ has a disconnected term $\sim U^\theta \hat{\sigma}$ and therefore 
we can redefine the normalization of the double-particle operator so as to keep separated the 
disconnected and connected contributions.

The only subtlety compared with the previous section is in the following OPE channel,
\begin{equation}
    \left(\cO_{p+q-2}\times \cO_p \right)\quad \cap \quad \left( \cO_{2} \times [\cO_{2}\cO_q]\right),
\end{equation}
since the half-BPS exchange at $\gamma=\gamma_m+2=2+q$ is the double-particle $[\cO_2\cO_q]$.
The system of equations for $a,b,c$ then reads,
\begin{equation}\label{eq:abcdouble}
    \begin{aligned}
        a+b &= \CC^{(1)}_{[2q],p+q-2,p} \CC^{(\frac12)}_{p,2,p}, \\
        b+c &= \CC^{(1)}_{[2q],p,p+q-2} \CC^{(\frac12)}_{p+q-2,2,p+q-2}, \\
         a+c &= \CC^{(1)}_{p+q-2,p,[2q]} \CC^{(\frac12)}_{[2q],2,[2q]}. \\
    \end{aligned}
\end{equation}
and it depends on three types of three-point couplings. 

However, the $O(c^{-\frac12})$ three-point function involving two double-particles 
distributes into single-particle three-point functions as follows,
\begin{equation}
    \CC^{(\frac12)}_{[2q],2,[2q]} =\begin{tikzpicture}[ baseline=(current bounding box.center), scale=0.65, dot/.style={circle,fill,inner sep=1.5pt} ] 
    \node[dot,label=above:$2$] (T) at (1,1.8) {}; 
    \node[dot,label=left:$2$] (L) at (0,0) {}; 
    \node[dot,label=right:$2$] (R) at (2,0) {}; 
    \node[dot] (V) at (1,0.6) {};
    \node[dot,label=left:$q$] (A) at (0,-0.6) {};
    \node[dot,label=right:$q$] (B) at (2,-0.6) {}; 
    \draw (T)--(V) (L)--(V)--(R); \draw (A)--(B); 
    \end{tikzpicture} 
    + 
    \begin{tikzpicture}[ baseline=(current bounding box.center), scale=0.65, dot/.style={circle,fill,inner sep=1.5pt} ] \node[dot,label=above:$2$] (T) at (1,1.8) {}; 
    \node[dot,label=left:$q$] (L) at (0,0) {}; 
    \node[dot,label=right:$q$] (R) at (2,0) {}; 
    \node[dot] (V) at (1,0.6) {};
    \node[dot,label=left:$2$] (A) at (0,-0.6) {};
    \node[dot,label=right:$2$] (B) at (2,-0.6) {}; 
    \draw (T)--(V) (L)--(V)--(R); 
    \draw (A)--(B); 
    \end{tikzpicture} = \CC^{(\frac12)}_{2,2,2} + \CC^{(\frac12)}_{q,q,2}.
\end{equation}
This follows from large-$c$ factorisation of the five-point function $\langle \cO_2 \cO_q \cO_2 \cO_2 \cO_q \rangle $.

The other double-particle data,
\begin{equation}\label{eq:C1}
{\cal C}^{(1)}_{[2q],p,p+q-2},
\end{equation}
is computed from the double-particle limit of a (single-particle) four-point function, as explicitly performed in section \ref{sec:doublelimit}. This requires both the single-particle $\FF^{(1)}_{2, q, p , p+q-2}$ \eqref{eq:Fsingle} calculated above, as well as the reduced correlator ${\cal H}^{(1)}_{2, q, p , p+q-2}$.
We independently reproduce the $\theta=2$ result using the chiral algebra in section \ref{chiralalgebra}.

All in all, we solve \eqref{eq:abcdouble} leaving \eqref{eq:C1} unfixed, 
finding
\begin{equation}\label{abc_doublep}
a=2\sqrt{2}  {\cal C}^{(1)}_{[2q],p,p+q-2}, \qquad b=\sqrt{2} (p-2){\cal C}^{(1)}_{[2q],p,p+q-2}, \qquad c=\sqrt{2} q{\cal C}^{(1)}_{[2q],p,p+q-2},
\end{equation}
which for $\theta=2$ matches an exact calculation using the chiral algebra in Appendix \ref{sec:AppWalgebra}.

\subsection{The double-particle limit at four-points}\label{sec:doublelimit}

In this section, we compute the three-point coupling ${\cal C}^{(1)}_{[2q],p,p+q-2}$, the last unknown needed to close 
the system of equations in \eqref{abc_doublep}. We  obtain this coupling following the approach of \cite{Aprile:2026uxe} and using the half-BPS double-particle limit on the four-point functions,
\begin{equation}\label{doubleparticle_4pts}
  \lim_{X_0\rightarrow X_1}\Bigg[ \lim_{Y_0\rightarrow Y_1}\frac{ \langle {\cal O}_2(Z_0) {\cal O}_q(Z_1) {\cal O}_{p}(Z_2) {\cal O}_{p+q-2}(Z_3)\rangle}{ |{\cal O}_2| |{\cal O}_q| |{\cal O}_p| {\cal O}_{p+q-2}| }\Bigg] = \prod_{i\neq j} g_{ij}^{d_{ij}}{\cal C}_{[2q],p,p+q-2}
\end{equation}
where 
\begin{equation}
d_{12}=2 \qquad\qquad d_{13}=q\qquad\qquad d_{23}=p-2\,,
\end{equation}
and more precisely: we first take the coincident limit on the $R$-symmetry 
variables $Y_0\rightarrow Y_1$ in order to project on the half-BPS sector, and then the coincident limit $X_0\rightarrow X_1$. 
Crucially, the double-particle limit is taken on the full correlator, 
\begin{equation}\label{how_to_3ptcoupling_withdp_1}
\!\!\!\langle {\cal O}_2 {\cal O}_q {\cal O}_{p} {\cal O}_{p+q-2}\rangle
=  {\cal P}_{2,q,p,p+q-2} \Bigg[ {\cal F}_{2,q,p,p+q-2} + U^{2\theta} D_{\theta}\Big[ {\cal I}(U,V,\hat{\sigma},\hat{\tau}) {\cal H}_{2,q,p,p+q-2}(U,V)\Bigg]
\end{equation}
where, a priori, both ${\cal F}_{2,q,p,p+q-2}(U,V,\hat\sigma,\hat\tau)$ and ${\cal H}_{2,q,p,p+q-2}(U,V)$ contribute.

In four-dimensional ${\cal N}=4$ SYM theory, the result of taking the double-particle limit  \eqref{doubleparticle_4pts} 
depends only on the free-theory contribution as a consequence of the partial non-renormalisation theorem. 
This was also explicitly checked in \cite[Appendix B]{Aprile:2026uxe}.
In the six-dimensional $\cN=(2,0)$ theory, by contrast, both ${\cal F}_{2,q,p,p+q-2}(U,V,\hat\sigma,\hat\tau)$ and ${\cal H}_{2,q,p,p+q-2}(U,V)$ contribute. 
Our computation therefore provides a particularly instructive illustration of the technique.

As a preliminary step, note that we can always expand our correlator \eqref{how_to_3ptcoupling_withdp_1} in a basis of six propagator structures. 
This is obvious for ${\cal F}_{2,q,p,p+q-2}(U,V,\hat\sigma,\hat\tau)$ but it will also be true for the total contribution seeded by ${\cal H}_{2,q,p,p+q-2}$. Then, given the explicit form of the prefactor,  
\begin{equation}
{\cal P}_{2,q,p,p+q-2}= |{\cal O}_2| |{\cal O}_q| |{\cal O}_p| {\cal O}_{p+q-2}| g_{01}^2 g_{13}^{q-2} g_{23}^p ,
\end{equation}
we readily obtain the relation
\begin{equation}\label{how_to_3ptcoupling_withdp_2}
\lim_{Y_0\rightarrow Y_1} g_{01}^2 g_{13}^{q-2} g_{23}^p  \Bigg[  a  U^{\theta}\hat{\sigma}+ c \frac{U^{\theta} \hat{\tau} }{V^\theta}  
+ b_{11} \frac{U^{2\theta} \hat{\sigma}\hat{\tau}}{V^\theta} + b_{20} U^{2\theta}\hat{\sigma}^2 + b_{02} \frac{U^{2\theta}\hat{\tau}^2}{V^{2\theta}} \Bigg]= \prod_{i\neq j} g_{ij}^{d_{ij}} \Bigg[ b_{11}+ b_{20}+b_{02}\Bigg],
\end{equation}
where we assumed $a,c,b_{ij}$ to be generic functions of $U,V$. At this point, the first part of the double-particle limit follows from matching the $b_{ij}$ functions with those obtained from \eqref{how_to_3ptcoupling_withdp_1}. 

Computing the double-particle limit on ${\cal F}_{2,q,p,p+q-2}(U,V,\hat\sigma,\hat\tau)$ using \eqref{how_to_3ptcoupling_withdp_2}
is straightforward, since the $b_{ij}$ are constants. From our result in \eqref{result_forFsingleparticle}, it follows that
\begin{equation}
b^{\cal F}_{02}=b^{\cal F}_{20}=0,\qquad\qquad b_{11}^{\cal F} \equiv b^{\cal F} \neq 0,
\end{equation}
with
\begin{equation}
    \begin{aligned}
         b^{\cal F}\Bigg|_{\theta=1}&=  (p-1)\sqrt{2 p q(p+q-2)}, \\
b^{\cal F}\Bigg|_{\theta=2}&=
\frac{ \sqrt{2} \frac{\Gamma[p+q-1]}{\Gamma[p-1]\Gamma[q-1]} \Gamma[-1+2p]\Gamma[-1+2q]}{
(\Gamma[-1+2p]\Gamma[-1+2q]\Gamma[-1+2(p+q-2)])^\frac{1}{2}
}\bigg[\frac{1}{(-1+q)} \bigg].
    \end{aligned}
\end{equation}

To compute the double-particle limit on the total contribution coming from ${\cal H}_{2,q,p,p+q-2}$ 
we expand the $R$-symmetry structures of ${\cal I}(U,V,\hat{\sigma},\hat{\tau})$ 
and identify the $b_{ij}$. We find that 
\begin{equation}\label{how_to_3ptcoupling_withdp_3}
b^{\cal H}_{11} +b^{\cal H}_{20}+b^{\cal H}_{02} = \Bigg[V^{\theta} D_{\theta} (x_1+x_2-2)  + D_{\theta} (1-x_1)(1-x_2)  + V^{2\theta}D_{\theta} \Bigg]  U{\cal H}_{2,q,p,p+q-2}(U,V),
\end{equation} 
where the object in square parenthesis is the sum of three operators acting on 
$U{\cal H}_{2,q,p,p+q-2}$. The $b_{ij}$ are now functions of $U,V$ and therefore 
we have to further take the limit ${X_0\rightarrow X_1}$. This corresponds to $x_1,x_2\rightarrow0$, 
equivalently $V\rightarrow 1$ and $U\rightarrow 0$. Under $V\rightarrow 1$ 
the operators in \eqref{how_to_3ptcoupling_withdp_3} simplify, 
and we obtain the intermediate result
 \begin{equation}\label{how_to_3ptcoupling_withdp_4}
\lim_{X_0\rightarrow X_1}  b^{\cal H}_{11} +b^{\cal H}_{20}+b^{\cal H}_{02} 
\equiv b^{\cal H} = \lim_{x_i\rightarrow0} \Bigg[ D_\theta (x_1 x_2) \Bigg] U{\cal H}_{2,q,p,p+q-2}(U,V). 
\end{equation}
To proceed, we need an explicit expression for the reduced correlator ${\cal H}_{2,q,p,p+q-2}$. 
We derive this expression and compute the relevant limit in the following intermezzo (section \ref{intermezzo}). 
Here, we simply quote the result and highlight the key difference between the four-dimensional ${\cal N}=4$ SYM 
and the six-dimensional $(2,0)$ theory:
\begin{equation}
b^{\cal H}\Bigg|_{\theta=1}=0\qquad\qquad b^{\cal H}\Bigg|_{\theta=2}\neq 0
\end{equation}

Then, putting together the ${\cal F}$ and ${\cal H}$ contribution we finally find
\begin{equation}\label{dp_result1}
{\cal C}^{(1)}_{[2q],p,p+q-2}\Bigg|_{\theta=1}= (p-1)\sqrt{2 p q(p+q-2)},
\end{equation}
and
\begin{equation}\label{dp_result2}
{\cal C}^{(1)}_{[2q],p,p+q-2}\Bigg|_{\theta=2}=
\frac{ \sqrt{2} \frac{\Gamma[p+q-1]}{\Gamma[p-1]\Gamma[q-1]} \Gamma[2p-1]\Gamma[2q-1]}{
(\Gamma[2p-1]\Gamma[2q-1]\Gamma[2(p+q-2)-1])^\frac{1}{2}
}\Bigg[ \frac{2pq+p+4q-7}{(p-1)(q-1)(2q+1)} \Bigg].
\end{equation}

\subsubsection{Intermezzo: the reduced correlator ${\cal H}_{2,q,p,p+q-2}$}\label{intermezzo}

In the four-dimensional ${\cal N}=4$ SYM theory, the reduced correlator 
${\cal H}_{2,q,p,p+q-2}$ at tree-level was computed for the first time in \cite{Uruchurtu:2011wh}. 
In the six-dimensional $(2,0)$ theory, a Mellin space formula for tree-level 
correlators with arbitrary external charges was given \cite{Alday:2020lbp}, up to unfixed normalizations. 
For our case, we derive the latter independently following the 
procedure described in Appendix \ref{app:subtlety}. Checking that all these results are consistent 
with each other we finally extract ${\cal H}_{2,q,p,p+q-2}$ also in 
the six-dimensional theory.  We quote here below the relevant formulae:
\begin{equation}\label{how_to_3ptcoupling_withdp_5}
{\cal H}^{(1)}_{2,q,p,p+q-2}(U,V)= {\cal N}^{(1)}_{p,q} \oint  \frac{ds dt}{(2\pi i)^2} U^s V^t 
\frac{ \pmb{\Gamma}_{p,q}}{ (s+1)(t+1)(u+1) },
\end{equation}
where $u=-s-t-4$, and we introduced  
\begin{equation}\label{eq:Gammakernel}
\pmb{\Gamma}_{p,q}= \frac{ \Gamma[\theta(p-2)+(\theta-1)-s]\Gamma[-s]\Gamma[(\theta-1)-t]\Gamma[-t]\Gamma[\theta(q-2)+(\theta-1)-u]\Gamma[-u]}{\Gamma[p-1]\Gamma[q-1]}.
\end{equation}
The denominator is analogous to the sphere gamma functions introduced in \cite{Aprile:2020luw}. 
We note that it takes the same form independently of $\theta$, as might 
be expected since it encodes the $R$-symmetry selection rules.
Finally, the normalizations are
\begin{equation}
{\cal N}^{(1)}_{p,q}\Bigg|_{\theta=1}= \sqrt{2(q)(p)(p+q-2)},\qquad\qquad {\cal N}^{(1)}_{p,q}\Bigg|_{\theta=2}= - \frac{4 \Gamma[p+q-1]}{(\prod_{i=1}^{4} \Gamma[-1+2 (p_i)])^{\frac{1}{2}}}.
\end{equation}
In our notation, the case $\theta=1$ is known from \cite{Aprile:2018efk}.

It is useful to understand how the poles in \eqref{how_to_3ptcoupling_withdp_5} 
map onto the OPE. Following the notation of \cite{Aprile:2019rep}, which is similar to the 
one introduced in the previous sections, let us first consider the above-window, and hence the first double poles 
of the Gamma functions in \eqref{eq:Gammakernel}. 
In the s-channel, the lightest operators exchanged above-window
have twist $\theta(2p+q-2)$. This is a particular case of the general formula $\theta(\max(p_1+p_2,p_3+p_4))$. 
The prefactor in the s-channel carries twist $\theta(q-2)$, corresponding to $\theta\gamma_m$. Taking into account $\Upsilon_{\theta}$, 
which carries twist $4+2(\theta-1)$, we find an equation for the power of $U$ corresponding to the first double pole in the s-channel. 
The result is $s=\theta p -(\theta+1)$ which agrees with the value obtained from the \eqref{eq:Gammakernel}. 
The same argument can be repeated in the $t$ and $u$ channel. The resulting formula, valid in all three channels $x=s,t,u$, is 
\begin{equation}\label{equation_doublepoles}
\theta \emph{M}_{\vec{p}} = \theta \gamma_{m} + 2x + 4 + 2(\theta-1), \qquad\qquad \emph{M}_{\vec{p}}\equiv\max(p_1+p_2,p_3+p_4),
\end{equation}
with solution $x=\frac{\theta}{2}(\emph{M}_{\vec{p}}-\gamma_m) - (\theta+1)$.

On the other hand, the simple poles $s=-1$, $t=-1$, $u=-1$ correspond to contributions at the unitarity bound in the corresponding channel.

Returning to our task, what is important to understand is the series expansion of $U{\cal H}^{(1)}_{2,q,p,p+q-2}$ which we want to plug in \eqref{how_to_3ptcoupling_withdp_4} to obtain the limit. From the Mellin amplitude in \eqref{how_to_3ptcoupling_withdp_5} we see that the relevant pole in the $s$ plane is the simple pole at $s=-1$. We thus expect that
\begin{equation}
\lim_{x_i\rightarrow 0 }U{\cal H}^{(1)}_{2,q,p,p+q-2}= b^{\cal H} ,
\end{equation}
where $b^{\cal H}$ is a constant.
With this information, the limit in \eqref{how_to_3ptcoupling_withdp_4} reduces 
to the computation of the leading order term of $[D_{\theta}(x_1x_2)]b^{\cal H}$ 
in the limit $x_i\rightarrow 0$. Note now that for $\theta=1$ the result vanishes 
trivially, because $D_{1}=1$ and $x_1 x_2\rightarrow 0$, therefore
\begin{equation}
b^{\cal H}\Bigg|_{\theta=1}=0.
\end{equation}
Instead, when $\theta=2$ 
the result of acting with $D_2$ yields $[D_{2}(x_1x_2)]b^{\cal H}= 3 b^{\cal H}$ so that
\begin{equation}
b^{\cal H}\Bigg|_{\theta=2}= \frac{ 12 \frac{\Gamma[p+q-1]}{\Gamma[p-1]\Gamma[q-1]}}{(\prod_{i=1}^4 \Gamma[-1+2p_i])^\frac{1}{2}}\frac{\Gamma[-1+2p]\Gamma[-1+2q]}{(-1+p)(1+2q)}.
\end{equation}
With the two results above, the derivation of 
\eqref{dp_result1}-\eqref{dp_result2} is straightforward.   

\subsection{Agreement with the chiral algebra}
\label{chiralalgebra}

An alternative definition of composite half-BPS operators in the 6d (2,0) 
theories comes from their underlying chiral algebras $\WW_\mathfrak{g}$ 
proposed in \cite{Beem:2014kka}. In this proposal, distinguished 6d 
half-BPS operators $\OO_p$ with dimension $\Delta=2p$ map to 2d chiral algebra operators as 
\begin{align}
    \OO_{p}\leftrightarrow W_p,
\end{align}
where $W_2\equiv T$ is the Virasoro stress tensor and $W_{p>2}$ are the 
additional 2d chiral currents with holomorphic weight $h=p$, needed to 
generate $\WW_\mathfrak{g}$. We emphasize that $\{W_{p_i}\}$ do not exhaust 
the list of quasi-primaries in $\WW_\mathfrak{g}$ and for general half-BPS 
operators $O_{\Delta}$ we have the more general map 
\begin{align}
    O_{\Delta}\leftrightarrow \hat{O}_{h=\frac{\Delta}{2}},
\end{align}
where three point coefficients in 6d are mapped to chiral algebra structure constants as 
\begin{align}
\CC_{p_1,p_2,p_3}\leftrightarrow C_{\hat{O}_{p_1}\hat{O}_{p_2}}^{\hat{O}_{p_3}}.
\end{align}
To organize the space of quasi-primaries with fixed weight $h=p$, it will be convenient to choose $W_{p>2}$ to be Virasoro primary operators. These strong generator operators are contrasted by composite operators such as $\hat{O}_{h=p_1+p_2,W_{p_1}W_{p_2}}$, which arise as the quasi-primaries involving the normal ordered product $:W_{p_1}W_{p_2}:$ which appears in the first regular term in the OPE $W_{p_1}\times W_{p_2}$. For $p\geq4$ there is a degenerate space of such operators, however it is always possible to make these orthogonal to the stronger generators $W_p$. For instance, the composite operators $\left[\OO_{2}\OO_{q}\right]$ considered in this work arise as the 6d uplift of the quasi-primary operators
\begin{align}
    \hat O_{q+2,TW_q}(z)=&\; \NN_{TW_q}\left(TW_q(z)-\frac{3}{2 (2 q+1)}\sqrt{\frac{2}{c}}W_q''(z)\right),
\end{align}
where, as we will explain in what follows, it will be convenient to choose the normalization
\begin{align}\label{eq:2qnorm}
    \NN_{TW_q}=\frac{\delta_{2,q}+1}{\delta_{2,q}+1+\frac{1}{c}\frac{2q(8q-5)}{2q+1}},\hspace{0.5cm}\rightarrow\hspace{0.5cm}\Big\langle\hat O_{q+2,TW_q}\Big|\hat O_{q+2,TW_q}\Big\rangle=&\;\frac{\left(\delta_{2,q}+1\right)^2}{\delta_{2,q}+1+\frac{1}{c}\frac{2q(8q-5)}{2q+1}}\nonumber\\
    =&\;1+\delta_{2,q}+O\left(c^{-1}\right).
\end{align}
Note that we will unit-normalize all remaining operators\footnote{In particular, our Virasoro stress tensor is rescaled to be $T=\sqrt{\frac{2}{c}}T_\text{canonical}$.}. Thanks to the Virasoro primary property of $\{W_{k}\}$, the operators $\hat O_{q+2,TW_q}$ are automatically orthogonal to the strong generator $W_{q+2}$ and can be made orthogonal to all other quasi-primaries of weight $h=q+2$.

That the strong generator operators $\OO_p\leftrightarrow W_p$ coincide with single-particle states in $\AdS_7\times \mathrm{S}^4$ is supported by the matching of three-point function coefficients of strong generators with sub-extremal three-point couplings of the KK-modes $\AdS_7$, where the authors of  \cite{Beem:2014kka} used $\WW_{N-1}$ at large $N$ to determine
\begin{align}
    \CC_{p_1,p_2,p_3}^{\left(\frac{1}{2}\right)}=\frac{2^{p_1+p_2+p_3-1}}{\left(\pi^3\right)^\frac{1}{2}}\Gamma\left[\frac{p_1+p_2+p_3}{2}\right]\frac{\Gamma\left[\frac{p_2+p_3-p_1+1}{2}\right]\Gamma\left[\frac{p_3+p_1-p_2+1}{2}\right]\Gamma\left[\frac{p_1+p_2-p_3+1}{2}\right]}{\sqrt{\Gamma\left[2p_1-1\right]\Gamma\left[2p_2-1\right]\Gamma\left[2p_3-1\right]}},
\label{eq:11dsugraC}
\end{align}
which is in agreement with the 11d supergravity computations of \cite{Bastianelli:1999en,Corrado:1999pi}. Moreover, one can use the OPEs of $\WW_\mathfrak{g}$ and the twisted correlator uplift of \cite{Woolley:2025qbd} (also outlined in Appendix \ref{sec:AppWalgebra})  to determine that extremal four-point correlators vanish when at least one of the operators is a composite operator. It then follows that the remaining 6d half-BPS operators uplift from composite quasi-primaries in $\WW_\mathfrak{g}$ and correspond to threshold bound states in the bulk.

Thus, the chiral algebra construction can be used to compute the protected contributions to $n$-point correlators. For instance, we use the mode expansions 
\begin{align}
    T(z)=\sum_{n\in\mathbb{Z}}z^{-n-2}L_n,\hspace{2cm}W_q(z)=\sum_{n\in\mathbb{Z}}z^{-n-q}\left(W_q\right)_n,
\end{align}
combined with the Virasoro algebra and the Virasoro primary property of $W_q$ to determine that
\begin{align}\label{eq:dp_resultchiralq}
    C_{q,2}^{[2q]}=\bra{W_q}T(1)\ket{\hat O_{q+2,TW_q}}=\delta_{2,q}+1,
\end{align}
which is independent of $c$, thanks to the normalization in \eqref{eq:2qnorm}. The $\delta_{2,q}$ signals an enhancement when $W_q$ is the (quasi-primary) Virasoro stress tensor $T$. Moreover, we can make contact with Equation \eqref{dp_result2} by using Virasoro symmetry to determine that the $W_p\times \hat O_{q+2,TW_q}$ OPE exchanges the $W_{p+q-2}$ current with coefficient\footnote{Note that we do not require structural knowledge of the $\mathcal{W}$-algebra to obtain this result. The constraints imposed by the $\cal{W}$-algebra Jacobi identities on $C_{q,p}^{p+q-2}$ can be conveniently calculated using the \texttt{Mathematica} package \texttt{OPEdefs} introduced in \cite{Thielemans:1991uw,Thielemans:1994er}.}
\begin{align}\label{eq:dp_resultchiralpq}
    C_{[2q],p}^{p+q-2}=&\;\bra{\hat O_{q+2,TW_q}}W_p(1)\ket{W_{p+q-2}}\nonumber\\
    =&\;\frac{\delta_{2,q}+1}{\delta_{2,q}+1+\frac{1}{c}\frac{2q(8q-5)}{2q+1}}\left(
\delta _{2,p} (\delta _{2,q}+1) +\frac{  (2 p q+p+4 q-7)}{2 q+1}\sqrt{\frac{2}{c}}C_{q,p}^{p+q-2}\right),
\end{align}
in terms of the $\cal{W}$-algebra structure constant $C_{q,p}^{p+q-2}\leftrightarrow \CC_{q,p,p+q-2}$. Using Equation \eqref{eq:11dsugraC}, we find that the leading order contributions are\footnote{Our normalization in \eqref{eq:2qnorm} removes factors of $\sqrt{\delta_{2,q}+1}$ from these expressions and causes $C_{[2q],p}^{p+q-2\;(1)}$ to vanish for $p=2$. Moreover, any other choice would have induced a $U^{2\theta}\sigma^2$ term at $O\left(c^{-3/2}\right)$ in the $\mathcal{F}_{[2q],p+q-2,2,p}$ that we compute in Appendix \ref{sec:AppWalgebra} which would conflict with our distinction between disconnected and connected contributions in section \ref{formulae_freet_db}.}
\begin{align}
   &\; C_{[2q],p}^{p+q-2\;(0)}=\delta _{2,p} \left(\delta _{2,q}+1\right),\label{eq:eq:dp_resultchiralpqO1/2}\\
   &\;C_{[2q],p}^{p+q-2\;(1)}=\frac{2q\left(5-8 q\right) \delta _{2,p}}{(2 q+1)}+\frac{2^{2 (p+q-2)+\frac{1}{2}} (2 p q+p+4 q-7) \Gamma \left[p-\frac{1}{2}\right] \Gamma \left[q-\frac{1}{2}\right] \Gamma \left[p+q-1\right]}{\pi  (2 q+1) \sqrt{\Gamma \left[2 p-1\right]\Gamma \left[2 q-1\right]\Gamma \left[2 p+2 q-5\right]} },\label{eq:eq:dp_resultchiralpqO3/2}
\end{align}
where $C_{[2q],p}^{p+q-2\;(0)}$ is an $O(1)$ enhancement in the case of three-point correlators involving multiple Virasoro stress tensors. The quantity $C_{[2q],p}^{p+q-2\;(1)}$ reproduces Equation \eqref{dp_result2}, where for $\text{N}^2$E configurations we needed $p>2$, and it vanishes when $p=2$.

\section{An evaluation formula for superconformal block coefficients}\label{sec:cauchy}

The double-particle predictions require explicit formulae for the block expansion 
of the ${\cal F}$ contributions, corresponding to the free-like part of the correlators.
A brute-force approach would involve decomposing free monomials into superconformal 
blocks and fitting the resulting coefficients as functions of the spin 
at fixed twist. 
For correlators beyond N$^2$E extremality, this procedure 
quickly becomes cumbersome and computationally inefficient. 
As explained in the previous section, although we bootstrap N$^2$E double-particle correlators, 
their predictions require data from single-particle correlators of higher extremality.
Therefore, an efficient way to calculate these block coefficients is essential to our programme.

A direct solution to this problem was developed in 
\cite{Doobary:2015gia,Aprile:2021pwd,Aprile:2025nta}, which derived an evaluation formula for the block coefficients 
of free-like monomials across a broad class of theories, including
\begin{equation}\label{eq:groups}
        \SU(m,m|2n) \ \ (\theta=1), \qquad 
    \OSp(4m^*|2n) \ \ (\theta=2), \qquad \OSp(4n|2m) \ \ (\theta=\tfrac{1}2).
\end{equation}
The understanding of this formula relied 
 on beautiful mathematics connecting the theory of symmetric and interpolation
polynomials \cite{okounkov1997shifted,Okounkov1998BCInterpolationMacdonald,Rains2005BCnSymmetricPolynomials,SergeevVeselov2009BCInfinity} and the theory of 
Heckman-Opdam hypergeometric functions \cite{Heckman1987RootSystemsI,Heckman1987RootSystemsII}, 
see also \cite{Isachenkov:2016gim}.  
In the following, we briefly review the derivation presented in \cite{Aprile:2021pwd} 
and state the main formula in \eqref{eq:cauchyformula}. We then demonstrate explicitly how to evaluate this formula in appendix \ref{app:superspace}.


\subsection{Block coefficients as Jacobi polynomials}

In the following, we keep $(m,n;\theta)$ generic, since this allows the same construction 
to describe several theories at the same time. In fact, we will show that the dependence on $m,n$ is minimal.
Theories with $(m,n)=(m,0)$ include ordinary conformal theories, whereas 
the case $(m,n)=(0,n)$ describes theories on compact manifolds. Theories with 
$m,n \neq 0$ are supersymmetric, and in particular,
the theories with $(m,n)=(2,2)$ are the maximally-supersymmetric conformal 
theories in 4, 6 and 3 dimensions, respectively (with the ordering \eqref{eq:groups}).

Four-point blocks can be defined as eigenfunctions 
of the quadratic Casimir for \eqref{eq:groups}
with given boundary conditions, following the approach of \cite{Dolan:2003hv}. 
For any $(m,n;\theta)$, this definition goes through in a neat way by considering the BC$_{m,n}$ root 
system with parameter $\theta$. We refer to \cite{Aprile:2021pwd} for details. The resulting eigenfunctions, i.e.~the $(m,n;\theta)$ 
blocks, admit the expansion  
\begin{equation}\label{eq:superblock}
    B_{\gamma\ula}(x|y;\theta)= \left(\frac{g_{24} g_{13}}{g_{12} g_{34}}\right)^{\frac{\gamma-\gamma_m}{2}} \sum_{\umu \geq \ula } (T_{\gamma,\theta,p_{12},p_{43}})_{\ula}^{\umu}\, P_{\umu} (x|y;\theta) := \left(\frac{g_{24} g_{13}}{g_{12} g_{34}}\right)^{\frac{\gamma-\gamma_m}{2}} F_{\gamma,\ula}(x|y;\theta).
\end{equation}
where the  $P_{\umu}$ are the super-Jack polynomials of \cite{SergeevVeselov2005GeneralisedDiscriminants}.
The leading one, i.e. for $\umu=\ula$, $P_{\ula} (x|y;\theta)$, is associated with the exchanged superprimary, 
while the sum over $\umu \geq \ula$ accounts for its descendants.  
We refer to appendix \ref{app:blocks} for a summary and explicit expressions for the super-Jacks and their coefficients.

Note that, when $m=0$, theories live on a compact space, their blocks are themselves polynomial, given by the dual 
Jacobi polynomials introduced in \cite{Aprile:2021pwd}. 
In the nomenclature of \cite{Aprile:2021pwd}, 
the $F_{\gamma,\ula}$ are then known as dual super-Jacobi functions for all $(m,n;\theta)$.
We show the explicit form of the blocks above in appendix \ref{app:blocks}.

A crucial property of the $(m,n;\theta)$ blocks, shared also by the super-Jack 
polynomials is \emph{stability}, 
\begin{equation}\label{eq:stability}
F_{\gamma,\ula}(x_1,\ldots x_m,0|y_1,\ldots,y_n)=F_{\gamma,\ula}(x_1,\ldots x_m|y_1,\ldots,y_n,0)=F_{\gamma,\ula}(x_1,\ldots x_m|y_1,\ldots,y_n).
\end{equation}
This property allows us to prove that the super-Jack coefficients 
$(T_{\gamma,\theta,p_{12},p_{43}})_{\ula}^{\umu}$ are theory-independent, 
i.e.~they do not depend on the particular values of $(m,n)$. Instead, 
they depend only on the Young diagrams $\ula,\umu$.  
This can be seen explicitly in the superconformal blocks
derived in \cite{Doobary:2015gia,Aprile:2021pwd}. A corollary is that 
the $(T_{\gamma,\theta,p_{12},p_{43}})_{\ula}^{\umu}$ are the same coefficients 
as for ordinary non-supersymmetric theories, either in $m$ 
(conformal, non-compact) or $n$ (compact).

Returning to our problem, we want to decompose an arbitrary $(m,n;\theta)$ 
free-theory four-point function into its corresponding blocks \eqref{eq:superblock}. 
The set of all possible free-like propagator 
structures  of such a correlator is given by the sum
\begin{equation}\label{eq:freecorr}
    \langle \cO_{p_1} \cO_{p_2} \cO_{p_3} \cO_{p_4} \rangle = g_{12}^{\frac12 (p_1+p_2)} g_{34}^{\frac12 (p_3+p_4)} \bigg(\frac{g_{14}}{g_{24}}\bigg)^{\frac12 p_{12}}\bigg(\frac{g_{14}}{g_{13}}\bigg)^{\frac12 p_{43}}
    \sum_{\gamma=\gamma_{m}}^{\gamma_{M}} \sum_{k=0}^{\frac12(\gamma -\gamma_{\text{min}})}a_{\gamma k}\left(\frac{g_{13} g_{24}}{g_{12} g_{34}}\right)^{\frac{\gamma}{2}} \left(  \frac{g_{14} g_{23}}{g_{13}g_{24}}\right)^k,
\end{equation}
where $\gamma_{m}$ and $\gamma_{M}$ are defined in \eqref{gammam} and \eqref{gammaM}, and
we assumed $p_{12},p_{43}\ge 0$ without loss of generality.
Note that $\gamma$ is equal to the number of propagators going 
from  the pair ${\cal O}_{p_1}{\cal O}_{p_2}$ to the pair 
${\cal O}_{p_3}{\cal O}_{p_4}$ and $k$ counts the number of such diagrams. 
Thus, each pair $\gamma,k$ labels a four-point propagator structure, 
and $a_{\gamma,k}$ is a numerical coefficient that depends on the theory, e.g., a color factor. 

We are interested in computing the block decomposition of a single four-point 
propagator structure labelled by $\gamma,k$, and then assembling the total result.  
Given the form of the blocks \eqref{eq:superblock}, we want to solve for the coefficients  
$M_{k,\gamma,\ula;p_{12},p_{43}}$ in
\begin{equation}
\left(  \frac{g_{14} g_{23}}{g_{13}g_{24}}\right)^k= \sum_{\ula} M_{k,\gamma,\ula;p_{12},p_{43};\theta} F_{\gamma,\ula}(x|y;\theta).
\end{equation}
Using the expression for the cross-ratios \eqref{eq:crossratios} in terms of superpropagators \eqref{two_point_conventions}, we can rewrite the above equation as
\begin{equation}\label{eq:cauchy}
\frac{\prod_i^n (1-y_j)^k}{\prod_i^m (1-x_i)^{k\theta}} = \sum_{\ula} M_{k,\gamma,\ula;p_{12},p_{43};\theta} F_{\gamma,\ula}(x|y;\theta).
\end{equation}
Crucially, the stability of $F_{\gamma,\ula}(x|y;\theta)$, see \eqref{eq:stability}, implies  
that the block coefficients $M_{k,\gamma,\ula}$ are also independent of $m,n$. 
Therefore, we can solve for them
by considering the compact case ($m=0$) where the blocks $F_{\gamma,\ula}$ are dual Jacobi polynomials. 

In fact, equation \eqref{eq:cauchy} is a particular case of a Cauchy identity in disguise: 
\begin{equation}\label{eq:cauchylike}
    \frac{\prod_{r,i}^{\beta,n} (1-Y_r y_j)}{\prod_{r,i}^{\beta,m} (1-Y_r x_i)^{\theta}} = \sum_{\ula} J^{(\beta)}_{\ula} (\{Y_r\};\theta,p_{\pm})  F_{\gamma(\beta), \ula}(x|y;\theta),
    \qquad\qquad \gamma(\beta)=2\beta+p_++p_-,
\end{equation}
where $p_{\pm}=\frac{1}{2}|p_{43}\pm p_{12}|$ 
and where $J^{(\beta)}_{\ula} (\{Y_r\};\theta)$ are Jacobi polynomials \cite{Koornwinder2015OkounkovBC}. 
As explained in  \cite{Aprile:2021pwd}, equation \eqref{eq:cauchylike} is itself a particular case of 
the ``superconformal Cauchy identity"
which pairs super-Jacobi polynomials of Young diagram $\ula$ with 
the dual super-Jacobi polynomials with complemented Young diagram.
Matching \eqref{eq:cauchylike} to the expansion of the propagator structure \eqref{eq:cauchy} 
we find,
\begin{equation}\label{eq:proprewriting}
 Y_r=\left\{\begin{aligned}
     &1, \quad r=1,\dots,k,\\
    &0, \quad r=k+1,\dots,\beta,
 \end{aligned}\right.\qquad\qquad 2\beta+p_++p_- = \gamma.
\end{equation}
We shall refer to the set $\{Y_r\}$ as a binary string. 

Solving for $\beta$ we recognize that $\beta=\frac{1}{2}(\gamma-\gamma_{m})$, with $\gamma_m$ \eqref{gammam}. Thus, using the Cauchy-like identity \eqref{eq:cauchylike} we obtain
\begin{equation}\label{eq:cauchyformula}
    M^{}_{k,\gamma, \ula; p_{12},p_{43};\theta} = J^{(\beta)}_{\ula} (\{Y_r\};\theta, p_{\pm}),\qquad\qquad \beta\equiv \frac{\gamma-\gamma_m}{2}.
\end{equation}
For the reader's convenience we give the relevant formulae for the Jacobi polynomials 
in Appendix \ref{app:superspace}. These formulae reduce the computation 
to a sum over semistandard Young tableaux, which is scalable in the size of the Young diagrams 
and can be implemented efficiently on a computer. The formula for $\theta=1$ can be 
written in a particularly explicit form by rearranging the determinantal representation of Jacobi polynomials, 
as shown in \cite{Aprile:2025nta}. This work provides the first application of
\eqref{eq:cauchyformula} to holographic correlators in $\theta=2$, i.e. 6 dimensions.

\subsection{Application to our predictions}\label{subsec:applyingcauchy}
We illustrate below how we use \eqref{eq:cauchyformula} to obtain block coefficients 
relevant for the predictions of sections \ref{2_p_sect}, \ref{2_pqm2_sect} and \ref{2q_2_sect}. 
In the following, the dependence on $\theta$ is omitted to avoid clutter. Let us also define,
\begin{equation}\label{M_asinMicheleformula}
 M^{p_1p_2p_3p_4}_{k,\gamma,\ula} \equiv M^{}_{k,\gamma, \ula,p_{12},p_{43}}\,,
\end{equation}
to make the individual external charges and the OPE channels more explicit, though the block coefficients depend only on the differences of the charges of both OPEs. Furthermore, note that these coefficients also do not depend on the nature of the external operators; they depend only on their total charges.

\paragraph{Disconnected contributions.}
As a first application, the disconnected contributions ${\cal C}^{(0)}$ appearing in the denominators of our predictions (see, e.g.~\eqref{eq:predictionlog}) are always given by
\begin{equation}
{\cal C}^{(0)}_{r,s,r,s}\big|_{\gamma,\ula}= M_{0,\gamma,\ula}^{r,s,r,s} + \delta_{r,s} M_{r,\gamma,\ula}^{r,s,r,s}.
\end{equation}

\paragraph{Long sector contributions.}
In the long sector, for example of the ${\cal O}_2\times {\cal O}_{p+q-2}$, with predictions given by \eqref{summary_long_2pqm2}
we will substitute the following combinations
\begin{align}
{\cal C}^{(\frac{1}{2})}_{[2q],p,2,p+q-2}\big|_{\gamma,\ula}&= 
 {\cal C}^{(\frac{1}{2})}_{q,p,p+q-2} M^{q+2,p,2,p+q-2}_{0,\gamma,\ula}, \\
 {\cal C}^{(\frac{1}{2})}_{[2q],p,q,p}\big|_{\gamma,\ula}&=
 {\cal C}^{(\frac{1}{2})}_{2,p,p} M^{q+2,p,q,p}_{0,\gamma,\ula} + \delta_{p,q} 
 {\cal C}^{(\frac{1}{2})}_{2,p,p} M^{q+2,p,q,p}_{q-1,\gamma,\ula}.
\end{align}
The  discussion in the ${\cal O}_2\times {\cal O}_p$ channel is identical.

\paragraph{Unitary bound contributions.}
Consider the channel 
$[{\cal O}_2{\cal O}_q]\times {\cal O}_2$, and the prediction 
in \eqref{eq:prednonlog}, where various inputs are needed. First, 
the correlator $\CC^{}_{[2q],2,2,q}$ in 
\eqref{corr_2q22q} will give the contribution, 
\begin{equation}
\CC^{(\frac12)}_{[2q],2,2,q}\big|_{\gamma=q+2,[2+l]}= \CC ^{(\frac12)}_{q,2,q} M^{q+2,2,2,q}_{0,q+2,[2+l]}
+\CC ^{(\frac12)}_{2,2,2} M^{q+2,2,2,q}_{1,q+2,[2+l]}.
\end{equation}
Similarly,
\begin{equation}
{\cal F}^{(\frac32)}_{[2q],2,p,p+q-2}\big|_{\gamma=q+2,[2+l]}=
\sqrt2\, q\, {\cal C}^{(1)}_{[2q],p,p+q-2}\, M^{q+2,2,p,p+q-2}_{0,q+2,[2+l]}
+2\sqrt2\, {\cal C}^{(1)}_{[2q],p,p+q-2}\, M^{q+2,2,p,p+q-2}_{1,q+2,[2+l]}\,.
\end{equation}
where ${\cal C}^{(1)}_{[2q],p,p+q-2}$ is given in \eqref{dp_result1}-\eqref{dp_result2}. 
Finally, the ${\cal F}$ part of the correlator $\CC^{(1)}_{2,q,p,p+q-2}$ will contribute as
\begin{equation}
{\cal F}^{(1)}_{2,q,p,p+q-2}\big|_{\gamma=q+2,[2+l]} =  \sqrt2 (p-1) \CC^{(\frac12)}_{p,q,p+q-2} M^{2,q,p,p+q-2}_{1,q+2,[2+l]}.
\end{equation}
This completes all the free-like contributions entering \eqref{eq_prednonlog2}.

\section{Results}\label{sec_results}

Having computed all predictions from the double-particle bootstrap, 
see section \ref{sec:doublebootstrap}, we insert them into the relevant 
(reduced) superconformal block expansions and perform the sum over spin for 
all the allowed twist values discussed in section  \ref{2_p_sect}, 
\ref{2_pqm2_sect} and \ref{2q_2_sect}. 
These results, together with crossing symmetry provide the input 
to obtain the reduced correlator $\HH^{(\frac32)}$.
The result is shown to be given as a sum of $D$-functions for all $q$ and $p$.
However, even for a given $q$, obtaining a compact position-space 
expression with unfixed $p$ is complicated, as already pointed out 
in the context of $\cN=4$ SYM in \cite{Aprile:2026uxe}.

We thus turn directly to Mellin space \cite{Penedones:2010ue} to present our results.
We refer to appendix \ref{app:blocks} for explicit formulae for the superconformal blocks, and to appendix \ref{app:subtlety} for the technical details of the block resummation at the unitarity bound, which as we will see fixes most of the Mellin amplitude.

\subsection{Anatomy of the Mellin amplitude}\label{subsec:Mellinansatz}

In general, the Mellin representation of the four-point reduced correlator takes the form
\begin{equation}
   {\cal H}_{\vec{p}} = \oint \frac{ds dt}{(2\pi i)^2} U^s V^t {\bf \Gamma}_{\vec{p}}(s,t) \MM_{\vec p}(s,t),
\end{equation}
where ${\bf \Gamma}_{\vec{p}}$ is a kernel made up of Gamma-functions, 
and $\MM_{\vec p}(s,t)$ is the Mellin amplitude of the reduced correlator. 
Mellin amplitudes are closely related to flat-space scattering amplitudes, as first described in \cite{Penedones:2010ue}.
In particular, the poles of $\MM_{\vec p}$ are in correspondence with the exchanged operators, 
and their residues are determined by the corresponding OPE data.

In the rest of this section, we choose the $s$-channel to be the orientation
\begin{equation}\label{eq:thecorrelator}
    \langle [\cO_2 \cO_q] \cO_{p+q-2} \cO_2 \cO_p\rangle,
\end{equation}
so that $\vec{p}=(2+q,p+q-2,2,p)$. Furthermore, since this is a two-parameter family of correlators, 
we will replace the label $\vec p$ by $q,p$. Our task will be to bootstrap the tree-level Mellin amplitude
\begin{equation}
\MM^{(\frac{3}2)}_{q,p}(s,t)\,.
\end{equation}
The dictionary between the results in section \ref{sec:doublebootstrap}, 
and the predictions for our Mellin amplitude is
\begin{equation}\label{eq:stutable}
    \begin{tabular}{c|c|c}
        Mellin channel & OPE channel & Relevant predictions \\ \hline
       $s$\text{-channel} & $\left([\cO_2\cO_q]\times \cO_{p+q-2}\right) \cap\left(\cO_2\times \cO_{p}\right)$& \text{section \ref{2_p_sect}}  \\
         $t$\text{-channel} & $\left([\cO_2\cO_q]\times \cO_{p}\right) \cap\left(\cO_2\times \cO_{p+q-2}\right)$& \text{section \ref{2_pqm2_sect}}\\
         $u$\text{-channel} & $\left([\cO_2\cO_q]\times \cO_{2}\right) \cap\left(\cO_p\times \cO_{p+q-2}\right)$& \text{section \ref{2q_2_sect}}
    \end{tabular}
\end{equation}
The Mellin variables $s,t,u$ satisfy $s+t+u=-4$.

The kernel ${\bf \Gamma}_{q,p}$ is purely kinematical and thus does not depend on the specific form of the operators. 
For correlators of single-particle operators, crossing symmetry fixes a unique choice of kernel, 
whose Gamma functions are sometimes referred to as Mack Gamma functions \cite{Mack:2009mi}. 
Inspired by this, we make the following choice, 
\begin{equation}\label{eq:kernel}
    {\bf \Gamma}_{q,p}(s,t)  = \frac{\Gamma [-s] \Gamma [-t] \Gamma [-u] \Gamma [-t+2 \theta -1]
   \Gamma [-u+(p-2) \theta -1] \Gamma [-s+q \theta -1]}{\Gamma[q] \Gamma[p-2]}.
\end{equation}
The denominator is analogous to the sphere gamma functions introduced in \cite{Aprile:2020luw}. 
As in the single-particle case in \eqref{eq:Gammakernel}, it is independent of $\theta$ and depends 
only on the total charge of the external operators. As 
in that case, it encodes the $R$-symmetry selection rules.

Now, since our goal is to bootstrap $\MM^{(\frac{3}2)}_{q,p}(s,t)$, we explain below how 
the above choice of kernel relates to the actual OPE predictions in each channel.

In each channel $x=s,t,u$, the double poles arising from ${\bf \Gamma}_{q,p}(s,t)$ begin at
\begin{equation}\label{fromtwist_to_x}
    x=\frac{\theta}{2}(M_{\vec{p}}-\gamma_m) - (\theta+1),\qquad\qquad M_{\vec{p}}=\max(p_1+p_2,p_3+p_4).
\end{equation}
This value of $x$ equals the threshold of triple-particle operators in the $s$ and $t$ channels, but not in the $u$ channel. 
Let us consider the $s$-channel first. We have $M_{\vec{p}}=p+2q$ and $\gamma_m=p-2$, so that $s=\theta q-1$.
However, in our discussion in section \ref{2_p_sect}, and in particular section \ref{subsec:long_s}, 
we found that the following double-particle twists should also correspond to double poles,  
\begin{equation}
    \tau_{\rm long}=\theta(p+2)+2i,\qquad i=0,\ldots,\theta(q-1)-1.
\end{equation} 
This set of twists translates into the following set of poles:\footnote{Use \eqref{fromtwist_to_x} 
with the replacement $\emph{M}_{\vec{p}}\rightarrow \tau_{\rm long}/\theta$.}
\begin{equation}
s=i+\theta-1,\qquad i=0,\ldots,\theta(q-1)-1.
\end{equation}
Since the s-channel simple poles in \eqref{eq:kernel} start at $s=0$, 
we deduce that the Mellin amplitude contains additional simple poles 
so that in combination with ${\bf \Gamma}_{q,p}$ the Mellin integrand has double poles at $s=i+\theta-1$. 
 Repeating the above line of reasoning, we can draw the same conclusion 
 for the $t$ channel: the  Mellin amplitude contains additional simple poles at
\begin{equation}
t=j+\theta-1,\qquad j=0,\theta-1. 
\end{equation}
In the $u$-channel the Mellin integrand has no additional double-poles. 

Finally, in all three channels we have the information about the unitarity bound
which corresponds to a simple pole at $s=-1$ in the $s$-channel, $t=-1$ in the $t$-channel and $u=-1$ in the $u$-channel.
We note that, in the ${\cal N}=4$ SYM theory, the authors of \cite{Aprile:2026uxe} found 
that all the families of Mellin amplitudes involving one double-particle operator 
that they studied were fully determined by the predictions at the unitarity bound, 
with the remaining OPE predictions providing additional, non-trivial consistency checks. 
Our result for $\cN=4$ SYM, presented in section \ref{sec:4dresult}, further supports this observation.
Then, in section \ref{section:6d}, we discuss the generalisation of this to the six-dimensional $(2,0)$ theory.

Taking into account the predictions from each OPE channel \eqref{eq:stutable} and their pole structure outlined above, we make the following ansatz for the tree-level Mellin amplitude,
\begin{equation}\label{eq:refinedM}
    {\cal M}^{(\frac32)}_{q,p}(s,t) = \frac{\tilde{c}_{q,p}}{(s+1)(t+1)(u+1)}+
    \sum_{j=0}^{\theta-1} \sum_{i=0}^{\theta(q-1)-1} \frac{c_{q,p}^{i,j}}{(s-(i+\theta-1))(t-(j+\theta-1))(u+1)}.
\end{equation}
The first term is analogous to the tree-level contribution found in N$^2$E 
correlators of four single-particle operators, while the double sum introduces 
new structures. The ranges of $i$ and $j$ in the sum are determined by the long 
sector in the $s$- and $t$-channels, respectively, as explained above. 
Although there is no one-to-one correspondence, the appearance 
of these additional structures is precisely what we expected from our 
considerations of five-point functions and the double-particle limit, as explained in the introduction to section \ref{sec:doublebootstrap}.

Note that we explicitly leave out terms of the form
\begin{equation}\label{leaveout}
    \frac{1}{(s+1)(t-j)(u+1)}, \qquad  \frac{1}{(s-i)(t+1)(u+1)}.
\end{equation}
This is justified by considering the resummation of the predictions at the unitarity bound. 
In the $s$- and $t$-channels, they have the following schematic form
\begin{equation} 
    \begin{aligned} 
        \text{s-channel:}&\qquad  \frac{\#}{V}+O(1) +V^{2\theta-1}\log V\,\big(\# + O(V)\big),\\
        \text{t-channel:}&\qquad  \frac{\#}{V}+O(1) +V^{\theta q-1}\log V\,\big(\# + O(V)\big). 
    \end{aligned} 
\end{equation}
In the $s$-channel, the term $V^{2\theta-1}\log V(\# + O(V))$ arises from the combination of the 
simple pole at $s=-1$ and the double poles at $t\ge2\theta-1$. The latter are already supplied 
by ${\bf \Gamma}_{q,p}$, and therefore they are accounted for by the first term in our ansatz, $\frac{1}{(s+1)(t+1)(u+1)}$. 
This explains why we do not need to include terms of the form $\frac{1}{(s+1)(t-j)(u+1)}$ \eqref{leaveout}. 
The same reasoning applies in the $t$-channel for the second term in \eqref{leaveout}.

The unitarity bound in the $u$-channel captures the novel features associated with the external
double-particle operator and requires the inclusion of all terms in the ansatz \eqref{eq:refinedM}.

It is instructive to note that the analysis presented above, which relies solely on considerations at the unitarity bound, already provides a reduced ansatz that can be used as an input for the bootstrap.

A more brute-force alternative would consist in considering the Gamma functions with double poles in the window, leading to
\begin{equation}
\frac{ \Gamma[(\theta-1)-s]\Gamma[-s]\Gamma[(\theta-1)-t]\Gamma[-t]{\Gamma [-u+(p-2) \theta -1]\Gamma[-u]}}{\Gamma[q]\Gamma[p-2]}
\end{equation}
and subsequently constructing a polynomial Mellin amplitude by considering 
a generic polynomial ${\cal P}(s,t)$ divided by the simple-pole structure $(s+1)(t+1)(u+1)$.
The degree of ${\cal P}(s,t)$ is, of course, constrained by the OPE data. Nevertheless, 
this approach introduces a larger number of unknown coefficients than the reduced structure described in \eqref{eq:refinedM}.

\subsection{Mellin amplitude with $\left[\OO_2\OO_q\right]$ in the 4d $\NN=4$ SYM theory}\label{sec:4dresult}

We shall first report on our study of the correlators ${\cal H}^{(\frac{3}{2})}$ 
in ${\cal N}=4$ SYM theory. Our main motivation for doing so is that a similar 
bootstrap analysis was successfully 
carried out in \cite{Aprile:2026uxe}. Our computation builds on these previous 
results and 
serves as a warm-up for the $(2,0)$ theory computation.

Let us now describe the implementation of our Mellin bootstrap algorithm.
The OPE predictions are given in the following schematic form, 
\begin{align}
\mathcal{H}^{(\frac32)}_{s-\textrm{channel}}\Bigg|_{\rm{OPE}} &=  (x_1x_2)^{-1} g^{(s)}_{-1}(x_1)  
+  \log(U)\Bigg[ (x_1x_2)^{0} g^{(s)}_{0}(x_1) + \ldots +  (x_1x_2)^{q-2} g^{(s)}_{q-2}(x_1)\Bigg], \\
\mathcal{H}^{(\frac32)}_{t-\textrm{channel}}\Bigg|_{\rm{OPE}} &=  (x_1x_2)^{-1} g^{(t)}_{-1}(x_1) 
+  \log(U) \Bigg[ (x_1x_2)^{0} g^{(t)}_{0}(x_1) \Bigg], \\
\mathcal{H}^{(\frac32)}_{u-\textrm{channel}}\Bigg|_{\rm{OPE}} &=  (x_1x_2)^{-1} g^{(u)}_{-1}(x_1) .
\end{align}
In each channel, the functions $g^{(\#)}(x_1)$ are obtained from summing over the conformal blocks with the predicted coefficients
and rewriting the sum as an expansion in powers of $x_2$. This corresponds to the OPE limit in which
\begin{equation}\label{expansion_blocks_small_UV}
U=x_1x_2\rightarrow 0,
\qquad
V=(1-x_1)(1-x_2)\rightarrow 1.
\end{equation}
However, evaluating the residues in Mellin space yields 
a simultaneous small-$U$, small-$V$ expansion of the correlator, 
which is not the regime \eqref{expansion_blocks_small_UV} above.
Thus, to match the two expansions, we introduce the variables
\begin{equation}
x_1=1-\chi_1,
\qquad
x_2=\chi_2,
\end{equation}
and we perform a resummation of the series in $x_1$. In all cases, the results are finite expressions written 
in terms of rational functions in the basis of unity and $\log(1-x_1)$. 
Using these resummations, the expansion  for small $\chi_1$ and $\chi_2$ 
of the OPE predictions can be carried out systematically.

With all the necessary machinery in place, matching the OPE 
predictions to the residue expansion of the Mellin amplitude yields 
a system of equations for the coefficients entering our ansatz.

This system contains infinitely many equations arising from the small-$V$ expansion for each power of $U$ arising from the small-$U$ expansion.
This provides several non-trivial tests of our ansatz. First, in a given channel, the small-$V$, small-$U$ expansion offers a test 
through the matching of infinitely many constraints. 
Second, we can test crossing symmetry by imposing the OPE predictions in all three independent channels.

Remarkably, we find that the three crossing-related orientations provide an overconstrained 
system of equations. The consistency of these constraints therefore constitutes 
a highly non-trivial check of both our ansatz and the reliability of our algorithm.

The final result is 
\begin{equation}
    \mathcal{H}^{(\frac32)}_{q,p}= 
     {\cal N}^{(\frac{3}{2})}_{p,q}
\oint \frac{ds dt}{(2\pi i)^2} U^s V^t {\bf \Gamma}_{q,p}(s,t) {\cal M}^{(\frac32)}_{q,p}(s,t)
\end{equation}
where the kernel ${\bf \Gamma}_{q,p}$ is given by \eqref{eq:kernel} with $\theta=1$, 
the normalization is
\begin{equation}
 {\cal N}^{(\frac{3}{2})}_{p,q}=\sqrt{ (2q)(p+q-2)(2)p}
\end{equation}
and the Mellin amplitude:
\begin{equation}\label{eq:SYMresult}
    {\cal M}^{(\frac32)}_{q,p}(s,t) =(p-1)\left(\frac{1}{(s+1)(t+1)(u+1)}+ \sum_{i=1}^{q-1} \frac{c_{q,p}^{i}}{(s-i+1)t(u+1)}\right),
\end{equation}
with the coefficients
\begin{equation}
        c_{q,p}^{i}=  \frac{(-1)^i\Gamma[q]}{\Gamma[2+i]\Gamma[-i+q]} - 
          \frac{ \delta_{i,q-1} }{q} \\
\end{equation}
We observe that the minimal information that fixes the solution is the protected sector. Here in particular, 
the totality of the coefficients $c_{q,p}^i$ is determined by the $u$-channel alone, 
i.e.~the channel which contains the new information about the double-particle insertion.

 One new feature, compared to the other families of correlators studied in $\mathcal N=4$ SYM, is 
 the appearance of a Kronecker delta that depends only on $q$, and hence on the choice of 
 constituent operators entering the composite external operator.
  Following \cite{Aprile:2026uxe}, we also compared our bootstrap result with 
the double-particle limit taken on the five-point correlators constructed in \cite{Fernandes:2025eqe}
and  found perfect agreement. 
Finally, we interpret the Gamma functions appearing in $c_{q,p}^{i}$ as originating from the sphere gamma functions of the five-point correlator.

\subsection{Mellin amplitude with $\left[\OO_2\OO_q\right]$ in the 6d $\NN=(2,0)$ theory}
\label{section:6d}

Our Mellin amplitude bootstrap in the six-dimensional $\cN=(2,0)$ 
theory follows the same general logic as in the previous section. 
The dynamics of the two theories is, of course, different, and so are the details 
of the bootstrap calculation. 

The OPE predictions are again given schematically by
\begin{align}
\mathcal{H}^{(\frac32)}_{s-\textrm{channel}}\Bigg|_{\rm{OPE}} &=  (x_1x_2)^{-1} h^{(s)}_{-1}(x_1)  
+  \log(U)\Bigg[ (x_1x_2)^{1} h^{(s)}_{1}(x_1) + \ldots +  (x_1x_2)^{2q-2} h^{(s)}_{2q-2}(x_1)\Bigg], \\
\mathcal{H}^{(\frac32)}_{t-\textrm{channel}}\Bigg|_{\rm{OPE}} &=  (x_1x_2)^{-1} h^{(t)}_{-1}(x_1) 
+  \log(U) \Bigg[ (x_1x_2)^{1} h^{(t)}_{1}(x_1)+ (x_1x_2)^{2} h^{(t)}_{2}(x_1) \Bigg], \\
\mathcal{H}^{(\frac32)}_{u-\textrm{channel}}\Bigg|_{\rm{OPE}} &=  (x_1x_2)^{-1} h^{(u)}_{-1}(x_1).
\end{align}
From the outset, however, we find that the intermediate stages of the 
calculation are considerably more cumbersome. 
For instance, obtaining closed-form expressions 
for the conformal block coefficients as functions of the spin $l$ and the charges $p$ and $q$ proves 
substantially more challenging. 
We therefore adopt a more pragmatic strategy: helped by the evaluation formula for the 
free-like block coefficients of section \ref{sec:cauchy}, we simply tabulate all the block coefficients numerically 
and use them to set up the corresponding resummations. The resummed expressions 
display a remarkably clear structure, already foreshadowing the structure of the final Mellin amplitude.

The final result is 
\begin{equation}
    \mathcal{H}^{(\frac32)}_{q,p}= 
     {\cal N}^{(\frac{3}{2})}_{p,q}
\oint \frac{ds dt}{(2\pi i)^2} U^s V^t {\bf \Gamma}_{q,p}(s,t) {\cal M}^{(\frac32)}_{q,p}(s,t),
\end{equation}
where the kernel ${\bf \Gamma}_{q,p}$ is given by \eqref{eq:kernel} with $\theta=2$, 
the normalization is
\begin{equation}
 {\cal N}^{(\frac{3}{2})}_{p,q}=- \frac{2 \Gamma[p+q-1]}{(\Gamma[-1+2p]\Gamma[-1+2q]\Gamma[-1+2(p+q-2)])^{\frac{1}{2}}},
\end{equation}
and the Mellin amplitude is
\begin{equation}\label{eq:6dresult}
   {\cal M}_{q,p}(s,t) =(2p-3)\left(\frac{\tilde{c}^{}_{q,p}}{(s+1)(t+1)(u+1)}+\sum_{j=1}^2 \sum_{i=1}^{2(q-1)} \frac{c_{q,p}^{i,j}}{(s-i)(t-j)(u+1)}\right),
\end{equation}
with coefficients
\begin{equation}
    \begin{aligned}
        \tilde{c}^{}_{q,p}&=\frac{((p+1)(2q+1)+2q-8)}{3(4q^2-1)},\\
        c_{q,p}^{i,j}&= \frac{(-1)^{i+j-1}ij \,\Gamma[2q-1]}{\Gamma[3+i+j]\Gamma[2q-i-1]}
        \Bigg[ \left(2p+(i-1)(j-1)\right)
        +2(p+q-2)\Bigg[ \frac{\delta_{i,2q-3}}{(2q-3)}+\delta_{i,2q-2}\Bigg]+(j-1)\delta_{i,2q-2}
        \Bigg].
    \end{aligned}
\end{equation}
Apart from the doubling $j=1,2$, 
which has a clear interpretation from the OPE, 
it is quite surprising that the final result takes such a simple form.

We note however that the predictions in the protected sector do \emph{not} suffice to fix the Mellin 
amplitude, unlike in the $\cN=4$ SYM theory. In particular, the $u$-channel predictions alone do not determine 
all coefficients, and the $s$- and $t$-channel predictions at the unitarity bound only probe the residues at $s=-1$ and $t=-1$. 
Therefore they determine $\tilde c_{q,p}$, consistently with the $u$-channel, 
but are blind to the $c^{i,j}_{q,p}$. The remaining freedom is fixed by 
the $\log U$ predictions. 
We have found that a minimal choice is to consider the two $\log U$ predictions from the $t$-channel. 
Nevertheless, the full set of $\log U$ predictions overconstrains the ansatz. The fact that our 
amplitude is uniquely determined therefore provides a non-trivial consistency 
check of our bootstrap approach.

It is suggestive to interpret the Gamma functions appearing in the coefficients $c_{q,p}^{i,j}$ 
as originating from the sphere gamma functions of a five-point correlator, although we do 
not have an explicit expression for such a correlator. The coefficients $c_{q,p}^{i,j}$ 
also contain Kronecker deltas that depend only on $q$, and hence on the choice of 
constituent operators entering the composite external operator. 
Both features are also present in our results for $\mathcal N=4$ SYM.

\section{Conclusion and outlook} \label{outlook}

The main result of this paper has been the derivation of the tree-level Mellin amplitudes for the correlators
\begin{equation}\label{corr_conclusion}
\langle [\cO_2 \cO_q] \cO_{p+q-2} \cO_2 \cO_p\rangle, \qquad q\geq2,p\geq3
\end{equation}
both in the ${\cal N}=4$ SYM theory \eqref{eq:SYMresult} 
and in the $(2,0)$ six-dimensional theory \eqref{eq:6dresult}. 
Our computation relied on the so-called 
double-particle bootstrap, described in detail in section \ref{sec:doublebootstrap}, which, 
to the best of our knowledge, has not previously been applied to M-theory.
We would like to emphasize below two aspects of our computation which we found particularly interesting.

\begin{itemize}
    \item On the kinematical side, a key advantage of our approach has been the use of the 
formalism developed in \cite{Aprile:2021pwd}, which allowed us to treat 
the superconformal blocks in both theories on the same footing. 
A crucial input from that work is the superconformal Cauchy identity
and the evaluation formula derived from it, which 
makes all free-theory computations essentially immediate, reducing the mechanical work involved in determining  conformal block coefficients to the direct evaluation of a BC Jacobi polynomial.
\item On the dynamical side, there are substantial differences between the two theories. 
For example, the double-particle limit of four-point correlators in the $(2,0)$ theory 
receives contributions from both ${\cal F}$ and ${\cal H}$, whereas only ${\cal F}$ 
contributes in $\mathcal N=4$ SYM, as a consequence of the partial non-renormalization 
theorem. Moreover, the details of the OPE predictions entering the bootstrap calculations 
are different. Nevertheless, we found that the final Mellin amplitudes exhibit a 
strikingly similar structure. This was particularly remarkable given that, at 
intermediate stages of the M-theory computation, no particularly useful patterns emerged 
in the OPE coefficients. 
\end{itemize}
\noindent
The simplicity of the final Mellin amplitude in \eqref{eq:SYMresult}-\eqref{eq:6dresult} is therefore especially satisfying.

Extending the perspective initiated by the seminal work of Penedones in \cite{Penedones:2010ue}, 
we interpret our findings in this paper as evidence for a form of Mellin-space 
universality for tree-level amplitudes involving a single multi-particle operator.

For four-point correlators of single-particle operators, $\langle {\cal O}_{p_1} \cO_{p_2} \cO_{p_3} \cO_{p_4}\rangle$, 
the authors of \cite{Alday:2020dtb} showed that 
a unified Mellin amplitude exists across different dimensions. Although we have not investigated 
the ABJM theory in this work, the extension of our analysis to AdS$_4\times$S$^7$ appears particularly natural.
This would extend the bootstrap investigations of \cite{Chester:2018aca,Binder:2019mpb,Agmon:2019imm,Alday:2020dtb,Alday:2022rly} in another interesting direction. 
For the kinematics, a similar formalism to the one employed in section \ref{sec:doublebootstrap} is already 
available from \cite{Aprile:2021pwd}, with the only modification being the value of the parameter $\theta=\frac{d-2}{2}$. 
Regarding the OPE, the generalization of the double-particle bootstrap to this setting appears 
conceptually straightforward.\footnote{The remaining obstacles are mainly computational, involving the 
resummation of the OPE data and its subsequent matching to the Mellin amplitude.}

As discussed in the Introduction,  our four-point correlators \eqref{corr_conclusion}, and their generalisation,
\begin{equation}
    \langle [\cO_{r_1} \cO_{r_2}] \cO_{p_2} \cO_{p_3} \cO_{p_4}\rangle,
\end{equation}
have a geometric interpretation 
in the holographic setting:  they provide response functions of light probes to smooth deformations 
of the vacuum LLM geometry \cite{Lin:2004nb} when the deformation is small. 
In ${\cal N}=4$ SYM, the authors of \cite{Aprile:2024lwy,Aprile:2025hlt} were able to 
match the bootstrap results with a direct gravitational computation. 
They also observed a close analogy between the Green's function 
representation arising from holography, as the multi-particle level of the correlator is increased, 
and the ladder integral representation familiar from integrability. 
Motivated by this analogy, they introduced the terminology gravitational 
ladders to describe families of tree-level connected correlators with 
multi-particle insertions and increasing multi-particle level.
For all these reasons it would be particularly interesting to 
reproduce our bootstrap results directly from bubbling geometries in M-theory. 
A key motivation for pursuing this direction is that the double-particle bootstrap alone does not uniquely determine correlators involving two double-particle operators, or, more generally, arbitrary multi-particle insertions. Additional input from the gravitational description is therefore needed to fully constrain these correlators.

Another interesting direction would be to explore the double-particle limit in M-theory, 
which was successfully investigated in ${\cal N}=4$ SYM theory \cite{Aprile:2026uxe}, 
but remains unexplored in the $\cN=(2,0)$  theory. This would require generalizing the bootstrap approach 
of \cite{Alday:2020lbp} to five-point functions. We find it conceivable that such a generalization could be achieved.
Having access to these results would make it particularly interesting to study the flat-space limit of \cite{Penedones:2010ue}, 
generalized to correlators with a double-particle insertion and, more generally, with a single 
multi-particle operator insertion of the form $[\cO_{r_1} \cO_{r_2} \ldots {\cal O}_{r_n}]$.

\section*{Acknowledgments}

We are very grateful to Daniele Dorigoni, Paul Heslop,  Joao Penedones, 
Rodolfo Russo and Congkao Wen for helpful discussions. 
We would also like to thank the organizers of the 
workshop \href{https://indico.global/event/14566/overview}{“New Ways to Higher Points”}, where this project was initiated,
and the organizers of the \href{https://indico.global/event/17419/}{“UK-Japan Workshop 
on Quantum Gravity"} for hospitality during the final stages of our collaboration. 

F.~Aprile is supported by the Ramon y Cajal program through the fellowship 
RYC2021-031627-I funded by MCIN/AEI/10.13039/501100011033 and by the European 
Union NextGeneration EU/PRTR and by the MSCA programme - HeI - \href{https://cordis.europa.eu/project/id/101182937}{DOI 10.3030/101182937}. HPR is supported by a Science and Technology 
Facilities Council (STFC) studentship. MW was supported by a Science and 
Technology Facilities Council (STFC) studentship as well as grants 
ERC-2021-CoG - BrokenSymmetries 101044226 and EXC 2121 Quantum Universe – 390833306.

\begin{appendix}

\section{Superconformal blocks}\label{app:blocks}
We write explicit formulae for the superconformal blocks used in this paper. Their expansion into super-Jack polynomials is key for deriving the evaluation formula for block coefficients (see section \ref{sec:cauchy}), and simplifying the block resummations leading to the final result (see section \ref{sec_results} and appendix \ref{app:subtlety}). We refer to \cite{Aprile:2021pwd} for the general theory. 
Here we will focus on the maximally supersymmetric case, thus fixing $m=2$, $n=2$ and leaving $\theta$ unfixed. 

We write a superconformal block $B_{\gamma\ula}$ for an exchanged superconformal multiplet $O^\gamma_{\ula}$ \eqref{tableQN1} as
\begin{equation}\label{eq:blockappendix}
    B_{\gamma\ula}(x_1,x_2,y_1,y_2;\theta)= \left(\frac{g_{24} g_{13}}{g_{12} g_{34}}\right)^{\frac{\gamma-\gamma_m}{2}} \sum_{\umu \geq \ula } (T_{\gamma,\theta,p_{12},p_{43}})_{\ula}^{\umu}\, P_{\umu} (x_1,x_2,y_1,y_2;\theta),
\end{equation}
where $P_{\umu}$ are the super-Jack polynomials, denoted by $J^{(2,2)}_{\umu}$ in \cite{Aprile:2021pwd}.

\paragraph{Super-Jack polynomials.}
As explained in section \ref{subsec:OPE}, N$^2$E correlators exchange Young diagrams of up to two rows, i.e., $\ula=[\lambda_1,\lambda_2]$. In this case, the relevant super-Jack polynomials are labelled by 
\begin{equation}
    \umu=[\mu_1,\mu_2], \qquad \mu_1 \geq \lambda_1 \geq \mu_2 \geq \ula_2.
\end{equation}
In the $\theta=1$ case, they are super-Schur polynomials, for which explicit formulae are given in \cite[section 3.5.1]{Doobary:2015gia}. In the $\theta=2$: 

The two-row super-Jack polynomials are given by
\begin{align}\label{eq:jack6d}
& 
P_{[\mu_1,\mu_2]}\big|_{\theta=2}
= D_2 \Bigg[\prod_{i,j}(x_i-y_j)  \times \\
&  \frac{1}{(\mu_1+3)(\mu_2+1)}\Bigg[  \frac{ U^{\mu_2-1}(x_1^{3+\mu_1-\mu_2}-x_2^{3+\mu_1-\mu_2}) }{(x_1-x_2)^3} 
-\frac{3+\mu_1-\mu_2}{1+\mu_1-\mu_2} \frac{U^{\mu_2} (x_1^{1+\mu_1-\mu_2}-x_2^{1+\mu_1-\mu_2})}{(x_1-x_2)^3} \Bigg]\Bigg].\notag
\end{align}

The one-row super-Jacks, given below for convenience,
\begin{equation}\label{onerowjack}
P_{[\mu]}\big|_{\theta=2}
=  D_2\Bigg[ \frac{ (x_1-y_1)(x_1-y_2) ( - \frac{x_1^{\mu+1}}{\mu+1}) - (x_2-y_1)(x_2-y_2) ( - \frac{x_2^{\mu+1}}{\mu+1})  }{(x_1-x_2)}\Bigg].
\end{equation}

\paragraph{Coefficients.} The coefficients $T_{\gamma;\theta;p_{12},p_{43}}$ entering \eqref{eq:blockappendix} for the relevant two-row case are given by a recursion formula \cite{Aprile:2021pwd} which we quote below. The external charges enter through the parameters
\begin{equation}
    \alpha=\frac{\gamma-\max(p_{12},p_{43})}{2}, \qquad \quad \beta=\frac{\gamma-\min(p_{12},p_{43})}{2}.
\end{equation}
First, the leading coefficient, i.e. when $\umu=\ula$, is given by
\begin{equation}
    (T_{\gamma;\theta;\alpha,\beta})^{\ula}_{\ula}= 1.
\end{equation}
Then, the recursion formula is as follows,
\begin{equation}\label{eq:Tgamma}
    (T_{\gamma;\theta;\alpha,\beta})^{[\mu_1,\mu_2]}_{[\lambda_1,\lambda_2]} = \frac{A^{(1)}_{[\mu_1-1,\mu_2]} f^{(1)}_{[\mu_1-1,\mu_2]} (T_{\gamma;\theta;\alpha,\beta})^{[\mu_1-1,\mu_2]}_{[\lambda_1,\lambda_2]} + A^{(2)}_{[\mu_1,\mu_2-1]} f^{(2)}_{[\mu_1,\mu_2-1]} (T_{\gamma;\theta;\alpha,\beta})^{[\mu_1,\mu_2-1]}_{[\lambda_1,\lambda_2]}}{\sum_i (\mu_i-\lambda_i)(\lambda_i+\mu_i+\theta \gamma - 2 \theta(i-1)-1)},
\end{equation}
where a term is omitted if $[\mu_1-1,\mu_2]$ or $[\mu_1,\mu_2-1]$ are not partitions, and 
\begin{equation} 
f_{[\nu_1,\nu_2]}^{(1)} = \frac{ (\nu_1-\nu_2+1)(\nu_1-\nu_2+2\theta) }{ (\nu_1-\nu_2+\theta)(\nu_1-\nu_2+\theta+1) }, 
\qquad 
f_{[\nu_1,\nu_2]}^{(2)} = \frac{ (\nu_1-\nu_2)(\nu_1-\nu_2+2\theta-1) }{ (\nu_1-\nu_2+\theta-1)(\nu_1-\nu_2+\theta) }, 
\end{equation}
\begin{equation}
    A_{[\nu_1,\nu_2]}^{(i)}=(\nu_i-1- \theta(i-1-\alpha))(\nu_i-1-\theta(i-1-\beta)).
\end{equation}

\paragraph{Reduced blocks.}
While the above formulae are needed for the derivation of the evaluation formula for block coefficients of section \ref{sec:cauchy}, the block resummations of the predictions from section \ref{sec:doublebootstrap} are done at the level of the reduced correlator $\HH$. For this, we write formulae for the reduced blocks $\tilde{B}_{\gamma\ula}$ which directly decompose $\HH$, i.e., are stripped off of the superconformal prefactor $\Upsilon_{\theta}$ \eqref{eq:D2}:
\begin{equation}
    \HH_{\vec{p}}=\sum \CC^{\HH}_{\vec{p}}\big|_{\gamma,\ula} \tilde{B}_{\gamma,\ula}(x_1,x_2,y_1,y_2;\theta).
\end{equation}
From the form of the super-Jack polynomials \eqref{eq:jack6d} it is straightforward to write the $\tilde{B}_{\gamma\ula}$ as an expansion of `reduced super-Jack polynomials' $\tilde{P}_{\umu}$,
\begin{equation}
    \tilde{B}_{\gamma\ula} (x_1,x_2,y_1,y_2;\theta)= \sum_{\umu \geq \ula }(T_{\gamma,\theta,p_{12},p_{43}})_{\ula}^{\umu} \tilde{P}_{\umu} (x_1,x_2,y_1,y_2;\theta),
\end{equation}
where crucially the coefficients are the same as in the full correlator, namely \eqref{eq:Tgamma}.

\section{Protected correlators in 6d from $\WW_\mathfrak{g}$}
\label{sec:AppWalgebra}
Following the chiral algebra construction of \cite{Beem:2013sza,Beem:2014kka}, the contribution $\mathcal{F}_{\vec{p}}$ can be determined using an uplift of a 2d chiral algebra correlator of chiral quasi-primaries with weight $h_i=p_i$. In particular, taking 6d chiral algebra twist $y_1,y_2\rightarrow x_2$, we have that 
\begin{align}
\mathcal{G}_{\vec{p}}\left(x_1x_2,(1-x_1)(1-x_2);\frac{1}{x_2^2},\frac{(1-x_2)^2}{x_2^2}\right)=&\;
\mathcal{F}_{\vec{p}}\left(x_1x_2,(1-x_1)(1-x_2);\frac{1}{x_2^2},\frac{(1-x_2)^2}{x_2^2}\right)
\equiv x_1^{p_{34}}\mathcal{F}_{\vec{p}}\left(x_1\right),
\end{align}
Following the ``holomorphic bootstrap" method of \cite{Headrick:2015gba}, we can use crossing symmetry and the observation that the only singularities of $\mathcal{F}_{\vec{p}}(x_1)$ in the complex $x_1$-plane  are at $x_1=0,1,\infty$, to obtain closed-form expressions for $\mathcal{F}_{\vec{p}}(x_1)$. In particular, for four-point functions of chiral quasi-primaries of weight $h_i=p_i$, we have that 
\begin{align}
    \mathcal{F}_{\vec{p}}(x_1)=G_{p_1p_2}G_{p_3p_4}+\sum_{n=1}^{p_1+p_4}\alpha^{\vec{p}}_{n}x_1^n+\sum_{n=1}^{p_2+p_3}\beta^{\vec{p}}_{n}\left(\left(1-x_1\right)^{-n}-1\right),
\label{eq:closedformF}
\end{align}
in terms of coefficients $\alpha^{\vec{p}}_{n}$ and $\beta^{\vec{p}}_{n}$ which appeared explicitly in \cite{Woolley:2025qbd}. These coefficients depend on chiral algebra data appearing in the singular part of the OPEs involving the external operators $\hat{O}_{p_i}$. As an example, we can exactly compute the protected correlator contributing to $\langle \left[\OO_2\OO_q\right]\OO_{p+q-2}\OO_2\OO_p\rangle$ and reproduce the $O(c^{-3/2})$ contribution computed in section \ref{formulae_freet_db}. Given the configuration $\vec{p}=([2q],p+q-2,2,p)$, the twisted correlator is 
\begin{align}
\mathcal{F}_{[2q],p+q-2,2,p}(x_1)=\sum_{n=1}^{p+q+2}\alpha^{\vec{p}}_{n}\;x_1^n+\sum_{n=1}^{p+q}\beta^{\vec{p}}_{n}\left(\left(1-x_1\right)^{-n}-1\right).
\end{align}
Assuming an orthonormal basis of quasi-primaries (other than $\hat O_{q+2,TW_q}$ normalized as in Equation \eqref{eq:2qnorm}), the coefficients $\alpha^{\vec{p}}_{n},\beta^{\vec{p}}_{n}$ in this configuration take the form
\begin{align}
    \alpha^{\vec{p}}_{n}=&\;\sum_{h=0}^{p+q+2-n}C_{p,p+q-2}^{O_h}C_{2,[2q]}^{O_h}\frac{\left(h+q-2\right)_{p+q+2-h-n}\left(h-q\right)_{p+q+2-h-n}}{\left(p+q+2-h-n\right)!\left(2h\right)_{p+q+2-h-n}},\\
    \beta^{\vec{p}}_{n}=&\; \sum_{h=0}^{p+q-n}C_{[2q],p}^{O_h}C_{2,p+q-2}^{O_h}
    \sum_{l=0}^{p+q-h-n}\frac{\left(h+p-q-2\right)_{l}\left(h+4-p-q\right)_{l}\left(-p-2\right)_{p+q-h-n-l}}{\left(p+q-h-n-l\right)!l!\left(2h\right)_{l}},
\end{align}
which involve sums over quasi-primaries $O_h$ of weight $h$ appearing in OPE contractions of two crossed four-point function configurations. The constrained form of the OPE between $T$ and the Virasoro primary $W_{p+q-2}$ localizes the sums in $\beta^{\vec{p}}_{n}$ to the $W_{p+q-2}$ contribution such that we simply have
\begin{align}
    \beta^{\vec{p}}_{1}=&\;-\frac{(p+q-2)p+2q}{p+q-2}C_{[2q],p}^{p+q-2}C_{2,p+q-2}^{p+q-2}, \hspace{1cm}\beta^{\vec{p}}_{2}=C_{[2q],p}^{p+q-2}C_{2,p+q-2}^{p+q-2},\hspace{1cm}
    \beta^{\vec{p}}_{n\geq3}=0.
\end{align}
The determination of the coefficients $\alpha^{\vec{p}}_n$ is more involved however, in that the range of $n$ and the sum over contributing quasi-primaries $O_h$ grows with $p$. For this reason it will be convenient to abandon the interpretation of $\alpha_{\vec{p},n}$ in terms of a dressed sum over exchanged quasi-primary data and instead recognize these coefficients as the three-point coupling
\begin{align}
    \alpha^{\vec{p}=([2q],p+q-2,2,p)}_n=\bra{W_p}W_{p+q-2}(1)\ket{L_{n-p}\hat O_{q+2,TW_q}},
\end{align}
between $W_p$, $W_{p+q-2}$, and the operator $L_{n-p}\hat O_{q+2,TW_q}$, i.e. the term at order $z^{p-n-2}$ in the $T(z)\times\hat O_{q+2,TW_q}(0)$ OPE, which will generically not be a quasi-primary. Using the Virasoro algebra\footnote{We stress that in our conventions we have $L_n=\sqrt{\frac{2}{c}}\left(L_\text{canonical}\right)_n$, in which case the Virasoro mode algebra is
\begin{align}
    \left[L_m,L_n\right]=\sqrt{\frac{2}{c}}(m-n)L_{m+n}+\frac{1}{6}m(m^2-1)\delta_{m+n,0}.\nonumber
\end{align}.} and the quasi-primary property of $\hat O_{q+2,TW_q}$, we evaluate this expression to find
\begin{align}
\alpha^{\vec{p}}_{n>p+2}=&\;0, \\
\alpha^{\vec{p}}_{p+2}=&\;\bra{W_p}W_{p+q-2}(1)\ket{L_{2}\hat O_{q+2,TW_q}}=\left(\delta_{2,q}+1\right)C_{q,p}^{p+q-2}, \\
\alpha^{\vec{p}}_{p+1}=&\;\bra{W_p}W_{p+q-2}(1)\ket{L_{1}\hat O_{q+2,TW_q}}=0, \\
\alpha^{\vec{p}}_{p}=&\;\bra{W_p}W_{p+q-2}(1)\ket{L_{0}\hat O_{q+2,TW_q}}=(q+2)\sqrt{\frac{2}{c}}\;C_{[2q],p}^{p+q-2}, \\
\alpha^{\vec{p}}_{0<n<p}=&\;\left((p-n)(p+q-2)+q+2-p\right)\sqrt{\frac{2}{c}}\;C_{[2q],p}^{p+q-2}.
\end{align}
With this, the sum over $\alpha^{\vec{p}}_{n}\;x_1^n$ and the two terms involving $\beta_{\vec{p},n}$ combine to give
\begin{align}
    x_1^{2-p}\mathcal{F}_{[2q],p+q-2,2,p}\left(x_1\right)=\left(\delta_{2,q}+1\right)C_{q,p}^{p+q-2}\;x_1^4+\sqrt{\frac{2}{c}}C_{[2q],p}^{p+q-2}\;\frac{x_1^2}{\left(1-x_1\right)^2}\left(p-4x_1+(q+2)x_1^2\right)
\end{align}
which we uplift to 6d to obtain
\begin{align}
\mathcal{F}_{[2q],p+q-2,2,p}\left(U,V;\hat{\sigma},\hat{\tau}\right)=&\;\left(\delta_{2,q}+1\right)C_{q,p}^{p+q-2}\;U^4\hat{\sigma}^2+\sqrt{\frac{2}{c}}C_{[2q],p}^{p+q-2}\left(2U^2\hat{\sigma}+(p-2)\frac{U^2}{V^2}\hat{\tau}+q\frac{U^4}{V^2}\hat{\sigma}\hat{\tau}\right).
\end{align}
Using Equations \eqref{eq:dp_resultchiralq}-\eqref{eq:eq:dp_resultchiralpqO3/2} we see that the large-$c$ expansion can be divided into disconnected and connected terms as 
\begin{align}
\mathcal{F}^{\left(\text{disc}\right)}_{[2q],p+q-2,2,p}\left(U,V;\hat{\sigma},\hat{\tau}\right)=&\;\left(\delta_{2,q}+1\right)\left(C_{q,p}^{p+q-2\;\left(\frac{1}{2}\right)}U^4\hat{\sigma}^2+\delta_{2,p}\sqrt{\frac{2}{c}}\left(2U^2\hat{\sigma}+(p-2)\frac{U^2}{V^2}\hat{\tau}+q\frac{U^4}{V^2}\hat{\sigma}\hat{\tau}\right)\right),\\
\mathcal{F}^{\left(\frac{3}{2}\right)}_{[2q],p+q-2,2,p}\left(U,V;\hat{\sigma},\hat{\tau}\right)=&\;\sqrt{2}C_{[2q],p}^{p+q-2\;(1)}\left(2U^2\hat{\sigma}+(p-2)\frac{U^2}{V^2}\hat{\tau}+q\frac{U^4}{V^2}\hat{\sigma}\hat{\tau}\right),
\end{align}
where the first connected contribution $\mathcal{F}^{\left(\frac{3}{2}\right)}_{[2q],p+q-2,2,p}$ matches the result in section \ref{formulae_freet_db} with $p>2$ and vanishes for $p=2$. Note that the normalization in Equation \eqref{eq:2qnorm} was tuned such that $C_{q,2}^{[2q]}=\delta_{2,q}+1$ is exactly $O(1)$. Any other choice would have induced a $U^{2\theta}\sigma^2$ term in $\mathcal{F}^{\left(\frac{3}{2}\right)}_{[2q],p+q-2,2,p}$, spoiling our separation between disconnected and connected contributions.

\section{Evaluating Jacobi polynomials}\label{app:superspace}

In this appendix we give a self-contained and explicit expression for \eqref{eq:cauchyformula}, given by \eqref{eq:finalformula}, as first derived in \cite{Aprile:2021pwd}. We recall this is a universal formula for all free-theory block coefficients in any theory with space-time symmetry
\begin{equation}\label{eq:supergroupsappendix}
    \SU(m,m|2n)\ \ (\theta=1), \qquad \OSp(4m^*|2n) \ \ (\theta=2), \qquad \OSp(4n|2m) \  \ (\theta=\tfrac12).
\end{equation}
For convenience, let us rewrite the starting point \eqref{eq:cauchyformula} here,
\begin{equation}\label{eq:maincauchyapp}
     M^{}_{k,\gamma, \ula; p_{12},p_{43};\theta} = J^{(\beta)}_{\ula} (\{Y_r\};\theta, p_{\pm}),
\end{equation}
and let us recall the notation: $\gamma,\ula$ label the exchanged superconformal multiplet in analytic superspace (see section \ref{subsec:OPE}), $k$ labels the different free-theory diagrams, and
\begin{equation}
    \beta=\frac12\left(\gamma-\max(|p_{12}|,|p_{43}|)\right), \qquad p_{\pm}=\frac{\theta}{2}|p_{12}\pm p_{43}|.
\end{equation}
Furthermore, $J^{(\beta)}_{\ula}$ is a Jacobi polynomial evaluated on a binary string $\{Y_r\}$ given by \eqref{eq:proprewriting}
\begin{equation}\label{eq:Yr}
     Y_r=\left\{\begin{aligned}
     &1, \quad r=1,\dots,k,\\
    &0, \quad r=k+1,\dots,d,
 \end{aligned}\right.\qquad\qquad d=\frac12 (\gamma-p_+-p_-).
\end{equation}

\paragraph{Jack expansion.}
To evaluate the Jacobi polynomial in \eqref{eq:maincauchyapp}, we expand into Jack polynomials $P_{\umu}$ as follows
\begin{equation}\label{eq:jacobi}
    J^{(\beta)}_{\ula}(\{Y_r\};\theta,p_{\pm}) = \sum_{\umu \leq \ula} (S^{(\beta)}_{\theta,p_{\pm}})^{\umu}_{\ula} P_{\umu} (\{Y_r\};\theta).
\end{equation}
The Jack polynomials evaluated on $\{Y_r\}$ \eqref{eq:Yr} are given by simple combinatorics as
\begin{equation}\label{eq:binaryjack}
    P_{\umu} (\{Y_r\};\theta) = \prod_{1\leq i<j\leq k}\frac{(\theta(j-i+1))_{\mu_i-\mu_j}}{(\theta(j-i))_{\mu_i-\mu_j}},
\end{equation}
where $\mu_i$ denotes the $i$th row (length) of $\umu$.

The coefficients $S_{\theta,p_{\pm}}^{(\beta)}$ do not admit an 
analytic expression for all exchanged Young diagrams $\ula$ (they do for two rows or less) and are instead given by a recursion formula.
This is because, in solving for them, one obtains an expression in terms of a non-trivial class of polynomials that generalise Jack polynomials.
These are the $\mathrm{BC}_N$ interpolation polynomials of \cite{Okounkov1998BCInterpolationMacdonald}, hereafter $P^*_{\umu}(\#;\theta,u)$.
More specifically, let 
\begin{equation}
    \tilde{p}_{\pm} := \theta p_{\pm} + \theta-1,
\end{equation}
then,
\begin{equation}\label{eq:Sinprogress}
    (S^{(\beta)}_{\theta,\tilde{p}_{\pm}})_{\ula}^{\umu}=\left.\frac{J_{\ula}(\{0_r\};\theta,\tilde{p}_{\pm})P_{\umu}^{*(\beta)}(\ula;\theta,u)}{J_{\umu}(\{0_r\};\theta,\tilde{p}_{\pm})P_{\umu}^{*(\beta)}(\umu;\theta,u)} \right|_{u=\frac{\tilde{p}_++\tilde{p}_-+1}{2}-\theta\beta}.
\end{equation}
We note that deriving this expression is highly non-trivial and refer to sections 7 and 8 of \cite{Aprile:2021pwd} for a full discussion. In particular, it involves deriving it in a compactified setting (by fixing $m=0$ in the supergroups \eqref{eq:supergroupsappendix}) and identifying the appropriate lift to the non-compact (conformal) $m\neq 0$ case.

All factors in \eqref{eq:Sinprogress} except $P_{\umu}^{*(\beta)}(\ula;\theta,u)$ admit a closed form expression.
To write them, we define the following combinatorial factors,
\begin{align}
C_{\ula}^{0}(\omega;\theta)
&=
\prod_{(i,j)\in\ula}
\left(j-1-\theta(i-1)+\omega\right),
\\[10pt]
C_{\ula}^{-}(w;\theta)
&=
\prod_{(i,j)\in\ula}
\left(\lambda_i-j+\theta(\lambda'_j-i)+w\right),
\\[10pt]
C_{\ula}^{+}(w;\theta)
&=
\prod_{(i,j)\in\ula}
\left(\lambda_i+j-\theta(\lambda'_j+i)+w\right),
\end{align}
where $\lambda_i$ is the length of the $i$ row of the Young diagram, while $\ula'$ denotes the transposed Young diagram, thus $\lambda'_j$ is the length of the $j$-column of the original Young diagram $\ula$. 
Then, the factors in \eqref{eq:Sinprogress} are given in terms of the above as
\begin{equation}
    P_{\umu}^{*(\beta)}(\umu;\theta,u)=
    C_{\umu}^{+}(2u-1+2\theta \beta;\theta)
    C_{\umu}^{-}(1;\theta),
\end{equation}
and
\begin{equation}
    J_{\ula}^{(\beta)}(\{0_r\};\theta,\tilde{p}_{\pm})
    =
    \frac{
    (-)^{|\ula|}
    C_{\ula}^{0}(\tilde{p}_++1+\theta(\beta-1);\theta)
    C_{\ula}^{0}(\theta \beta;\theta)
    C^-_{\ula}(1; \theta)}{
    P_{\ula}^{*(\beta)}(\ula;\theta,u)
    C^-_{\ula}(\theta; \theta)}.
\end{equation}

All in all, inserting into \eqref{eq:Sinprogress} we obtain
\begin{equation}\label{eq:SuptoP}
\left(S_{\theta,\tilde{p}_{\pm}}^{(\beta)}\right)_{\ula}^{\umu}
={}
(-)^{|\ula|-|\umu|}
\frac{C_{\umu}^-(\theta;\theta)}{C_{\umu}^-(1;\theta) C_{\ula}^-(\theta;\theta)}
\frac{C_{\ula/\umu}^{0}
(\tilde{p}_++1+\theta(N-1),\theta N;\theta)}{C_{\ula}^{+}(2u-1+2\theta \beta;\theta)
    }
P_{\umu}^{*(\beta)}(\ula;\theta,u),
\end{equation}
where $u=\frac{\tilde{p}_++\tilde{p}_-+1}{2}-\theta \beta$ as before, and we defined the following shorthand notation
\begin{equation}
    C_{\ula/\umu}^0= C_{\ula}^0/C_{\umu}^0\,, \qquad \quad C^0_{\ula}(A,B;\theta)=C^0_{\ula}(A;\theta)C^0_{\ula}(B;\theta).
\end{equation}
Equations \eqref{eq:SuptoP} and \eqref{eq:binaryjack} fix the evaluation formula in \eqref{eq:jacobi} up to the interpolation polynomial  $P_{\umu}^{*(\beta)}$, which we determine below.

\paragraph{Recursion relation for interpolation polynomials.}
Many classical bases of symmetric polynomials indexed by Young diagrams, such as Jack polynomials, admit a recursion of the form,
\begin{equation}\label{eq:recursiongeneral}
    P_{\ula}(x_1,\dots,x_{m-1},x_{m})=\sum_{\unu \prec \ula} \psi_{\ula,\unu}\left( \prod_{(i,j)\in \ula / \unu} f_{ij}(x_{m})\right) P_{\unu}(x_1,\dots,x_{m-1}),
\end{equation}
where $\unu$ is a partition with $m-1$ rows satisfying 
\begin{equation}
    \unu= \{ [\nu_1,\dots,\nu_{m-1}]\ : \ \lambda_{m} \leq \nu_{m-1} \leq \lambda_{m-1} \leq \dots \leq \lambda_2 \leq \nu_1 \leq \lambda_1\}, \qquad \quad (\Leftrightarrow\unu \prec \ula).
\end{equation}
This is a decomposition of the $m$-variable basis elements into the $(m-1)$-variable basis, and is thus governed by combinatorial branching rules. 
 
To obtain a recursion of the form of \eqref{eq:recursiongeneral} for the interpolation polynomial $P^*_{\umu}$ entering \eqref{eq:SuptoP}, we need to introduce a symmetric version of it. This is achieved by a linear coordinate shift as follows, 
\begin{equation}\label{eq:shift}
     P^*(\vec{x}; \theta, u) = P^{\text{ip}}(\vec{x}+\theta \vec{\delta}+ u; \theta ,u),
\end{equation}
with 
\begin{equation}
    \vec{x}=(x_1,\dots,x_m),\qquad \quad \vec{\delta}=(m-1,\dots, 1,0),
\end{equation}
and recall $u=\frac{\tilde{p}_++\tilde{p}_-+1}{2}-\theta \beta$.
Under this shift, $P^{\ip}$ is symmetric and thus admits a recursion of the form \eqref{eq:recursiongeneral}.
In particular, the combinatorial coefficients are
\cite{okounkov1997shifted}
\begin{equation}\label{eq:psi}
    \psi_{\ula,\unu}(\theta,u) = \prod_{1 \leq i \leq j < m} \frac{ \bigl(\nu_i-\lambda_{j+1}+1+\theta(j-i)\bigr)_{\lambda_i-\nu_i} }{ \bigl(\nu_i-\lambda_{j+1}+\theta(j-i+1)\bigr)_{\lambda_i-\nu_i} } \frac{ \bigl(\nu_i-\nu_j+\theta(j-i+1)\bigr)_{\lambda_i-\nu_i} }{ \bigl(\nu_i-\nu_j+1+\theta(j-i)\bigr)_{\lambda_i-\nu_i} } \, ,
\end{equation}
and the one-variable functions associated with each box $(i,j)\in\ula/\unu$ are
\cite{Okounkov1998BCInterpolationMacdonald}
\begin{equation}\label{eq:fij}
    f_{ij}(x_{m};\theta,u) = x_m^2-((j-1)-\theta(i-1)+\theta(m-1)+u)^2.
\end{equation}

\paragraph{Final formula.}
We quote here the final formula derived in this appendix:
\begin{equation}\label{eq:finalformula}
    M^{}_{k,\gamma, \ula; p_{12},p_{43};\theta} = \sum_{\umu \leq \ula} \left(S_{\theta,\tilde{p}_{\pm}}^{(\beta)}\right)_{\ula}^{\umu}\prod_{1\leq i<j\leq k} \frac{(\theta(j-i+1))_{\mu_i-\mu_j}}{(\theta(j-i))_{\mu_i-\mu_j}},
\end{equation}
where the coefficients are given by \eqref{eq:SuptoP} which depend on the interpolation polynomials $P^*_{\umu}$ defined by the recursion \eqref{eq:recursiongeneral} with \eqref{eq:psi} and \eqref{eq:fij} (noting the coordinate shift \eqref{eq:shift}).

\section{Resummations at the unitarity bound}\label{app:subtlety}

In this appendix, we show how the predictions at the unitarity bound 
are turned into constraints on the reduced correlator ${\cal H}$. 
In ${\cal N}=4$ SYM, this statement is obvious 
because there is an algebraic relation between superconformal blocks
corresponding to exchanged operators at the 
unitary bound \cite{Doobary:2015gia,Dolan:2006ec}, known as recombination. More precisely,  
take $\tau=2a+b+2$ then,
\begin{equation}
{B}^{\cal L}_{\gamma,[\lambda,1^{\lambda'-1}]}={B}^{}_{\gamma,[\lambda,1^{\lambda'-1}]} + {B}^{}_{\gamma+2,[\lambda-1,1^{\lambda'}]}  \sim  
 \frac{ \prod_{ij}(x_i-y_j) }{(y_1 y_2)^2} {\cal H}|_{\tau=\gamma, l=\lambda-2,a=\lambda'-1,b=\gamma-2\lambda'},
\end{equation}
where ${\cal L}$ denotes an exchanged long multiplet. The right-hand-side can 
be directly rewritten as a prediction for the reduced correlator. 

In the six-dimensional $\cN=(2,0)$ theory, however, there is no such simple relation.
The brute-force approach to imposing the OPE predictions 
at the unitarity bound is to apply $\Upsilon_2$ to ${\cal H}_{\vec{p}}$ 
and consider the \emph{super}conformal block expansion of the full correlator, namely
\begin{equation}
 \FF^{}_{\vec{p}} + \Upsilon_{2}\circ\CH^{}_{\vec{p}}\, .
 \end{equation}
 This procedure seems to undermine the usefulness of introducing a reduced correlator. 
 However, as we show below, there is a simple way around this.
 We focus on this six-dimensional case and therefore set $\theta=2$ 
 throughout the remainder of this section.

We recall the predictions at the unitarity bound fix the entire Mellin amplitude \eqref{eq:refinedM} in $\cN=4$ SYM and most of it in the six-dimensional $\cN=(2,0)$ theory so this discussion is key. We also note that the discussion below is independent of the large-$c$ expansion.

We wish to insert the OPE prediction data, see e.g.~\eqref{eq:prednonlog}, 
into the superconformal block expansion of the correlator.
The semi-short sector corresponds to one-row Young diagrams $\ula=[\lambda]$ with $\lambda\geq 2$ \eqref{tableQN1}, in which case, the superconformal blocks read \cite{Aprile:2021pwd}:
\begin{equation}
    B_{\gamma,\ula}(x_1,x_2,y_1,y_2)= \frac{ (x_1 x_2)^2}{y_1 y_2} \sum_{\mu\geq \lambda} (T_\gamma)_{[\lambda]}^{[\mu]} P_{[\mu]}(x_1,x_2,y_1,y_2),
\end{equation}
where $T_\gamma$ are the coefficients of the (one-row) super-Jack polynomials $P_{[\mu]}$, which are normalised as $(T_{\gamma})_{[\lambda]}^{[\lambda]}=1$ and are given by the recursion \eqref{eq:Tgamma}. 
We can write super-Jack polynomials 
explicitly as \eqref{onerowjack},
\begin{equation}\label{short_JacK}
    P_{[\mu]} (x_1,x_2,y_1,y_2)=  D_2\Bigg[ \frac{(x_1-y_1)(x_1-y_2)\big(-\frac{x_1^{\mu+1}}{\mu+1}\big)- (x_1\leftrightarrow x_2)}{x_1-x_2}\Bigg].
\end{equation}
Recall that super-Jack polynomials are defined to solve the
superconformal Ward identities, so \eqref{short_JacK} is an explicit solution for one-row diagrams.

In the above form, inserting any unitarity bound data $\CC_{\vec{p}}|_{\gamma,[\lambda]}$, in our case \eqref{eq:prednonlog}, into the superconformal block expansion of the correlator yields
\begin{equation}\label{eq:resumedshortpred}
\left. \FF^{}_{\vec{p}} + \Upsilon_{2}\circ\CH^{}_{\vec{p}} \right|_{\text{semi-short}}= 
\frac{(x_1 x_2)^2}{y_1 y_2} D_2 \Bigg[ \frac{(x_1 -y_1)( x_1-y_2) h(x_1) - (x_1\leftrightarrow x_2)}{x_1-x_2} \Bigg],
\end{equation}
where $h(x)$ is given by
\begin{equation}\label{eq:hx}
    h(x)= \sum_{\lambda\geq 2} \left.\CC_{\vec{p}}\right|_{\gamma,[\lambda]} \sum_{\mu \geq \lambda} (T_\gamma)^{[\mu]}_{[\lambda]} \left(-\frac{x^{\mu+1}}{\mu+1} \right) :=\sum_{\lambda\geq 2} \left.\CC_{\vec{p}}\right|_{\gamma,[\lambda]} h_\lambda(x).
\end{equation}
The main feature of $h$ we need is
\begin{equation}
    h(x) = O(x^{3}), \qquad \text{as } x\to 0.
\end{equation}
The fact that $h(x)$  starts at order $x^3$ follows from considering 
$\ula=[2+l]$ and can be seen by using the explicit Jack polynomials in \eqref{short_JacK}.

We want to understand the relationship between $h(x)$ in \eqref{eq:hx}  
and the reduced correlator. To do so let us focus on the poles at
the unitarity bound that appear in the Mellin amplitude 
representation of $\HH^{}_{\vec{p}}$, which give a result of the following form,
\begin{equation}\label{eq:Hschematic}
    \HH_{\vec{p}} = \frac{f(V)}{U} + O(U^0).
\end{equation}
Let us make the ansatz
\begin{equation}\label{eq:fVansatz}
    f(V)= \sum_{i\geq 0} c_i x_1^i + x_2\sum_{i\geq 0} d_i x_1^i + O(x_1^2).
\end{equation}
Applying $\Upsilon_2$ to \eqref{eq:Hschematic} 
we would like to rewrite the above result as in \eqref{eq:resumedshortpred}. 
We match the order $x_2^2$ and $x_2^3$ contributions of \eqref{eq:resumedshortpred} as follows,
\begin{equation}\label{eq:lemma}
\Upsilon_2\circ\CH_{\vec{p}}=\frac{(x_1x_2)^2}{y_1 y_2}\,D_2\!\left[ \frac{( x_1 -y_1)( x_1-y_2) h(x_1) - (x_1\leftrightarrow x_2)}{x_1-x_2}\right]+O(x_2^4).
\end{equation}
This is an equation we can solve for the coefficients $c_i$ and $d_i$ \eqref{eq:fVansatz} entering the above at orders $x_2^2$ and $x_2^3$, respectively. 
Before we do so, it is worth stating the key arguments that lead to \eqref{eq:lemma}:
\begin{itemize}
    \item First note the requirement that $h(x)=O(x^3)$. Indeed, $h(x)=x$ and $h(x)=x^2$ correspond to quarter-BPS exchanges and half-BPS exchanges, respectively. For such exchanges, there is no regular $f(V)$ \eqref{eq:fVansatz} which could reproduce the superblock expansion \eqref{eq:resumedshortpred} (when allowing $\lambda <2$). Thus, $\cal H$ never contributes to the half- and quarter-BPS multiplets at $\gamma=\gamma_m+2$; these are entirely accounted for by $\FF$, as used
in section \ref{sec:F}.

\item The $\gamma=\gamma_m+4$ sector, which contributes two-row (mostly long) blocks starts at order $(x_2)^4$. Indeed, blocks in this sector carry the prefactor $U^\theta \hat{\sigma}^2=(x_1 x_2)^4 (y_1y_2)^2$. Similarly, we note the $\gamma=\gamma_m$ sector starts at order $x_2^0$, but only contributes half-BPS operators, which are not included in $\HH$ as explained above. Therefore, the orders $(x_2)^2$ and $(x_2)^3$ in \eqref{eq:lemma} are entirely given by the one-row blocks, and thus $h(x)$ in \eqref{eq:hx}, at $\gamma=\gamma_m+2$.
\end{itemize}

Solving \eqref{eq:lemma} at both orders we  find
\begin{equation}
    c_i = \frac{h^{(i+3)}(0)}{(i+3)!}, \qquad \quad d_i = \frac{h^{(i+3)}(0)}{(i+3)!},
\end{equation}
which is nothing but the Taylor expansion of $h(x)$,
so 
\begin{equation}\label{eq:UBresult}
    f(V)= \frac{1}{x_1^3} h(x_1) + \frac{x_2}{x_1^3} h(x_1) + O(x_2^2).
\end{equation}
Therefore, the prediction from the semi-short OPE data of \eqref{eq:prednonlog} can be turned directly into a prediction for $\HH^{}$ in \eqref{eq:resumedshortpred} without applying $\Upsilon_2$. 
In the notation of section \ref{sec_results}, 
\begin{equation}
    \CH^{(\frac32)}_{u\textrm{-channel}}\big|_{\rm OPE}=\frac{1}{x_1 x_2}h^{(u)}_{-1}(x_1)
\end{equation}
with
\begin{equation}
    h^{(u)}_{-1}(x)=\frac{1}{x^3}\sum_{l\geq 0} \left. \CC_{\vec{p}}\right|_{\gamma,[2+l]}h_{[2+l]}(x).
\end{equation}
This function is resummed as a function of $x$.

In practice, the resummations are performed in \texttt{Mathematica}. Let us briefly describe the process. For each spin $l$ the coefficients $\CC_{\vec{p}}|_{\ula=[2+l]}$ are obtained from
\eqref{eq:prednonlog} using the evaluation formula of section \ref{sec:cauchy} for
the free-like inputs; see in particular section \ref{subsec:applyingcauchy} for explicit formulae.
We then compute the Taylor coefficients $k_n$  of
\begin{equation}
    h(x)=\sum_l \CC_{\vec{p}}|_{\gamma,[2+l]}h_{[2+l]}(x)=\sum_{n} k_n x^n,
\end{equation}
up to some high order. Since $h_{[2+l]}=O(x^{3+l})$, the
coefficient $k_n$ only receives contributions from $l\le n$, and the truncation is exact.
For fixed $q$ and $p$ we find that
\begin{equation}
    \frac{(n+4)_{2q}}{(n+1)}k_n 
\end{equation}
is a polynomial in $n$, which can be
fitted, 
and resummed in closed form. The result
is a rational function of $x$ in the basis $\{1,\log(1-x)\}$, as anticipated in
section \ref{sec_results}.

\end{appendix}

\bibliographystyle{JHEP}
\bibliography{multiparticleevenD}

\end{document}